\documentclass[11pt]{article}

\usepackage[papersize={8.5in,11in}, margin=1in]{geometry}

\usepackage{times}
\usepackage{url}
\usepackage{hyperref}
\hypersetup{colorlinks=true, citecolor=blue, linkcolor=red, urlcolor=blue}
\usepackage[utf8]{inputenc}
\usepackage{caption}
\usepackage{graphicx}
\usepackage{amsmath,amssymb,amsthm}
\usepackage{xspace}
\usepackage{amsfonts}
\usepackage{booktabs}
\usepackage{comment}
\usepackage[numbers]{natbib}
\usepackage{color}
\usepackage{xfrac}
\usepackage{tikz}
\usepackage{pgfplots}
\pgfplotsset{compat=1.18}
\usetikzlibrary{patterns,arrows.meta}
\usepackage{pgfplotstable}
    \pgfplotsset{
    name nodes near coords/.style={
        every node near coord/.append style={
            name=#1-\coordindex,
            alias=#1-last,
        },
    },
    name nodes near coords/.default=coordnode
    }

\usepackage{subcaption}
\usepackage{enumitem}

\usepackage{float}
\floatstyle{ruled}
\newfloat{algorithm}{tbp}{loa}
\floatname{algorithm}{Algorithm}
\makeatletter
\newcounter{ALG@line}
\newcount\ALG@indent
\newlength{\algorithmicindent}
\newcommand{\ALG@emit}[1]{%
  \refstepcounter{ALG@line}%
  \item[{\makebox[2.2em][r]{\ifnum\ALG@numberfreq>0\arabic{ALG@line}:\fi}}]%
  \begingroup
    \@tempdima=\algorithmicindent
    \multiply\@tempdima by \ALG@indent
    \hspace*{\@tempdima}%
  \endgroup
  #1%
}
\newenvironment{algorithmic}[1][0]{%
  \def\ALG@numberfreq{#1}%
  \setcounter{ALG@line}{0}%
  \ALG@indent=0
  \begin{list}{}{%
    \setlength{\leftmargin}{3.25em}%
    \setlength{\labelwidth}{2.5em}%
    \setlength{\labelsep}{0.35em}%
    \setlength{\itemsep}{0.15ex}%
    \setlength{\parsep}{0pt}%
    \setlength{\topsep}{0.5ex}%
  }%
}{\end{list}}
\newcommand{\State}{\ALG@emit{}}
\newcommand{\Require}{\item[\textbf{Input:}]}
\newcommand{\Ensure}{\item[\textbf{Output:}]}
\newcommand{\If}[1]{\ALG@emit{\textbf{if} #1 \textbf{then}}\advance\ALG@indent by 1}
\newcommand{\ElsIf}[1]{\advance\ALG@indent by -1\ALG@emit{\textbf{else if} #1 \textbf{then}}\advance\ALG@indent by 1}
\newcommand{\Else}{\advance\ALG@indent by -1\ALG@emit{\textbf{else}}\advance\ALG@indent by 1}
\newcommand{\EndIf}{\advance\ALG@indent by -1\ALG@emit{\textbf{end if}}}
\newcommand{\While}[1]{\ALG@emit{\textbf{while} #1 \textbf{do}}\advance\ALG@indent by 1}
\newcommand{\EndWhile}{\advance\ALG@indent by -1\ALG@emit{\textbf{end while}}}
\newcommand{\For}[1]{\ALG@emit{\textbf{for} #1 \textbf{do}}\advance\ALG@indent by 1}
\newcommand{\EndFor}{\advance\ALG@indent by -1\ALG@emit{\textbf{end for}}}
\newcommand{\Return}{\textbf{return}\xspace}
\newcommand{\Comment}[1]{\hfill$\triangleright$\,\textit{#1}}
\makeatother
\makeatletter
\providecommand*{\theHALG@line}{}
\renewcommand*{\theHALG@line}{\thealgorithm.\arabic{ALG@line}}
\makeatother

\usepackage{setspace}
\makeatletter
\renewcommand\paragraph{\@startsection{paragraph}{4}{\z@}%
  {3.25ex \@plus 1ex \@minus .2ex}%
  {-1em}%
  {\normalfont\normalsize\bfseries}}
\makeatother

\newcommand{\MMS}{\mathsf{MMS}}

\newcommand{\bP}{\mathbf{P}}
\newcommand{\bv}{\mathcal{V}}

\newcommand{\cI}{\mathcal{I}}

\newcommand{\bagFill}{\mathtt{ALG}}

\newcommand{\largestwater}{\Delta}

\newcommand{\reduction}[1]{R^{\alpha}_{#1}}

\newcommand{\bW}[1][]{\mathbf{W}^{#1}}
\newcommand{\bB}[1][]{\mathbf{B}^{#1}}

\usepackage[capitalize]{cleveref}

\newtheorem{theorem}{Theorem}[section]

\newtheorem{lemma}[theorem]{Lemma}

\theoremstyle{definition}
\newtheorem{example}[theorem]{Example}
\theoremstyle{plain}
\newtheorem{definition}[theorem]{Definition}
\newtheorem{proposition}[theorem]{Proposition}

\title{A Strongly Polynomial-Time 7/9-Approximation Algorithm for Maximin Share Allocations}

\author{Xin Huang\thanks{{\tt 
 Xin Huang is supported by Wakate Grant Number 26K21169 and JST ERATO Grant Number JPMJER2301,
Japan}}  \\huangxin@inf.kyushu-u.ac.jp \\ Kyushu University, Fukuoka, Japan
	\and Shengwei Zhou\\s.arthur.zhou@gmail.com \\ Nanyang Technological University, Singapore}
 
\date{}

\begin{document}
\pagenumbering{roman}
\maketitle

\begin{abstract}
We study maximin share (MMS) allocations of indivisible goods under nonnegative additive valuations. 
We give a strongly polynomial-time algorithm that allocates to every agent a bundle worth at least $7/9$ of her MMS, improving the previous $10/13$ guarantee of~\cite{conf/soda/HeidariKSS26}. 
Our analysis uses a dynamic witness-allocation framework that tracks the remaining goods from each agent’s perspective throughout the allocation process.
We obtain the computational result by iteratively refining upper bounds on MMS values, using only polynomially many runs of the allocation algorithm.
 Finally, we formalize and verify our main results in Lean 4.\footnote{The accompanying code is available at \href{https://github.com/ARTHURCHOU/EconCSLib-7-9MMS}{here}.}
\end{abstract}
\clearpage
\tableofcontents
\clearpage
\pagenumbering{arabic}

\section{Introduction}
Fair allocation of indivisible goods is a fundamental problem in economics and theoretical computer science. 
We consider allocating $m$ indivisible goods among $n$ agents with nonnegative additive valuations, seeking an allocation that guarantees every agent a fair share.
Indivisibility can preclude classical fairness guarantees such as proportionality~\cite{steihaus1948problem} and envy-freeness~\cite{foley1967resource}, motivating alternative benchmarks.

A central benchmark is the \emph{maximin share} (MMS), introduced by Budish~\cite{conf/bqgt/Budish10}. 
An agent's MMS is the largest value she can guarantee by partitioning the goods into as many bundles as there are agents and receiving the least-valued bundle. Although exact MMS allocations need not exist even for additive valuations~\cite{journals/jacm/KurokawaPW18,conf/wine/FeigeST21}, approximate guarantees are possible: an $\alpha$-MMS allocation gives every agent a bundle worth at least an $\alpha$ fraction of her MMS.

A long line of work has sought the largest approximation factor achievable for all additive instances. 
Early work established a $2/3$ guarantee~\cite{journals/jacm/KurokawaPW18,journals/talg/AmanatidisMNS17}, followed by a $3/4$ guarantee~\cite{conf/sigecom/GhodsiHSSY18}. 
Garg and Taki~\cite{journals/ai/GargT21} gave a strongly polynomial-time $3/4$-MMS algorithm and proved an existence guarantee above $3/4$, which was subsequently improved by~\cite{conf/ijcai/AkramiGST23}. 
More recently, Akrami and Garg~\cite{conf/soda/AkramiG24} established a guarantee exceeding $3/4$ by a constant independent of the number of agents. \cite{conf/soda/HeidariKSS26} improved the guarantee to $10/13$ and gave a polynomial-time algorithm achieving $10/13-\varepsilon$ for every fixed $\varepsilon>0$. 
Table~\ref{tab:mms-comparison} summarizes these developments.

\begin{table}[ht]
\centering
\caption{Selected MMS guarantees for additive goods.}
\label{tab:mms-comparison}
\begin{tabular}{@{}lll@{}}
\toprule
Paper & Existence guarantee & Algorithmic guarantee \\
\midrule
\cite{journals/jacm/KurokawaPW18,journals/talg/AmanatidisMNS17} & $2/3$ & $(2/3-\varepsilon)$ PTAS\\
\cite{conf/sigecom/GhodsiHSSY18} & $3/4$ & $(3/4-\varepsilon)$ PTAS \\
\cite{journals/ai/GargT21} & $3/4+1/(12n)$ & $3/4$ strongly polynomial \\
\cite{conf/ijcai/AkramiGST23} & $3/4+\min\{1/36,3/(16n-4)\}$ & --\\
\cite{conf/soda/AkramiG24} & $3/4+3/3836$ & -- \\
\cite{conf/soda/HeidariKSS26} & $10/13$ & $(10/13-\varepsilon)$ PTAS \\
This paper & $7/9$ & $7/9$ strongly polynomial \\
\bottomrule
\end{tabular}%
\end{table}

Alongside approximation quality, computational tractability is a central challenge.
Computing the MMS value is NP-hard, by its connection to classical partitioning and scheduling problems~\cite{books/fm/GareyJ79}.
Thus, an efficient allocation algorithm cannot in general begin by computing all MMS values exactly.
Existing approaches range from fixed-ratio polynomial-time algorithms to PTAS-type threshold procedures.
It is therefore important to improve the approximation guarantee while retaining polynomial running time without access to exact MMS values.

By definition, each agent can partition the original goods into $n$ bundles, each worth at least her MMS.
A natural way to analyze successive allocations is to maintain, for each unassigned agent, an analogous partition of the remaining goods into one bundle per unassigned agent, each still worth at least her original MMS.
Related residual-partition guarantees are central to lone-divider arguments and recent work on the residual maximin share~\cite{Feige2025RMMS,AkramiEtAl2026RMMS}.
However, a bundle allocated to one agent may be highly valuable to another, leaving too little total value to preserve such a partition for her.
The methodological challenge is to establish the final approximation guarantee even when these MMS-level partitions can no longer be maintained.

In this paper, we improve the approximation guarantee to $7/9$ and attain it with a strongly polynomial-time algorithm.
Our result addresses both the existence of stronger MMS approximations and their efficient computation.
To overcome the methodological challenge above, we develop a dynamic witness-allocation framework that maintains a private auxiliary allocation for each remaining agent throughout the process.

\subsection{Our Results and Contributions}
We first show that a $7/9$-MMS allocation always exists.
We establish this result through an algorithm that takes the agents' MMS values as input and assigns bundles to agents one at a time.
We then obtain a strongly polynomial-time algorithm by searching over target thresholds, starting from the agents' proportional shares.
We prove that the search keeps each target at least the corresponding MMS value and requires only polynomially many calls to the allocation algorithm.

\medskip
\noindent
\textbf{Main Result} (Theorem~\ref{recon-comp:thm:search})\textbf{.}
\emph{For nonnegative additive valuations, a complete $7/9$-MMS allocation can be computed in strongly polynomial time.}
\medskip

For the threshold search, we establish a stronger individual guarantee for the allocation algorithm.
Let $\MMS_i$ be agent $i$'s MMS value and $\alpha_i$ her target threshold.
Whenever $\alpha_i\le\MMS_i$, the algorithm gives her a bundle worth at least $7\alpha_i/9$, regardless of the other agents' thresholds.
Consequently, if a call fails, every unassigned agent must have $\alpha_i>\MMS_i$.
We describe the main technical components below.

\paragraph{Private witness allocations.}
To address the methodological challenge above, we analyze the allocation process from the perspective of a fixed agent $i$ with $0<\alpha_i\le\MMS_i$.
Adapting a partition-maintenance idea from the chore-allocation analysis of Huang and Lu~\cite{conf/sigecom/HuangL21}, we initialize a private \emph{witness allocation} using copies of the goods grouped according to an MMS partition for $i$.
While $k$ agents remain unassigned, we maintain $k$ witness bundles, allowing some bundles to have value below $\alpha_i$.

To connect the witness allocation to the real instance, we maintain \emph{value dominance}:
\begin{equation*}
    v_i(w_r)\le v_i(u_r)\qquad(r\ge1),
\end{equation*}
where the witness items $w_1,w_2,\ldots$ and the remaining real goods $u_1,u_2,\ldots$ are separately ordered by nonincreasing value to $i$.
This condition lets us use witness values to derive lower bounds on the remaining real goods.
The witness allocations are proof objects; the algorithm does not compute an MMS partition or perform the witness updates.

\paragraph{Maintaining witnesses across rounds.}
Whenever another agent receives a bundle, we delete suitable witness items and reorganize the remainder into $k-1$ bundles while preserving value dominance.
The deleted items may lie in different witness bundles, and the reorganization can leave some surviving bundles with value below $\alpha_i$.
We call a bundle's shortfall below this target its \emph{deficit}.
To maintain the witness through subsequent rounds, we distinguish large, medium-sized, and small witness items.
We track large items individually and use the value retained in medium-sized and small items to support exchanges between bundles.
In such an exchange, a bundle can receive a large item from another witness bundle and transfer some of its retained medium-sized or small-item value in return.
These exchanges allow us to compensate for losses, but must leave both bundles able to support subsequent updates.
We therefore control deficits together with the number and distribution of large items and the smaller-item value retained in each bundle.
Maintaining these conditions together with value dominance proves that agent $i$ eventually receives a bundle worth at least $7\alpha_i/9$, regardless of the other agents' targets.

\paragraph{A strongly polynomial-time algorithm.}
Recall that we obtain the existence result by running the allocation algorithm with the agents' MMS values as target thresholds.
To avoid computing these values, we instead initialize the targets at the agents' proportional shares, which provide natural upper bounds on their MMS values.
If a run leaves some agents unassigned, we lower their targets and rerun the algorithm on the original instance.
Our guarantee implies that an agent can remain unassigned only if her target exceeds her MMS value.
The challenge is to lower her target while keeping it at least her MMS value, which is~unknown.

During each run, we record whether each tested bundle is worth at least $7/9$ of an agent's current target.
We use these records to choose strictly smaller targets for unassigned agents while keeping them at least their MMS values.
The key observation is that, with the rules for choosing bundles and recipients fixed, preserving the results of all these tests preserves the entire execution.
We combine this observation with the individual guarantee to prove that the threshold updates are safe.
This argument does not rely on the details of witness maintenance.
Strict threshold decrease alone does not guarantee polynomially many restarts.
We use the structure of the bundles tested by the allocation algorithm to bound the number of new threshold values that can arise during the search.
Together with strict threshold decrease, this gives $O(n^5m^4)$ calls to the allocation algorithm and $O(n^7m^5)$ running time.

\subsection{Related Work}
Given the extensive literature on fair allocation, we focus on the lines of work most closely related to our results.
For broader surveys, we refer the reader to \cite{journals/sigecom/AzizLMW22} and \cite{journals/ai/AmanatidisABFLMVW23}.
For additive goods, the closest results are the guarantees for general instances summarized above.

A separate line of work obtains stronger guarantees in restricted settings, such as a small number of agents or items, or structured valuation classes.
For example, exact MMS allocations always exist for two agents, and stronger approximation ratios are known for three and four agents \cite{journals/jacm/KurokawaPW18, journals/talg/AmanatidisMNS17, journals/tcs/GourvesM19, journals/corr/abs-2205-05363, conf/sigecom/GhodsiHSSY18, conf/ijcai/ShahkarG25}.
Existence results have also been established for instances with few items and for certain structured valuations \cite{journals/talg/AmanatidisMNS17, conf/wine/FeigeST21, Feigetwoitemvalues}.

\paragraph{Chore Allocation.}
The dual problem of allocating \emph{chores} (items with non-positive values) has also seen rapid development.
For chores, approximation ratios are stated in the standard cost-minimization convention: a $\rho$-approximation with $\rho\ge 1$ means that every agent receives cost at most $\rho$ times her MMS cost.
Aziz, Rauchecker, Schryen, and Walsh~\cite{conf/aaai/AzizRSW17} initiated the study of MMS allocations for chores by extending the MMS definition and proving the first approximation result with a ratio of $2$.
The approximation ratio for chores has since been improved to $4/3$~\cite{journals/teco/BarmanK20}, $11/9$~\cite{conf/sigecom/HuangL21}, and then $13/11$~\cite{conf/sigecom/HuangS23}.
A notable distinction in the chores literature is its explicit connection to bin packing and job scheduling, including the use of classical algorithms such as First Fit Decreasing (\texttt{FFD}) and \texttt{MultiFit}~\cite{conf/sigecom/HuangL21, conf/sigecom/HuangS23, conf/aaai/GargHS25}.

\paragraph{Generalizations and Variations.}
\begingroup\sloppy
Beyond the additive model, researchers have explored ordinal preferences~\cite{conf/ijcai/AmanatidisBM16,conf/ijcai/0002021, journals/mp/AzizLW24,conf/sigecom/FeigeH23} and weighted MMS for agents with asymmetric entitlements~\cite{journals/jair/FarhadiGHLPSSY19,conf/ijcai/0001C019, conf/ijcai/00020L24,seddighin2025fair,li2025constant}.
The problem has also been extended to online settings in which items or agents arrive over time~\cite{conf/icml/0002B023,journals/corr/abs-2507-14039,journals/corr/abs-2505-24503,conf/sigecom/KulkarniMS25}.
Related notions such as the \emph{AnyPrice Share} (APS)~\cite{journals/mor/BabaioffEF24,conf/sigecom/FeigeH23} have also been proposed and studied.
Babaioff and Feige~\cite{conf/stoc/BabaioffF25} introduced a unifying framework for share-based fairness that encompasses both MMS and APS under arbitrary entitlements.
\par\endgroup

\section{Preliminaries}\label{sec:preliminaries}
We denote a fair division instance by $\cI=(N,M,\bv)$.
Here $N=\{1,2,\ldots,n\}$ is the set of agents, $M=\{e_1,e_2,\ldots,e_m\}$ is the set of goods, and $\bv=(v_1,v_2,\ldots,v_n)$ is the vector of valuation functions.
For each agent $i\in N$, the function $v_i:2^M\to\mathbb{R}_{\ge0}$ gives her value for every subset of goods.
We focus on additive valuations: for every agent $i$ and every set $S\subseteq M$, we have $v_i(S)=\sum_{g\in S}v_i(g)$.
For convenience, we write $v_i(g)$ instead of $v_i(\{g\})$ for a singleton item.
We refer to a subset of goods as a bundle.

For a set of items $S$, a \emph{$k$-partition} of $S$ is a collection of $k$ disjoint bundles $\{P_1,\ldots,P_k\}$ whose union is $S$.
We denote by $\Pi_k(S)$ the set of all $k$-partitions of $S$.
An \emph{allocation} $\bB=(B_1,\ldots,B_n)$ is an $n$-partition of $M$, where $B_i$ is the bundle allocated to agent~$i$.

\begin{definition}[Maximin Share]
    The \emph{maximin share} of agent $i$ is
    \begin{align*}
        \MMS_i = \max_{\bP \in \Pi_n(M)}\min_{P_j \in \bP} v_i(P_j)
    \end{align*}
\end{definition}

For each agent $i\in N$, let $\bP^i$ be an $n$-partition of $M$ that attains her MMS value, i.e., $v_i(P)\ge\MMS_i$ for every $P\in\bP^i$.
We refer to any such partition $\bP^i$ as an \emph{MMS partition} of agent $i$.

\begin{definition}[$\alpha$-MMS]
    For $\alpha\le1$, an allocation $\bB$ is an \emph{$\alpha$-MMS allocation} if $v_i(B_i)\ge\alpha\cdot\MMS_i$ for every agent $i\in N$.
\end{definition}

Barman and Krishnamurthy~\cite{journals/teco/BarmanK20} showed that approximate MMS allocation on general instances can be reduced to approximate MMS allocation on \emph{ordered} instances, where all agents share a common item order.

\begin{definition}[Ordered instance]
     An instance is \emph{ordered} if all agents have the same nonincreasing ordering of the goods.
     Equivalently, there is an ordering $(e_1,e_2,\ldots,e_m)$ such that, for every agent $i\in N$,
     \[
        v_i(e_1)\ge v_i(e_2)\ge\cdots\ge v_i(e_m).
     \]
\end{definition}

\begin{lemma}[{\cite[Section~3.1]{journals/teco/BarmanK20}}]
\label{recon-comp:lem:ordered}
For any instance $\cI$, we can construct in polynomial time an ordered instance $\hat{\cI}=(N,\hat M,\hat{\bv})$, where $\hat{\bv}=(\hat v_i)_{i\in N}$, with the same MMS value for each agent.
We can convert any allocation $\hat{\bB}$ of $\hat{\cI}$ in polynomial time into an allocation $\bB$ of $\cI$ satisfying $v_i(B_i)\ge\hat v_i(\hat B_i)$ for every agent $i$.
\end{lemma}

Hence, throughout the analysis, we restrict attention to ordered instances.
We assume without loss of generality that the goods are ordered so that $v_i(e_1)\ge v_i(e_2)\ge\cdots\ge v_i(e_m)$ for every agent $i\in N$.
Fix this common order permanently.  Every residual order inherits it,
so equal-valued real goods keep the same tie order after deletions.

\subsection{Reduction Rules}
Reduction rules are a standard tool in MMS algorithms~\cite{journals/jacm/KurokawaPW18, journals/ai/GargT21, conf/soda/AkramiG24, conf/soda/HeidariKSS26}.
They identify agents that can be safely satisfied by small structured bundles, thereby reducing the problem size while preserving the target guarantee for the remaining agents.
In our threshold algorithm, reductions are applied to the current residual instance, not only as a preprocessing step.
Let $A\subseteq N$ be the set of active agents, let $U\subseteq M$ be the set of unallocated goods, and let $k=|A|$.
Since the instance is ordered, we index the residual goods as $u_1,u_2,\ldots$ in the common nonincreasing order for all agents in $A$.
Whenever an indexed position is needed beyond the number of remaining
real goods, we extend the indexed sequence by zero-valued dummy goods.
These dummies are not members of the finite set $U$ and are not counted
in $|U|$.  This convention does not affect any MMS value or
approximation guarantee.
The algorithm is given target thresholds $(\alpha_i)_{i\in N}$, and a reduction allocates a small bundle to an active agent $i$ only when that agent values the bundle at least $(7/9)\alpha_i$.

\begin{definition}[Residual reduction rules]
For a residual ordered instance $(A,U)$ with $k=|A|$ and thresholds $(\alpha_i)_{i\in A}$, the reduction rules $R_0^{\alpha}$, $R_1^{\alpha}$, $R_2^{\alpha}$, and $R_3^{\alpha}$ are defined as follows.
\begin{itemize}
    \item \textbf{$R_0^{\alpha}$:} If $v_i(\{u_1\}) \ge (7/9)\alpha_i$ for some $i \in A$, allocate $\{u_1\}$ to agent~$i$.
    \item \textbf{$R_1^{\alpha}$:} If $v_i(\{u_k,u_{k+1}\}) \ge (7/9)\alpha_i$ for some $i \in A$, allocate $\{u_k,u_{k+1}\}$ to agent~$i$.
    \item \textbf{$R_2^{\alpha}$:} If $v_i(\{u_{2k-1}, u_{2k}, u_{2k+1}\}) \ge (7/9)\alpha_i$ for some $i \in A$, allocate this triple to agent~$i$.
    \item \textbf{$R_3^{\alpha}$:} If $v_i(\{u_{3k-2}, u_{3k-1}, u_{3k}, u_{3k+1}\}) \ge (7/9)\alpha_i$ for some $i \in A$, allocate these four items to agent~$i$.
\end{itemize}
After any rule is applied, the chosen agent is removed from $A$, the chosen real goods are removed from $U$, and the residual goods are re-indexed.
\end{definition}

The ordered numerical consequences of failed reduction rules that are
used in the proof are collected once, in
Lemma~\ref{lem:reduction-bounds}.

\paragraph{Classical validity.}
In the classical analysis, a reduction is \emph{valid} if it preserves feasible partitions at the target thresholds.
Specifically, suppose every active agent $i$ can partition $U$ into $k$ bundles, each worth at least $\alpha_i$ to her.
After a valid reduction, every remaining agent can partition the remaining goods into $k-1$ bundles, each still worth at least her threshold.

\begin{lemma}[Reduction bounds]
\label{lem:reduction-bounds}
The inapplicability of $\reduction{1}$,
$\reduction{2}$, and $\reduction{3}$ gives, respectively,
\[
 v_i(u_r)<\frac{7}{18}\alpha_i\ (r\ge k+1),\qquad
 v_i(u_r)<\frac{7}{27}\alpha_i\ (r\ge2k+1),\qquad
 v_i(u_r)<\frac{7}{36}\alpha_i\ (r\ge3k+1).
\]
The inapplicability of $\reduction{2}$ also gives
\[
    v_i(u_{2k}+u_{2k+1})<\frac{14}{27}\alpha_i.
\]
\end{lemma}


\section{Algorithm and Threshold Guarantee}\label{sec:overview}
We now present the main algorithm.
The algorithm maintains the set $U$ of unallocated goods and the set $A$ of unsatisfied agents.
At the beginning of each round, it sets $k=|A|$ and re-indexes $U$ as $u_1,u_2,\ldots$ in the common nonincreasing order.

The pseudocode marks three branches: pair, triple, and bag filling.
The algorithm selects the pair branch if some active agent finds $\{u_1,u_{k+1}\}$ satisfactory.
Otherwise, it selects the triple branch if some active agent finds $\{u_1,u_{k+1},u_{2k+1}\}$ satisfactory, and the bag-filling branch if both tests fail.

Each branch first tests its listed reductions: $\reduction{0},\reduction{1},\reduction{2}$ in the pair branch, $\reduction{2}$ in the triple branch, and $\reduction{3}$ in the bag-filling branch.
An applicable reduction ends the current round immediately.
After every allocation, the algorithm updates $A$ and $U$, re-indexes the remaining goods, and tests all branch conditions again.
We defer $\reduction{3}$ until the bag-filling branch because its witness maintenance uses the bounds supplied by the failed triple test.


If no reduction applies, the pair and triple branches seek an efficient way to satisfy an active agent using $u_1$ together with one or two additional goods, respectively. Both branches favor less valuable additional goods whenever possible, preserving more valuable goods for later rounds. The bag-filling branch starts with $\{u_1,u_{k+1},u_{2k+1}\}$ and adds the remaining goods in order until some active agent finds the bundle satisfactory or all goods are exhausted.

\begin{algorithm}[H]
\caption{$\bagFill(\cI, (\alpha_i)_{i\in N})$}
\label{alg:pebble-and-water}

\begin{algorithmic}[1]
\Require Ordered instance $\cI = (N, M, \bv)$ and thresholds $(\alpha_i)_{i\in N}$.
\Ensure $(\textsc{Success},\mathcal B=(B_i)_{i\in N})$ or $(\textsc{Failure},F)$.

\State $A\gets N$; $U\gets M$; $B_i\gets\varnothing$ for every $i\in N$.

\While{$A\neq\emptyset$}
    \State $k\gets |A|$; 
    \If{$\exists i \in A, v_i(u_1 + u_{k+1}) \ge \frac{7}{9}\alpha_i$}\Comment{Pair Branch}
        \If{$\reduction{0}$, $\reduction{1}$, or $\reduction{2}$ applies}
            \State Apply the lowest-index applicable reduction to an
            eligible agent $i$.
            \State Update $A,U$, and \textbf{continue}.
        \EndIf
        \State $\ell\gets\max\{j:\exists i\in A,\ v_i(u_1+u_j)\ge \frac79\alpha_i\}$. 
        \State $T \gets \{u_1, u_\ell\}$. 

    \ElsIf{$\exists i \in A, v_i(u_1 + u_{k+1} + u_{2k+1}) \ge \frac{7}{9}\alpha_i$}\Comment{Triple Branch}
        \If{$\reduction{2}$ applies}
            \State Apply $\reduction{2}$ to an eligible agent $i$.
            \State Update $A,U$, and \textbf{continue}.
        \EndIf
        \State $\ell\gets\max\{j:\exists i\in A,\ v_i(u_1+u_j+u_{\max\{j+1,2k+1\}})\ge \frac79\alpha_i\}$. 
        \State $T \gets \{u_1, u_\ell, u_{\max\{\ell+1,\,2k+1\}}\}$.

    \Else\Comment{Bag-filling Branch}
        \If{$\reduction{3}$ applies}
            \State Apply $\reduction{3}$ to an eligible agent $i$.
            \State Update $A,U$, and \textbf{continue}.
        \EndIf
        \State $T \gets \{u_1, u_{k+1}, u_{2k+1}\}$; $j\gets 3k+1$.
            \While{$j\le |U|$ and $\forall i\in A, v_i(T)<\frac79\alpha_i$ }
                \State $T\gets T\cup\{u_j\}$; $j\gets j+1$.
            \EndWhile
         \If{$T$ is not satisfactory}
            \State \Return $(\textsc{Failure},A)$.
        \EndIf
    \EndIf
    
    \State Choose an arbitrary $i\in A$ such that $v_i(T) \ge \frac{7}{9}\alpha_i$.
    \State $A \gets A \setminus \{i\}$; $U \gets U \setminus T$; $B_i \gets T$.
\EndWhile

\State \Return $(\textsc{Success},\mathcal B)$.
\end{algorithmic}
\end{algorithm}

The following theorem gives an individual threshold guarantee, which implies the existence of a $7/9$-MMS allocation.

\begin{theorem}\label{thm:7/9}
    For any ordered additive instance $\cI=(N,M,\bv)$, any threshold vector $(\alpha_i)_{i\in N}$, and any agent $i\in N$, if $\alpha_i\le \MMS_i$, then Algorithm~\ref{alg:pebble-and-water} allocates to agent $i$ a bundle $B_i$ such that
    \[
        v_i(B_i)\ge \frac79\alpha_i.
    \]
    Consequently, whenever Algorithm~\ref{alg:pebble-and-water}
    returns $(\textsc{Failure},F)$, every agent $i\in F$ satisfies
    $\alpha_i>\MMS_i$.
\end{theorem}

The proof framework uses a private witness allocation for every agent
whose threshold is feasible. Section~\ref{sec:analysis} introduces
witness allocations, item classes, proof stages, notation, and closure tools.
Appendix~\ref{apdx:running-example} illustrates witness maintenance in a three-agent example.
Section~\ref{sec:stage-analysis-overview} assembles these tools into the proof of Theorem~\ref{thm:7/9}.
Stage~2 ends immediately before the first ordinary bag-filling round,
which begins Stage~3. Applying the theorem with $\alpha_i=\MMS_i$ for
every agent yields a $7/9$-MMS allocation.

\section{Witness Allocations and Proof Tools}
\label{sec:proof-architecture}
\label{sec:analysis}
\label{sec:preparation}

The key idea in the proof of Theorem~\ref{thm:7/9} is to maintain a private \emph{witness allocation} for each remaining agent whose threshold is feasible. This allocation tracks the changes made by Algorithm~\ref{alg:pebble-and-water} and certifies that the agent can still receive a satisfactory bundle. It is used only in the analysis and is not accessed by the algorithm, so its maintenance need not be computationally efficient. This section develops the common tools used to maintain these witnesses; a worked example is given in Appendix~\ref{apdx:running-example}.

Fix an agent $i$ with a feasible threshold $\alpha_i\le\MMS_i$.
If $\alpha_i>0$, scale $v_i$ so that $\alpha_i=1$ and hence $\MMS_i\ge1$; the case $\alpha_i\le0$ is immediate.
Throughout the proof, all witness values are measured by $v_i$.
The witness allocation $\bW$ starts from an MMS partition for agent $i$. Since each initial bundle has value at least $1$, we may decrease witness-item values so that every initial bundle has value exactly $1$.

The proof follows agent $i$ until she receives a satisfactory real bundle. During this process, the maintenance procedure retains the value and item-count bounds needed for the applicable case. Whenever the algorithm allocates a bundle to another agent, the procedure takes that real bundle and the current witness allocation as input, reorganizes the witness allocation, and removes a suitable witness bundle to reflect the real allocation.


To make witness maintenance flexible, we allow the witness allocation to differ from the real instance in two ways:
\begin{itemize}
    \item A witness item may be \emph{solid} or \emph{fluid}. A solid item is kept whole in one witness bundle. A fluid item's value may be divided among witness bundles, with its portions summing to the item's current value.
    \item We may decrease the value of a witness item.
\end{itemize}
In either state, each witness item appears once, at its full current value, in the ordered witness sequence. Changing its state or redistributing its fluid value does not change that value or create additional witness items.

\subsection{Validity of Witness Allocations}
\label{ssec:witness}
We first describe the relationship maintained between the witness allocation and the real residual instance. The idea is to match each witness item to a distinct real item of at least the same value. To state this condition formally, write the real residual goods as $u_1,u_2,\ldots$ and all current witness items as $w_1,w_2,\ldots$, each in nonincreasing order of value.

\begin{definition}
    A witness allocation $\bW$ is \emph{valid} if $v_i(w_j)\le v_i(u_j)$ for every index $j$.
\end{definition}

In each sequence, index $1$ refers to the largest item, index $2$ to the second largest, and so on. The sequences are re-indexed after each update.

To preserve validity when the algorithm allocates a real bundle, we use the following notion.

\begin{definition}[Dominance bundle]
\label{def:dominance-bundle}
    Given a real bundle $T$, a witness bundle $W$ is a \emph{dominance bundle} for $T$ if there is an injective map $f:T\rightarrow W$ such that, for every $t\in T$, the index of $f(t)$ in the witness sequence is no larger than the index of $t$ in the real sequence. Each item $f(t)$ must be present whole in $W$.
\end{definition}

The following two operations preserve validity.

\begin{proposition}[Allowed witness-side operations]
\label{prop:remove-dominance}
Suppose that the current witness allocation is valid. It remains valid after either of the following operations and subsequent re-indexing:
\begin{enumerate}[label=(\roman*)]
    \item Decrease the value of a witness item.
    \item When Algorithm~\ref{alg:pebble-and-water} allocates a real bundle $T$, delete a dominance bundle for $T$ from the witness allocation.
\end{enumerate}
\end{proposition}

The proof is given in Appendix~\ref{apdx:common-witness}.



\paragraph{Standard Reductions.}
As a simple application, consider a reduction $\reduction r$ that allocates a bundle to another agent, where $r\in\{0,1,2,3\}$ and $k\ge2$.
The corresponding witness target is
\[
    T_r^{*}:=\{w_{rk-r+1},\ldots,w_{rk+1}\}.
\]
When the indexed witness items are available as whole items, the following three-step update suffices. All indices refer to the order before the update; zero-valued dummies may be placed in witness bundles as needed.

\begin{algorithm}[H]
\caption{Standard reduction witness update}
\label{alg:standard-reduction}
\begin{algorithmic}[1]
\Require Witness allocation $\bW$ with $k\ge2$ bundles and reduction index $r$.
\Ensure $k-1$ witness bundles, with exactly $T_r^{*}$ removed.
\State \textbf{Identify:} choose $X\in\bW$ containing at least $r+1$ items among $w_1,\ldots,w_{rk+1}$.
\State \textbf{Exchange:} swap each item in $T_r^{*}\setminus X$ into $X$, giving its holder a distinct item of $X$ with index less than $rk-r+1$.
\State \textbf{Delete:} distribute $X\setminus T_r^{*}$ among the other bundles, then remove $X$ from $\bW$ and re-index.
\end{algorithmic}
\end{algorithm}

The Pigeonhole Principle guarantees a suitable $X$. Each missing target item can be exchanged with an earlier item in $X$, so no other bundle loses value. After the exchanges, $X$ contains the entire witness target, which is a dominance bundle for the real reduction bundle. The last step removes exactly this target and preserves validity by Proposition~\ref{prop:remove-dominance}; distributing the remaining items of $X$ can only increase the values of the surviving bundles.

\subsection{Item Classes and Proof Stages}

To simplify the analysis of the maintenance procedure, we classify witness items by value.

\begin{definition}[Value classes]
\label{def:witness-value-classes}
Before melting, a witness item $e$ is classified as
\begin{itemize}
    \item  \emph{Pebble} if $v_i(e)\ge 2/9$;
    \item \emph{Ice} if $4/27\le v_i(e)<2/9$;
    \item \emph{Water} if $v_i(e)<4/27$.
\end{itemize}
\end{definition}

These value classes are distinct from the solid and fluid states. Pebbles remain solid throughout the maintenance. Water may be kept solid and may be made fluid whenever a water transfer requires it. Ice may be melted, meaning that it becomes fluid and is treated exactly like ordinary water; we call it water from then on, even if it is later collected whole. Thus solid ice means ice that has not been melted.

Changing states and transferring water do not change the ordered witness-item values unless a value decrease is explicitly specified. In particular, $\Delta$ uses each non-pebble item's full current value. A fluid item used in a dominance bundle or a triple $C_r$ is first collected whole in that bundle.


We next classify the rounds of the algorithm according to the branch selected and whether a reduction is applied.

\begin{definition}[Rounds]
\label{def:rounds}
A \emph{round} is one iteration of the main while-loop of Algorithm~\ref{alg:pebble-and-water}. Each round is classified as
\begin{itemize}
    \item a \emph{pair round} if $\exists a\in A,\ v_a(u_1+u_{k+1})\ge\frac79\alpha_a$;
    \item a \emph{triple round} if it is not a pair round and $\exists a\in A,\ v_a(u_1+u_{k+1}+u_{2k+1})\ge\frac79\alpha_a$;
    \item a \emph{bag-filling round} if $\forall a\in A,\ v_a(u_1+u_{k+1}+u_{2k+1})<\frac79\alpha_a$.
\end{itemize}
A round is further classified as a \emph{reduction round} if it applies a reduction, and as an \emph{ordinary round} otherwise.
\end{definition}

For example, a reduction pair round is a pair round in which a reduction is applied.
Using this classification, we divide the execution into three consecutive proof stages.

\begin{definition}[Proof stages]
\label{def:proof-stages}
The stages are blocks of consecutive rounds:
\begin{itemize}
    \item \textbf{Stage~1} begins with the first round and ends immediately before the first non-pair round.

    \item \textbf{Stage~2} begins immediately after Stage~1 and ends immediately before the first ordinary bag-filling round.

    \item \textbf{Stage~3} begins with the first ordinary bag-filling round and continues until the algorithm terminates.
\end{itemize}
\end{definition}
If a round marking a transition to the next stage never occurs, the current stage continues until termination and the subsequent stages are empty. A stage may also be empty if its starting and ending boundaries coincide.

Every executed round belongs to exactly one stage. These stages are conventions used in the proof, rather than variables maintained by the algorithm; their boundaries do not depend on the witness allocation or the pebble count.

\subsection{Common Notation}
We collect here the notation used throughout the analysis. Readers may refer back to this list as needed.

\begin{itemize}
    \item $v_i(W)$ is the total value of the solid items in witness bundle $W$ plus the fluid value assigned to it.
    \item $\largestwater$ is the largest value of a current non-pebble witness item, including both ice and water. We set $\largestwater=0$ if no such item remains.
    \item $P(W)$ is the number of pebbles in a witness bundle $W$.
    \item $S(W)$ is the number of pebbles and solid ice items in $W$.
    \item $F(W)$ is the total water value assigned to $W$, including both solid water items and fluid water shares. It excludes pebbles and solid ice.
    \item $D(W)=1-v_i(W)$ is the deficit of $W$, measured relative to its initial value of $1$.
\end{itemize}
For the entire witness allocation, we write
\[
    P(\bW)=\sum_{W\in\bW}P(W),
    \qquad
    S(\bW)=\sum_{W\in\bW}S(W).
\]
Thus $P(\bW)$ and $S(\bW)$ count the remaining pebbles and the remaining pebbles together with solid ice, respectively. All these quantities refer to the current witness allocation of the fixed agent $i$ and are updated after each witness operation.

\subsection{Two Closure Cases}
\label{sec:general-tools}
We identify two situations in which witness maintenance admits a uniform continuation through all remaining rounds. These closure cases take priority over the stage-specific cases considered later: once either case is reached, the corresponding procedure handles the rest of the execution until agent $i$ is satisfied.

In both cases, the procedure reorganizes the witness allocation as needed so that a bundle containing $w_1$ dominates the real bundle allocated in the current round. Deleting this witness bundle preserves validity, and the remaining items can be reorganized to retain the same closure structure for the next round.

\subsubsection{The canonical case}
\label{subsec:canonical-definition}
The first closure case organizes the witness items around three interleaved sequences of ranks. A sufficient value margin in every bundle then allows the remaining non-pebble items to be used in the maintenance procedure.

\begin{definition}[Water-Level Condition]
\label{def:water-level-condition}
A witness allocation satisfies the \emph{Water-Level Condition},
abbreviated \textnormal{\textsc{WLC}}, if
\[
    v_i(W)>7/9+\largestwater
    \qquad\text{for every }W\in\bW.
\]
\end{definition}

We use \textnormal{\textsc{WLC}} as a common invariant in the maintenance analysis. Its margin of $\largestwater$ is useful when a construction stops immediately after adding a non-pebble item.
For the current active count $k$, write
\[
    L:=7/9+\largestwater,
    \qquad
    C_r:=\{w_r,w_{k+r},w_{2k+r}\}
    \quad (r\in[k]).
\]

\begin{definition}[Canonical witness allocation]
\label{def:canonical-witness-allocation}
A witness allocation $\bW=(W_1,\ldots,W_k)$ is \emph{canonical} if it satisfies \textnormal{\textsc{WLC}} and $C_r\subseteq W_r$ for every $r\in[k]$.
\end{definition}

The following lemma gives sufficient conditions for entering the canonical case.

\begin{lemma}[Canonical reorganization]
\label{lem:canonical-entry}
Suppose that $\bW$ is valid, satisfies \textnormal{\textsc{WLC}}, and
\[
    v_i(C_1)\le L,
    \qquad
    P(\bW)\le3k.
\]
Then $\bW$ can be reorganized into a valid canonical witness allocation.
\end{lemma}

Once a canonical allocation is available, the next lemma provides the continuation. The reorganization is part of the maintenance procedure; the three items of $C_1$ alone need not dominate every possible real allocation.

\begin{lemma}[Canonical dominance]
\label{lem:canonical-packet}
\label{lem:stage3-closure}
Suppose that $\bW$ is valid and canonical, with $P(\bW)\le3k$.
The current round cannot return failure.
In any round that allocates a real bundle to another agent, the witness allocation can be reorganized so that $W_1$ is a dominance bundle for that real bundle.
After deleting $W_1$ and re-indexing, the remaining items can be reorganized into a valid canonical witness allocation with $k-1$ bundles and at most $3(k-1)$ pebbles.
\end{lemma}

Indeed, the initial real triple together with all available real tail items has value at least $v_i(W_1)>7/9$ by validity, so an ordinary bag-filling round cannot fail.
The full reorganization and maintenance constructions are given in Appendix~\ref{apdx:common-witness}.
Its application to Stage~3 is explained in Section~\ref{sec:stage-analysis-overview}.

\subsubsection{The trivial dense case}
\label{subsec:dense-closure}
The second closure case applies when there are enough pebbles and solid ice items to form $k$ four-item witness groups, each worth more than $7/9$ to the fixed agent $i$. Here the item counts themselves provide the required value guarantee.

\begin{definition}[Trivial dense case]
\label{def:dense-witness-state}
A witness allocation $\bW$ with $k$ bundles is in the \emph{trivial dense case} if
\[
    P(\bW)\ge3k,
    \qquad
    S(\bW)\ge4k.
\]
\end{definition}

These inequalities allow us to form $k$ disjoint four-item groups, each containing at least three pebbles and one further item worth at least $4/27$. Every group therefore has value at least
\[
    3\cdot\frac29+\frac4{27}=\frac{22}{27}>\frac79.
\]

\begin{lemma}[Dense closure]
\label{lem:stage2-dense-certificate}
Suppose that $\bW$ is valid and the trivial dense case applies.
The witness items can be regrouped so that
\[
    \{w_r,w_{k+r},w_{2k+r},w_{3k+r}\}\subseteq W_r
    \qquad(r\in[k]).
\]
The four-item group containing $w_1$ dominates the real bundle allocated in the current round. If agent $i$ remains active, deleting this group and regrouping the remaining items yields a valid witness allocation in the trivial dense case with $k-1$ bundles.
\end{lemma}

The first group dominates the pair and triple allocations, as well as the corresponding reductions. If the bag-filling branch is selected, $\reduction{3}$ is applicable: its first three indexed items are pebbles and its fourth is worth at least $4/27$. This reduction is also dominated by the first group.
Deleting the group removes four items counted by $S(\bW)$ and either three or four pebbles. A fourth pebble is removed only when $P(\bW)\ge3k+1$, so both count inequalities continue to hold with $k$ replaced by $k-1$.

\section{Overview of the Stage Analysis}
\label{sec:stage-analysis-overview}
We now explain how the stage-specific maintenance procedures prove Theorem~\ref{thm:7/9}. Fix an agent $i$ whose threshold is feasible. While she remains active, we keep a valid witness allocation with $k$ bundles, together with the bounds needed for the current maintenance case. The stage-specific procedures preserve these conditions until a closure case applies; canonical or dense maintenance then provides a uniform continuation. Figure~\ref{fig:stage-analysis} summarizes the dependencies.

\begin{figure}[H]
\centering
\captionsetup{font=normalsize}
\begin{tikzpicture}[
    x=1cm,y=1cm,
    font=\normalfont\normalsize,
    casebox/.style={draw=black!65,rounded corners=2pt,fill=white,
        text width=3.15cm,minimum height=1.3cm,align=center,
        inner sep=4pt},
    resultbox/.style={casebox,fill=black!6,line width=.8pt},
    transition/.style={draw=black!75,line width=.65pt,
        -{Stealth[length=2mm,width=1.4mm]}},
    proofedge/.style={transition,dashed},
    edgelabel/.style={fill=white,inner sep=2pt,align=center}
]
\draw[draw=black!65,rounded corners=3pt,fill=black!2]
    (.15,-1.55) rectangle (16.1,-5.2);
\node[anchor=west,font=\bfseries] at (.4,-1.85) {Stage 2};
\draw[draw=black!40,rounded corners=2pt,fill=black!1]
    (6.05,-2.0) rectangle (15.85,-4.2);
\node[font=\bfseries] at (11.0,-2.22) {With-ice case};

\node[casebox,text width=4.1cm] (s1) at (8.15,0)
    {\textbf{Stage 1}\\Pair rounds\\Entrance conditions};
\node[casebox,text width=4.3cm] (noice) at (2.8,-3.3)
    {\textbf{No-ice case}\\Repair and maintenance\\
     Subsection~\ref{subsec:stage2-noice}};
\node[casebox] (clean) at (8.15,-3.3)
    {\textbf{Clean}\\Values $\ge1$\\
     Subsection~\ref{subsec:clean-ice}};
\node[casebox,text width=3.55cm] (critical) at (13.2,-3.3)
    {\textbf{Critical}\\Continues without ice\\
     Subsection~\ref{subsec:criticalstate}};

\coordinate (entrance) at (8.15,-1.15);
\draw[draw=black!75,line width=.65pt] (s1.south) -- (entrance);
\draw[transition] (entrance) -| (noice.north);
\draw[transition] (entrance) -- (clean.north);
\node[edgelabel] at (4.55,-.88) {no-ice entrance};
\node[edgelabel,anchor=west] at (8.35,-1.35) {with-ice entrance};

\draw[transition] (clean.east) --
    node[edgelabel,above] {tests fail} (critical.west);
\draw[transition] (clean.south) -- (8.15,-4.35)
    -- (2.8,-4.35) -- (noice.south);
\node[edgelabel] at (5.45,-4.75) {$\Delta<4/27$: repair};

\node[casebox,text width=3.55cm] (s3) at (2.8,-6.7)
    {\textbf{Stage 3}\\Canonical entry\\
     Lemma~\ref{lem:stage3-entry}};
\node[casebox,text width=4.3cm] (canonical) at (8.15,-6.7)
    {\textbf{Canonical continuation}\\
     Lemma~\ref{lem:canonical-packet}};
\node[casebox,text width=3.55cm] (dense) at (13.2,-6.7)
    {\textbf{Dense continuation}\\
     Lemma~\ref{lem:stage2-dense-certificate}};

\draw[transition] (2.8,-5.2) -- (s3.north);
\node[edgelabel,anchor=east,align=right,text width=2.4cm]
    at (2.55,-5.62) {first ordinary\\bag filling};
\draw[transition] (8.15,-5.2) -- (canonical.north);
\draw[transition] (13.2,-5.2) -- (dense.north);
\draw[transition] (s3.east) -- (canonical.west);

\node[resultbox,text width=6.6cm] (main) at (8.15,-8.45)
    {\textbf{Theorem~\ref{thm:7/9}}\\
     No failure while $i$ is active;\\
     $i$ receives value at least $7\alpha_i/9$};
\draw[proofedge] (s3.south) -- (main.north west);
\draw[proofedge] (canonical.south) -- (main.north);
\draw[proofedge] (dense.south) -- (main.north east);
\end{tikzpicture}
\caption{Maintenance cases and proof dependencies.
Solid arrows show case transitions; dashed arrows support the theorem.
Closures may take priority at any checkpoint.
The clean-to-no-ice handoff occurs before the next ordinary triple.}
\label{fig:stage-analysis}
\end{figure}

Stage~1 establishes the entrance conditions for the later analysis. Stage~2 begins at the first non-pair round and covers ordinary triples together with intervening pair and reduction rounds; such intervening rounds may also precede its first ordinary triple. Its procedures preserve the applicable bounds or enter a closure case. Stage~3 begins at the first ordinary bag-filling round. The following lemma reduces all subsequent rounds to canonical maintenance.

\begin{lemma}[Entry to Stage~3]
\label{lem:stage3-entry}
On entry to Stage~3, the witness allocation $\bW$ can be reorganized into a valid canonical witness allocation.
\end{lemma}
\begin{proof}[Proof sketch]
At the first ordinary bag-filling round, the failed triple test gives $v_i(C_1)<7/9$, and the failure of $\reduction{3}$ implies that at most $3k$ pebbles remain. Thus every item outside the triples $C_r$ can be treated as fluid water. The maintenance procedures below supply the value margin required by \textnormal{\textsc{WLC}}; in the critical case, the failed reduction additionally gives $\Delta<1/9$, while every bundle is worth at least $17/18$.

Each triple is therefore worth less than $L$, while the total witness value exceeds $kL$. Place each $C_r$ in its own bundle and distribute the remaining water to equalize the bundle values above $L$. This gives a valid canonical allocation without changing any item value. The complete verification of the maintenance cases is given in Appendix~\ref{apdx:stage3-entry}.
\end{proof}

\begin{proof}[Proof of Theorem~\ref{thm:7/9}]
Fix $i$ with $\alpha_i\le\MMS_i$. If $\alpha_i\le0$, nonnegativity gives the value guarantee, and the availability of $\reduction{0}$ precludes failure while $i$ remains active. Otherwise normalize $\alpha_i=1$. An MMS partition, with witness values decreased as necessary, gives a valid initial allocation of $n$ bundles worth $1$ each.

Induct on allocations to other agents. The procedures in Section~\ref{sec:stage1}, including Lemmas~\ref{lem:stage1-clean-ice}, \ref{lem:stage1-canonical}, and~\ref{lem:stage1-repair}, preserve validity through Stage~1. As specified at the start of Section~\ref{sec:stage2}, these initial pair and standard reduction updates continue until the first ordinary triple, unless closure or Stage~3 is reached earlier. At the first no-ice ordinary triple, Lemma~\ref{lem:stage2-repair} establishes the entrance conditions; the same repair is performed before an earlier no-ice $\reduction{2}$ that would reach equality. Appendix~\ref{apdx:clean-noice-handoff} covers a later clean-to-no-ice handoff. Thereafter, the ordinary-triple procedures and Lemma~\ref{lem:stage2-other-complete} cover all round types. An equality output continues by Lemma~\ref{lem:stage2-equality}, and critical maintenance continues when ice disappears. Each completed update removes one witness bundle, preserves validity, and retains the continuing bounds or establishes the next case's entrance conditions. Whenever canonical or dense closure is reached, Lemma~\ref{lem:canonical-packet} or~\ref{lem:stage2-dense-certificate} supplies all subsequent updates.

Only an ordinary bag-filling round can fail. The first such round starts Stage~3, where Lemma~\ref{lem:stage3-entry} gives canonical entry and Lemma~\ref{lem:canonical-packet} rules out failure. Every successful round removes an agent, so within at most $n$ allocations agent $i$ receives value at least $7/9$. Rescaling proves the guarantee; its contrapositive gives the failure certificate.
\end{proof}

\section{Stage 1}
\label{sec:stage1}
Stage~1 consists only of pair rounds. Reduction rounds use Algorithm~\ref{alg:standard-reduction}. For ordinary pair rounds, we distinguish three cases according to the end-of-stage witness conditions.
The real execution and the boundary of Stage~1 are independent of witness maintenance. The constructions below also delete the same indexed witness items without changing their values, so the remaining item values and pebble count at that boundary agree. Since the witness is used only in the proof, we may choose the appropriate construction with these end-of-stage conditions in mind; its solid and fluid states are determined by that construction.

\subsection{More than $2k$ pebbles with ice at the end of Stage~1}
When Stage~1 ends with $P(\bW)>2k$ and ice remains, we use the following exchange and deletion. Throughout these updates, witness items remain whole and their values are unchanged.

\begin{algorithm}[H]
\caption{Pair maintenance when Stage~1 ends with more than $2k$ pebbles and ice}
\label{alg:stage1-pair}
\begin{algorithmic}[1]
\Require A real pair $\{u_1,u_\ell\}$ allocated to another agent, and $k$ witness bundles.
\State Let $X$ contain $w_1$ and let $Y$ contain $w_\ell$.
\If{$X=Y$}
    \State Distribute $X\setminus\{w_1,w_\ell\}$ among the other bundles.
\Else
    \State Exchange $w_\ell$ for all items of $X\setminus\{w_1\}$.
\EndIf
\State Delete $X$.
\end{algorithmic}
\end{algorithm}

The update removes exactly the two witness items corresponding to the real pair. If they lie in distinct bundles, the surviving bundle is
\[
    (X\setminus\{w_1\})\cup(Y\setminus\{w_\ell\}).
\]
Validity is preserved, although the value of this survivor may decrease.

\begin{lemma}[Clean maintenance with ice]
\label{lem:stage1-clean-ice}
If Stage~1 ends with $P(\bW)>2k$ and an ice item remains, Algorithms~\ref{alg:stage1-pair} and~\ref{alg:standard-reduction} preserve validity and keep every witness bundle worth at least $1$ throughout Stage~1.
\end{lemma}

The proof is given in Appendix~\ref{apdx:stage1}.

\subsection{At most $2k$ pebbles at the end of Stage~1}
When Stage~1 ends with $P(\bW)\le2k$, our goal is to enter the canonical case at that point. The following maintenance retains \textnormal{\textsc{WLC}} throughout the preceding pair rounds, so that no further stage-specific maintenance is needed for this case.

\begin{algorithm}[H]
\caption{Pair maintenance when Stage~1 ends with at most $2k$ pebbles}
\label{alg:stage1-nomorethan2k}
\begin{algorithmic}[1]
\Require A real pair $\{u_1,u_\ell\}$ allocated to another agent, and $k$ witness bundles.
\State Let $X$ contain $w_1$ and let $Y$ contain $w_\ell$.
\If{$w_\ell$ is not a pebble or $X=Y$}
    \State Apply Algorithm~\ref{alg:stage1-pair} and end this update.
\Else
    \State Treat all non-pebble items as water.
    \If{$P(Y)=1$ and $v_i(X\setminus\{w_1\})<v_i(w_\ell)$}
        \State Let $q$ be the least-valued pebble among bundles with at least two pebbles, and let $Z$ contain $q$.
        \State Exchange $q$ for $w_\ell$ and water of value $\min\{v_i(q),1/2\}-v_i(w_\ell)$ from $Y$.
    \EndIf
        \State Exchange $w_\ell$ for all items of $X\setminus\{w_1\}$.

\EndIf
\State Delete $X$.
\end{algorithmic}
\end{algorithm}

The payment balances the exchange when $v_i(q)\le1/2$. For larger $q$, the $1/2$ cap lets $Y$ retain the excess value instead of paying away too much water. In either case, $Y$ never loses value and keeps water worth at least its pre-exchange value minus $1/2$. The donor subsequently receives the remainder of $X$; when the cap applies, every pebble left in the donor is worth more than $1/2$.

Reductions continue to use Algorithm~\ref{alg:standard-reduction}. Treating non-pebble items as water does not obstruct these updates: pebbles remain whole, and any non-pebble target items can first be collected whole by redistributing water without changing any bundle's value. The water bounds proved in Appendix~\ref{apdx:stage1-canonical-proof} guarantee this collection. In particular, $\reduction{2}$ may still occur after water is introduced; its non-pebble targets are handled by this preliminary step.

\begin{lemma}[Canonical entry after pair rounds]
\label{lem:stage1-canonical}
Algorithm~\ref{alg:stage1-nomorethan2k} preserves \textnormal{\textsc{WLC}}. If Stage~1 ends with $P(\bW)\le2k$, the witness allocation admits a valid canonical reorganization.
\end{lemma}

The proof is deferred to Appendix~\ref{apdx:stage1-canonical-proof}.

\subsection{More than $2k$ pebbles without ice at the end of Stage~1}
When Stage~1 ends with $P(\bW)>2k$ and no ice, the ordinary-triple maintenance will require the deficit and water inequalities in~\eqref{eq:noice-high-invariants}. During Stage~1, we first seek the following bounds, which allow bundles with fewer than two pebbles:
\begin{equation}
\label{eq:no-ice-bundle-bounds}
\begin{aligned}
    &P(W)\le 2\ \Longrightarrow\ D(W)\le2/27;\\
    &P(W)=3\ \Longrightarrow\ D(W)\le1/27;\\
    &P(W)\ge4\ \Longrightarrow\ D(W)\le0.
\end{aligned}
\end{equation}
The repair in Subsection~\ref{subsec:stage2-repair} then gives every bundle at least two pebbles and establishes the remaining entrance conditions.

\begin{algorithm}[H]
\caption{Pair maintenance when Stage~1 ends with more than $2k$ pebbles and no ice}
\label{alg:stage1-noice}
\begin{algorithmic}[1]
\Require A real pair $\{u_1,u_\ell\}$ allocated to another agent, and $k$ witness bundles.
\State Let $X$ contain $w_1$ and let $Y$ contain $w_\ell$.
\If{$w_\ell$ is not a pebble or $X=Y$}
    \State Apply Algorithm~\ref{alg:stage1-pair} and end this update.
\Else
    \State Treat all non-pebble items as water.
    \If{$D(Y)>0$ and $P(Y)\le2$ and $v_i(X\setminus\{w_1\})<v_i(w_\ell)$}
        \State Choose a bundle $Z$ with $P(Z)\ge3$.
        \State Let $q$ be the least-valued pebble of $Z$.
        \State Exchange $q$ for $w_\ell$ and water of value $\max\{0,\min\{v_i(q),1/3\}-7/27\}$ from $Y$.
    \EndIf
        \State Exchange $w_\ell$ for all items of $X\setminus\{w_1\}$.
\EndIf
\State Delete $X$.
\end{algorithmic}
\end{algorithm}

In this branch, $v_i(w_\ell)<7/27$, so the payment never exceeds the value that $Y$ gains by replacing $w_\ell$ with $q$. When $v_i(q)\le1/3$, it compensates the donor enough that the remainder of $X$ limits the donor's loss to $1/27$. For larger $q$, the payment stops at $2/27$, protecting $Y$'s water; the donor instead retains at least two pebbles worth more than $1/3$ each.

Making non-pebble items fluid does not obstruct the standard reductions. Once an ordinary pair has a pebble partner, every later ordinary pair in Stage~1 does as well. The assumed exit condition $P(\bW)>2k$ forces this inequality to hold throughout the remaining Stage~1 updates, with $k$ denoting the current number of bundles. Rule $\reduction{0}$ cannot recur after an ordinary pair, and all target items of $\reduction{1}$ and $\reduction{2}$ are therefore solid pebbles. Thus Algorithm~\ref{alg:standard-reduction} applies unchanged; the count argument is given in Appendix~\ref{apdx:stage1-repair-proof}.

\begin{lemma}[Deficit bounds during Stage~1]
\label{lem:stage1-repair}
Algorithm~\ref{alg:stage1-noice} preserves~\eqref{eq:no-ice-bundle-bounds} throughout Stage~1 and the initial pair and standard reduction updates of Stage~2.
\end{lemma}

The proof is deferred to Appendix~\ref{apdx:stage1-repair-proof}.

\section{Stage 2}
\label{sec:stage2}
Stage~2 ends immediately before the first ordinary bag-filling round. Ordinary triple rounds, pair rounds, and reduction rounds can occur in any order during this stage. We first describe maintenance for ordinary triples and then state the interfaces required when another type of round intervenes. The closure cases in Section~\ref{sec:general-tools} take priority whenever their hypotheses hold. An update reaching $P(\bW)=2k$ either enters the canonical case immediately or continues by Lemma~\ref{lem:stage2-equality} until canonical entry or agent $i$ is satisfied. 

Before the first ordinary triple, continue the initial pair and standard reduction updates from Stage~1 while $P(\bW)>2k$, unless a closure case or Stage~3 is reached first. The Stage~1 proofs apply unchanged to these updates, retaining the weaker deficit and water bounds when some bundles still have fewer than two pebbles. There is one boundary precaution: before an initial no-ice $\reduction{2}$ with $P(\bW)=2k+1$, apply the repair below. Otherwise repair is deferred until the first no-ice ordinary triple. Subsequent pair and reduction rounds use the interfaces in Subsection~\ref{subsec:notordinarytriplerounds}.

For an ordinary triple, write the real bundle as $\{u_1,u_\ell,u_t\}$, where $t=\max\{\ell+1,2k+1\}$. The pair condition and $\reduction{2}$ have failed. While $P(\bW)>2k$, validity therefore gives
\begin{equation}
\label{eq:ordinary-triple-bounds}
    v_i(w_1)+v_i(w_{k+1})<7/9,
    \qquad v_i(w_1)<5/9,
    \qquad v_i(w_{2k+1})<7/27.
\end{equation}
The procedures below construct a dominance bundle for this real triple and leave $k-1$ valid witness bundles when another agent is served. 

\subsection{Ordinary Rounds Without Ice}
\label{subsec:stage2-noice}

We consider ordinary triple rounds with $P(\bW)>2k$ and $\Delta<4/27$. The entrance comes either from the Stage~1 maintenance in Lemma~\ref{lem:stage1-repair} or from the clean maintenance in Subsection~\ref{subsec:clean-ice}, followed by the repair below. The ordinary-triple procedures require at least two pebbles in every bundle, together with the applicable deficit and water inequalities stated below.

The procedures distinguish whether $v_i(w_1)$ is above or at most $1/3$. The pseudocode in Subsection~\ref{subsubsec:highanchor} and~ \ref{subsubsec:lowanchor} gives their main exchanges. The complete choices and proofs appear in Appendix~\ref{apdx:stage2-noice}.

\subsubsection{Repair}
\label{subsec:stage2-repair}
Before the first no-ice ordinary triple, we apply the following repair to bundles with fewer than two pebbles. It preserves the deficit bounds supplied by Stage~1 and establishes the remaining entrance conditions in~\eqref{eq:noice-high-invariants}. The same repair also applies when the clean maintenance passes to the no-ice case: we use it immediately before the first ordinary triple handled by the no-ice procedures. If an initial no-ice $\reduction{2}$ would reach equality earlier, repair is performed immediately before that reduction instead.

\begin{algorithm}[H]
\caption{Repair before no-ice maintenance}
\label{alg:stage2-repair}
\begin{algorithmic}[1]
\Require A valid checkpoint with $\Delta<4/27$, $P(\bW)>2k$, and $v_i(w_1)<5/9$, produced by the maintenance of Lemma~\ref{lem:stage1-repair} with every deficient singleton having at least $4/9$ water, or with every bundle worth at least $1$.
\While{some bundle $Z$ has $P(Z)<2$}
    \State Choose $Y$ with $P(Y)\ge3$, and let $a$ be its least-valued pebble.
    \State Move $a$ from $Y$ to $Z$.
    \State Transfer water of value $\max\{2/9, \min\{v_i(a),1/3\}-1/27\}$ from $Z$ to $Y$.
\EndWhile
\end{algorithmic}
\end{algorithm}

The $2/9$ lower bound supplies the donor with the water reserve needed after it loses a pebble. For $v_i(a)\le1/3$, the other term limits its loss to $1/27$, while the payment never exceeds the imported pebble's value, so $Z$ does not lose value. For larger $a$, the payment stops at $8/27$, leaving more water in $Z$; the donor's remaining pebbles and the payment already give the required value.

\begin{lemma}[Repair]
\label{lem:stage2-repair}
Algorithm~\ref{alg:stage2-repair} is executable, terminates, and preserves validity. Every output bundle has at least two pebbles and satisfies~\eqref{eq:no-ice-bundle-bounds}; each two-pebble bundle retains value at least $4/9$ after removing its larger pebble. At a high-anchor ordinary-triple checkpoint, the output satisfies~\eqref{eq:noice-high-invariants}.
\end{lemma}

The proof is deferred to Appendix~\ref{apdx:stage2-repair}. When the largest pebble is at most $1/3$, the subsequent ordinary-triple maintenance uses the low-anchor inequalities instead.

\subsubsection{High Anchor Maintenance}
\label{subsubsec:highanchor}
Suppose that $v_i(w_1)>1/3$, and let $X$ contain $w_1$.
If $P(X)\ge3$, any three pebbles of $X$ including $w_1$ dominate the real triple.
The difficult case is $P(X)=2$, especially when its other pebble is also large.
These exchanges may create a deficit; to support subsequent maintenance, we must prevent deficits from accumulating beyond $2/27$ in any two-pebble bundle over successive rounds.

At the first no-ice ordinary triple after this preparation, fix
\[
    \sigma=\min\{v_i(w_{k+1}),7/27\}.
\]
Do not recompute this value in subsequent rounds.
For each bundle $W$, let $a,b$ be its smallest and largest pebbles.
While the high-anchor case applies, maintain
\begin{equation}
\label{eq:noice-high-invariants}
\begin{aligned}
&P(\bW)>2k,\qquad P(W)\ge2,\qquad v_i(w_1)+\sigma<7/9;\\
&P(W)=2\quad\Longrightarrow\quad D(W)\le2/27;\\
&P(W)=3\quad\Longrightarrow\quad D(W)\le1/27;\\
&P(W)\ge4\quad\Longrightarrow\quad D(W)\le0;\\
&D(W)>0,\ P(W)=2\quad\Longrightarrow\quad F(W)+\max\{v_i(a)-\sigma,0\}\ge2/9;\\
&D(W)>0,\ P(W)=3\quad\Longrightarrow\quad F(W)+\max\{v_i(b)-\sigma,0\}\ge2/9.
\end{aligned}
\end{equation}
The last two inequalities allow a deficient bundle to retain the value needed for later exchanges partly in the indicated pebble, instead of requiring $2/9$ water in every such bundle.

A standard $\reduction{2}$ update reduces $k$ without necessarily deleting $w_1$.
Re-indexing can therefore put an earlier, larger pebble at position $2k$, with value above the fixed $\sigma$, even though no item value increases.
We use the special $\reduction{2}$ maintenance in Subsection~\ref{subsec:notordinarytriplerounds} and, when $\sigma<7/27$, additionally maintain
\begin{equation}
\label{eq:noice-indices}
    j>k\quad\Longrightarrow\quad
    v_i(w_j)\le\sigma\ \text{or}\ v_i(w_j)>5/18.
\end{equation}
This controls the pebble values beyond the first $k$ positions as the active count decreases.
When this index condition is imposed, failed $\reduction{2}$ at an ordinary triple gives $v_i(w_{2k})<5/18$, so~\eqref{eq:noice-indices} ensures $v_i(w_{2k})\le\sigma$.

For a two-pebble $X$, prepare a bundle $Y$ with at least three pebbles.
Exchange one prepared pebble for the water of $X$ when its other pebble is worth at most $\sigma$; otherwise exchange a pair for all of $X\setminus\{w_1\}$.
Both exchanges collect a dominance triple in $X$ and return its remaining value to $Y$.
The preparation must preserve the deficit and water bounds, including when few pebbles have value at most $\sigma$.

The following algorithm is an overview. The full preparation, exchanges, and deletion are given in Appendices~\ref{apdx:noice-high-general}--\ref{apdx:noice-high-boundary}.

\begin{algorithm}[H]
\caption{High-anchor maintenance without ice: overview}
\label{alg:noice-high-overview}
\begin{algorithmic}[1]
\Require No ice, $P(\bW)>2k$, $v_i(w_1)>1/3$, and the fixed value $\sigma$.
\State Let $X$ contain $w_1$.
\If{$P(X)\ge3$}
    \State Choose two other pebbles $q,r$ of $X$ and delete $w_1,q,r$.
    \State Distribute the remainder of $X$ among the surviving bundles.
\Else
    \State Prepare a suitable bundle $Y$ with at least three pebbles.
    \State Let $s$ be the other pebble of $X$.
    \If{$v_i(s)\le\sigma$}
        \State Exchange one pebble of $Y$ for the water in $X$.
    \Else
        \State Exchange two pebbles of $Y$ for $X\setminus\{w_1\}$.
    \EndIf
\EndIf
\State Remove $X$.
\end{algorithmic}
\end{algorithm}

\begin{lemma}[High-anchor maintenance above the boundary]
\label{lem:noice-high}
Under the stated high-anchor invariants, if $P(\bW)\ge2k+2$, Algorithm~\ref{alg:noice-high-overview} gives a valid update preserving~\eqref{eq:no-ice-bundle-bounds}. While the high-anchor case continues, it also preserves~\eqref{eq:noice-high-invariants} and, when required,~\eqref{eq:noice-indices}.
\end{lemma}

\begin{lemma}[High-anchor maintenance at the boundary]
\label{lem:noice-high-boundary}
If $P(\bW)=2k+1$, the stated entrance conditions admit a valid update leaving $2(k-1)$ pebbles. Every survivor has deficit at most $2/27$, and the output can immediately be reorganized into a valid canonical witness allocation.
\end{lemma}

At the boundary, only the final deficit bound is needed: it gives \textnormal{\textsc{WLC}}, and deleting $w_1$ gives the rank inequalities for canonical entry. No water or index condition is maintained after this exit. The constructions and proofs appear in Appendix~\ref{apdx:noice-high-general}--\ref{apdx:noice-high-boundary}. The guarantees concern each bundle after the round; an individual operation may increase its deficit by more than $1/27$.

\subsubsection{Low Anchor Maintenance}
\label{subsubsec:lowanchor}
Once $v_i(w_1)\le1/3$, the maintenance no longer needs $\sigma$ or~\eqref{eq:noice-indices}. In addition to~\eqref{eq:no-ice-bundle-bounds}, it maintains
\begin{equation}
\label{eq:noice-low-invariants}
    D(W)>0,\ P(W)=3
    \quad\Longrightarrow\quad v_i(b)+F(W)\ge v_i(w_1),
\end{equation}
where $b$ is the largest pebble of $W$. This inequality is supplied by the repair and is inherited when the largest witness value falls to at most $1/3$. Here $X$ always denotes the bundle currently containing $w_1$. The equal-value exchanges and the final water return create no new positive deficit and do not increase an existing positive deficit.

The following algorithm is an overview. The complete procedure, including the third-pebble choice, index corrections, and final water return, is given in Appendix~\ref{apdx:noice-low}.

\begin{algorithm}[H]
\caption{Low-anchor maintenance without ice: overview}
\label{alg:noice-low-overview}
\begin{algorithmic}[1]
\Require No ice, $P(\bW)>2k$, $v_i(w_1)\le1/3$, and real positions $1,\ell,t$.
\State Let $X$ contain $w_1$.
\If{every bundle has at least four pebbles}
    \State \Return the trivial dense case.
\EndIf
\If{$P(X)\ge4$}
    \State Exchange $w_1$ with a pebble in a bundle with at most three pebbles, using an equal-value water transfer.
\EndIf
\State Bring in the pebbles needed for dominance, one at a time, using equal-value exchanges.
\State Remove $X$.
\end{algorithmic}
\end{algorithm}


\begin{lemma}[No-ice ordinary-triple maintenance]
\label{lem:noice-ordinary-triple}
The applicable inequalities~\eqref{eq:noice-high-invariants}--\eqref{eq:noice-low-invariants} are retained whenever their cases continue to apply.  If the update reaches $P(\bW)=2k$, it immediately enters the canonical case.
\end{lemma}

The proofs and the complete procedures are given in Appendix~\ref{apdx:stage2-noice}.

\subsection{Ordinary Round With Ice}
\label{subsec:stage2-ice}
We now consider ordinary triple rounds with $P(\bW)>2k$ and $\Delta\ge4/27$. We first keep every bundle worth at least $1$, using the original maintenance until the critical state is reached. We then allow a deficit of at most $1/18$ in three-pebble bundles. Throughout this subsection, $X$ denotes the bundle containing $w_1$.

\subsubsection{Clean Maintenance}
\label{subsec:clean-ice}
We first try to complete each ordinary triple while keeping every surviving bundle worth at least $1$. If none of the clean constructions applies, their failed tests imply the structural conditions for the critical state. We then pass the same pending round to the maintenance in Subsection~\ref{subsec:criticalstate}.

The following algorithm is an overview. Its complete subroutines and correctness proofs are given in Appendices~\ref{apdx:clean-ice}--\ref{apdx:clean-divisible}.
\begin{algorithm}[H]
\caption{Clean maintenance with ice: overview}
\label{alg:stage2-clean-ice}
\begin{algorithmic}[1]
\Require An ordinary triple, $P(\bW)>2k$, and every witness bundle worth at least $1$.
\State Let $X$ contain $w_1$.
\If{$v_i(w_1)<13/27$}
    \State Complete the round by the clean procedure for $v_i(w_1)<13/27$ and return.
\ElsIf{a bundle $Y$ contains two ice items, or ice and a pebble of index greater than $k$}
    \State Choose those two items $a,b$ and delete $\{w_1,a,b\}$ by the two-bundle subroutine.
\ElsIf{a bundle $Y$ has fewer than two pebbles, or has two pebbles with one of index greater than $k$}
    \State Use the pebble--ice preparation and deletion subroutine on $X,Y$.
\ElsIf{a three-pebble bundle $B$ has no pebble among the first $k$}
    \State Use the three-bundle reorganization subroutine on $X,B$ and an ice-containing bundle.
\Else
    \State \Return the current witness allocation for critical maintenance of this same round.
\EndIf
\end{algorithmic}
\end{algorithm}

The two-bundle deletion merges the remainders when the target spans two bundles; an internal target only distributes gains. The preparation subroutine buys the missing pebble and ice for equal water value. The three-bundle subroutine reserves value $1$ for one output and assigns the remaining items and water to the other.

When none of the tests applies, the allocation satisfies the \emph{critical invariant}
\begin{equation}
\label{eq:critical-bounds}
\begin{aligned}
    &P(\bW)\ge3k,\qquad P(W)\ge2;\\
    &\text{each bundle contains at most one solid ice item}.
\end{aligned}
\end{equation}
Indeed, every two-pebble bundle contains two of the first $k$ items, and every three-pebble bundle contains at least one. If $n_2,n_3,n_4$ count bundles with two, three, or at least four pebbles, then
\[
    2n_2+n_3\le k=n_2+n_3+n_4,
    \qquad
    P(\bW)\ge3k+n_4-n_2\ge3k.
\]
The other two conditions follow directly from the failed tests. This proves the critical invariant at the clean entrance.

\begin{lemma}[Clean maintenance and critical entry]
\label{lem:stage2-clean-ice}
Algorithm~\ref{alg:stage2-clean-ice} either completes a valid witness update with every survivor worth at least $1$, enters a closure case, or hands the pending round to critical maintenance with~\eqref{eq:critical-bounds} and every bundle worth at least $1$. Outside the $v_i(w_1)<13/27$ branch, all retained ice items remain whole.
\end{lemma}

The complete subroutines and their proofs are given in Appendix~\ref{apdx:clean-ice}--\ref{apdx:clean-divisible}. Once the $v_i(w_1)<13/27$ branch is entered, its procedure also applies at later ordinary triples because the largest witness value cannot increase. Pair and reduction rounds are treated in Subsection~\ref{subsec:notordinarytriplerounds}.

\paragraph{Passing from clean maintenance to the no-ice case.}
Finish the current round first and apply a closure case whenever available. An output with $P(\bW)=2k$ is handled by Lemma~\ref{lem:stage2-equality}. If $P(\bW)>2k$ and $\Delta<4/27$, retain the clean bounds through intervening pair and reduction rounds. Immediately before the next ordinary triple, apply Algorithm~\ref{alg:stage2-repair} and enter Subsection~\ref{subsec:stage2-noice}, initializing $\sigma$ at this checkpoint when the high-anchor case applies. The repair may create deficits, all within the no-ice entrance bounds. If $\Delta\ge4/27$, continue the clean maintenance. Once critical maintenance has begun, continue that procedure instead.

\subsubsection{Critical State}
\label{subsec:criticalstate}
When Algorithm~\ref{alg:stage2-clean-ice} returns the critical state, we apply the following maintenance to the same pending ordinary triple.

\paragraph{Deficit bounds.}
We retain the critical invariant~\eqref{eq:critical-bounds} and additionally maintain
\begin{equation}
\label{eq:critical-deficit-bounds}
\begin{aligned}
&P(W)=2\ \text{or}\ P(W)\ge4
   \quad\Longrightarrow\quad D(W)\le0;\\
&D(W)>0
   \quad\Longrightarrow\quad
   P(W)=3,\quad D(W)\le1/18.
\end{aligned}
\end{equation}
Thus every bundle has value at least $17/18$. For the largest pebble $b$ of $W$, we also retain
\begin{equation}
\label{eq:critical-value-bound}
    D(W)>0,\quad P(W)=3,\quad W\text{ has no ice}
    \quad\Longrightarrow\quad v_i(b)+F(W)\ge4/9.
\end{equation}
This inequality is proved together with the maintenance; no distribution of the first $k$ items among the bundles is required.

\paragraph{Preparation for critical maintenance.}
Before each critical update, make every bundle with at least four pebbles ice-free. An ice item in such a bundle can be decreased to value $1/9$ and melted into water: the four pebbles and this water already have value at least $4(2/9)+1/9=1$. This leaves the pebble count unchanged and does not increase $\largestwater$. Repeat this preparation during the update whenever a bundle with at least four pebbles receives ice or an ice-containing bundle reaches four pebbles.

\paragraph{High Anchor.}
Suppose that $v_i(w_1)>1/3$. Any three pebbles including $w_1$ form a dominance bundle for the real ordinary triple. If $X$ has at least three pebbles, this permits a deletion from $X$. Otherwise a bundle with at least four pebbles supplies two items for the deletion, and its remainder receives $X\setminus\{w_1\}$.

The following algorithm summarizes the critical high-anchor update. The full redistribution rules and their verification are given in Appendices~\ref{apdx:critical-operations} and~\ref{apdx:critical-high}.

\begin{algorithm}[H]
\caption{Critical high-anchor maintenance: overview}
\label{alg:stage2-critical-high}
\begin{algorithmic}[1]
\Require Conditions~\eqref{eq:critical-bounds}--\eqref{eq:critical-value-bound} and $v_i(w_1)>1/3$.
\State Decrease ice in bundles with at least four pebbles to $1/9$ and melt it.
\State Let $X$ contain $w_1$.
\If{$P(X)\ge3$}
    \State Choose two other pebbles $q,r$ of $X$ and delete $w_1,q,r$.
    \State Distribute the remainder of $X$ among the surviving bundles.
\Else
    \State Choose $Y$ with at least four pebbles and its two least pebbles $q,r$.
    \State Move $X\setminus\{w_1\}$ into $Y$.
    \State Let $X=\{w_1,q,r\}$ be the dominance bundle, and delete $X$.
\EndIf
\end{algorithmic}
\end{algorithm}

In the first branch, the remaining pebbles are distributed whole. Any remaining ice is assigned to an ice-free survivor; if the recipient has at least four pebbles, the preceding value decrease applies. If an ice item requires redistribution and no such recipient exists, the trivial dense case is taken before this maintenance. The full redistribution and its proof are deferred to Appendix~\ref{apdx:stage2-ice}.

\begin{lemma}
\label{lem:stage2-critical-high}
Algorithm~\ref{alg:stage2-critical-high} either enters dense closure or completes the round with a valid witness allocation satisfying \eqref{eq:critical-bounds}--\eqref{eq:critical-value-bound}. Any newly deficient bundle has exactly three pebbles and deficit less than $1/18$; no previously deficient bundle incurs further loss.
\end{lemma}

\paragraph{Low Anchor.}
Suppose that $v_i(w_1)\le1/3$. We first give every bundle at least three pebbles by exchanges of equal value. Unless dense closure applies, we then choose an ice-free three-pebble bundle, place $w_1$ in it, and replace its other pebbles only when their indices are too large to dominate the real triple. Every such replacement is paid for by an equal value of water from the chosen bundle.

The following algorithm is an overview. The complete construction and its correctness proof are given in Appendix~\ref{apdx:critical-low}.
\begin{algorithm}[H]
\caption{Critical low-anchor maintenance: overview}
\label{alg:stage2-critical-low}

\begin{algorithmic}[1]
\Require Conditions~\eqref{eq:critical-bounds}--\eqref{eq:critical-value-bound}, $v_i(w_1)\le1/3$, and real positions $1,\ell,t$.
\State Decrease ice in bundles with at least four pebbles to $1/9$ and melt it.
\While{some bundle $Z$ has two pebbles}
    \State Choose an ice-free bundle $Y$ with at least four pebbles.
    \State Exchange its least pebble for any ice of $Z$ and enough water to equalize values.
\EndWhile
\If{there is no ice-free three-pebble bundle}
    \State \Return the trivial dense case.
\EndIf
\State Choose an ice-free three-pebble bundle $X$.
\State Use an equal-value exchange to place $w_1$ in $X$, if necessary.
\State Replace the other pebbles as needed for dominance, one at a time, using equal-value exchanges.
\State Remove $X$ and distribute its remaining water.
\end{algorithmic}
\end{algorithm}

\begin{lemma}
\label{lem:stage2-critical-low}
Every exchange in Algorithm~\ref{alg:stage2-critical-low} is feasible. The procedure either enters dense closure or completes the round with a valid witness allocation satisfying \eqref{eq:critical-bounds}--\eqref{eq:critical-value-bound}. No surviving bundle acquires a new positive deficit or an increase of an existing positive deficit.
\end{lemma}

Both critical procedures also apply when no ice remains. If the last ice disappears during an update, finish the current round and continue with the same critical maintenance; no transition to Subsection~\ref{subsec:stage2-noice} is needed. At a completed update with no ice, every bundle has value at least $17/18$ and every non-pebble item has value below $4/27$, so \textnormal{\textsc{WLC}} holds. The complete critical procedures and their proofs are given in Appendix~\ref{apdx:stage2-ice}.

\subsection{Pair and Reduction Rounds}
\label{subsec:notordinarytriplerounds}
Pair and reduction rounds can occur between ordinary triples. The closure cases retain priority. We give below the complete short pair update; the longer special procedures and their proofs are in Appendix~\ref{apdx:stage2-other}. Each procedure specifies its exchanges, deletion, and redistribution.

\paragraph{Pair rounds.}
In the clean and no-ice cases, a pebble whose index is no later than the real partner suffices. The following update is complete. Critical pair rounds use the same deletion, followed when necessary by the two-ice repair in Appendix~\ref{apdx:stage2-pair-complete}. That repair may create a three-pebble deficit, bounded by $1/18$.

\begin{algorithm}[H]
\caption{Pair maintenance in the clean and no-ice cases}
\label{alg:stage2-special-pair}
\begin{algorithmic}[1]
\Require An ordinary pair $\{u_1,u_\ell\}$, the current clean or no-ice inequalities, and $P(\bW)>2k$.
\State Set $q=w_{\min\{\ell,P(\bW)\}}$; let $X,Y$ contain $w_1,q$.
\If{$X=Y$}
    \State Distribute $X\setminus\{w_1,q\}$ among the other bundles, keeping pebbles and ice whole.
\Else
    \State Exchange $q$ for $X\setminus\{w_1\}$.
\EndIf
\State Delete $X=\{w_1,q\}$ and re-index.
\end{algorithmic}
\end{algorithm}

The anchor remainder has value at least $4/9$, whereas $v_i(q)<7/18$. An external merge therefore gains more than $1/18$. The proof checks the count-dependent bounds and water inequalities and, in the critical case, the placement of the retained ice items. This update deletes two pebbles and one bundle, so it does not reach $P(\bW)=2k$ from an input with $P(\bW)>2k$.

\paragraph{Reductions $\reduction{1}$ and $\reduction{2}$.}
Choose $C$ as in the first line below. If this choice requires the high-anchor $\reduction{2}$ index correction, use the special update instead. Otherwise apply the common deletion, redistributing only the remaining pebbles and discarding all source ice and water. The shared proof is in Appendix~\ref{apdx:stage2-reductions}.

\begin{algorithm}[H]
\caption{Common witness update for $\reduction{1}$ and $\reduction{2}$}
\label{alg:stage2-reductions}
\begin{algorithmic}[1]
\Require $r\in\{1,2\}$, $P(\bW)>2k$, and the current clean, no-ice, or critical bounds; the special $\reduction{2}$ branch below does not apply.
\State In the critical case, decrease ice in bundles with at least four pebbles to $1/9$ and melt it.
\State Choose $C$ containing $r+1$ of $w_1,\ldots,w_{rk+1}$.
\State Let $T$ consist of its first $r+1$ pebbles.
\State Distribute every other pebble of $C$ whole among the surviving bundles.
\State In the critical case, lower ice in recipients with at least four pebbles to $1/9$ and melt it.
\State Delete $C$, including $T$ and all its remaining ice and water, and re-index.
\end{algorithmic}
\end{algorithm}

\paragraph{Special $\reduction{2}$.}
Only the continuing no-ice high-anchor case can require a different update, to preserve~\eqref{eq:noice-indices}. If the selected triple contains no old top-$k$ item and old $w_k$ would violate that condition after re-indexing, exchange $w_k$ for the remainder of $C$ and delete $w_k$ together with the last two selected pebbles. The receiving bundle becomes nondeficient. The complete procedure and proof are in Appendix~\ref{apdx:noice-high-r2}.

If $P(\bW)=2k+1$ before $\reduction{2}$, first apply Algorithm~\ref{alg:stage2-repair} if the initial no-ice repair has not yet been performed. Here only its count, deficit, and larger-pebble remainder guarantees are needed. Every bundle then has at least two pebbles, so exactly one has three. The common update deletes that bundle and leaves every no-ice survivor unchanged with two pebbles and deficit at most $2/27$. It then continues by Lemma~\ref{lem:stage2-equality}; canonical entry need not be immediate. No high-anchor parameter needs to be frozen for this equality continuation. A clean equality output uses the preparation in the same lemma.

\begin{lemma}[Continuation to the canonical case]
\label{lem:stage2-equality}
Under the current clean or no-ice bounds, an output with $P(\bW)=2k$ admits valid maintenance until agent $i$ is satisfied or the witness becomes canonical. No positive deficit is created or increased.
\end{lemma}

The proof is given in Appendix~\ref{apdx:stage2-equality}: ordinary maintenance introduces no new positive deficit. The frozen high-anchor parameter and index condition are not needed during this continuation.

\paragraph{Special $\reduction{3}$.}
The failed triple test gives $v_i(w_1)<1/3$. In a clean or no-ice state satisfying \textnormal{\textsc{WLC}}, $P(\bW)\le3k$ permits canonical entry before the reduction. Otherwise all four target items are pebbles and the standard update suffices. In the critical case, the full procedure in Appendix~\ref{apdx:critical-r3} chooses the target container according to the holder of $w_{3k+1}$; this avoids creating a deficient four-pebble survivor.

\begin{lemma}[Intervening pair and reduction rounds]
\label{lem:stage2-other-complete}
Under their stated entrance conditions, the pair and reduction procedures preserve validity and the applicable inequalities while $P(\bW)>2k$, or enter a closure case. An output with $P(\bW)=2k$ continues to canonical entry by Lemma~\ref{lem:stage2-equality}, unless agent $i$ is satisfied first.
\end{lemma}

\section{A Strongly Polynomial-Time Algorithm}\label{sec:exact-search}
In this section, we use Theorem~\ref{thm:7/9} to compute a \(7/9\)-MMS allocation without computing MMS values.
We obtain a strongly polynomial-time algorithm by searching over the agents' thresholds.
Moreover, we justify the threshold updates using Theorem~\ref{thm:7/9} and the comparison outcomes of a fixed implementation, without using the details of witness maintenance.
We therefore believe that this approach may also be useful for other MMS algorithms and, more broadly, for other problems with similar threshold guarantees.

Our search starts from the agents' proportional shares \(v_i(M)/n\) and lowers failed agents' thresholds to values determined by comparisons in the failed execution. 
We keep each threshold at least the corresponding MMS value throughout the search and bound the number of updates using the structure of tail filling.
Throughout this section, we assume that each arithmetic operation and numerical comparison takes $O(1)$ time.

\begin{theorem}\label{recon-comp:thm:search}
For nonnegative additive valuations, we can compute a complete $7/9$-MMS allocation in strongly polynomial time.
The algorithm makes $O(n^5m^4)$ calls to Algorithm~\ref{alg:pebble-and-water} and runs in $O(n^7m^5)$ time.
\end{theorem}

\subsection{Initialization and threshold updates}

If $n>m$, we have $\MMS_i=0$ for every agent $i$, since every $n$-partition contains an empty bundle.
We then return any complete allocation.
For the rest of this section, we assume $n\le m$.
We initialize each agent's threshold at her proportional share:
\begin{equation*}
 \alpha_i=v_i(M)/n.
\end{equation*}
We have $\alpha_i\ge\MMS_i$, since every $n$-partition contains a bundle worth at most $v_i(M)/n$ to agent $i$.
We compute the initial thresholds in $O(nm)$ time.

In each call to Algorithm~\ref{alg:pebble-and-water}, we scan candidates in increasing order of $j$ when selecting the largest satisfactory index in the pair or triple branch.
If several agents find the selected bundle satisfactory, we assign it to the one with the smallest original index.
Whenever we test a bundle $T$, including each partial bundle constructed during tail filling, we evaluate $v_i(T)\ge7\alpha_i/9$ for every active agent $i$ and record $T$ for each agent who finds it unsatisfactory.
We evaluate these comparisons even if we have already found a satisfactory recipient.
We use each threshold $\alpha_i$ only in comparisons of the form $v_i(T)\ge7\alpha_i/9$.
If we change the thresholds while preserving every comparison outcome, we obtain the same tested bundles, selected recipients, and final outcome.

If Algorithm~\ref{alg:pebble-and-water} returns $(\textsc{Failure},F)$, we update the thresholds of the failed agents.
For each $i\in F$, we define $b_i$ as the largest value of $9v_i(T)/7$ among the bundles $T$ tested during this call that satisfy $v_i(T)<7\alpha_i/9$.
We have tested the final failed bundle for every $i\in F$, so this maximum exists.
We then set $\alpha_i=b_i$ for every $i\in F$ and restart.

\begin{algorithm}[H]
\caption{Threshold search for a $7/9$-MMS allocation}
\label{alg:threshold-search-main}
\begin{algorithmic}[1]
\Require An ordered instance $\cI$ with nonnegative additive valuations and $n\le m$.
\For{each agent $i$}
  \State Set $\alpha_i\gets v_i(M)/n$.
\EndFor
\While{true}
  \State Run Algorithm~\ref{alg:pebble-and-water} on $\cI$ with $(\alpha_i)_{i\in N}$ and record all bundles tested during this call.
  \If{it returns $(\textsc{Failure},F)$}
    \For{each $i\in F$}
      \State Compute $b_i\gets\max\{9v_i(T)/7:T\text{ was tested in this call and }v_i(T)<7\alpha_i/9\}$.
      \State Set $\alpha_i\gets b_i$.
    \EndFor
  \Else
    \State Assign any unallocated goods arbitrarily and return the complete allocation.
  \EndIf
\EndWhile
\end{algorithmic}
\end{algorithm}

We know from Theorem~\ref{thm:7/9} that $\alpha_i>\MMS_i$ for every failed agent, but we still need to determine how far we can safely lower her threshold.
The following lemma shows that we can replace $\alpha_i$ with $b_i$ while keeping it at least $\MMS_i$.

\begin{lemma}[Safe threshold updates]\label{lem:exact-search-safe-update}
For every failed call and every $i\in F$, we have $\MMS_i\le b_i<\alpha_i$.
\end{lemma}
The proof of Lemma~\ref{lem:exact-search-safe-update} is given in Appendix~\ref{apdx:exact-search-safe-update}.

We update all failed agents simultaneously using the thresholds from the completed call; the correctness of these updates is established in Appendix~\ref{apdx:exact-search-safe-update}.

\subsection{Number of calls and running time}

We bound the number of failed calls to Algorithm~\ref{alg:pebble-and-water} in two parts.
Before a call first reaches tail filling, every tested bundle contains at most four goods.
We use these comparisons to bound how often the active agents or remaining goods at the first tail-filling step can change across calls.
We then consider consecutive calls that reach this step with the same active agents and remaining goods.
For these calls, we use the structure of tail filling to bound the possible threshold values and hence the number of calls.
We combine these bounds to obtain a polynomial bound on the total number of failed calls.

\begin{lemma}\label{lem:exact-search-call-groups}
For each failed call, consider the active agents and remaining goods when that call first reaches tail filling.
Across the entire sequence of failed calls, these sets change at most $O(n^3m^2)$ times.
\end{lemma}

By starting a new group whenever either set changes, we partition the failed calls into $O(n^3m^2)$ consecutive groups.
Within each group, all calls first reach tail filling with the same active agents and remaining goods.
We next bound the number of failed calls within each group.
Consider one group, and let $K$ be the number of active agents when tail filling first begins.
For each failed call and each agent $i\in F$, we consider the largest value of $9v_i(T)/7$ over her unsatisfactory bundles tested from the first tail-filling step onward.

\begin{lemma}\label{lem:exact-search-bag-candidates}
For each agent $i$, we obtain at most $O(Km^2)$ possible values for this maximum across the calls in the group.
\end{lemma}

The proofs of Lemmas~\ref{lem:exact-search-call-groups} and~\ref{lem:exact-search-bag-candidates} are given in Appendices~\ref{apdx:exact-search-call-groups} and~\ref{apdx:exact-search-bag-candidates}, respectively.
We now combine the two lemmas to bound the total number of calls and the running time.

\begin{proof}[Proof of Theorem~\ref{recon-comp:thm:search}]
Consider an ordered instance with $n\le m$.
By Lemma~\ref{lem:exact-search-call-groups} and the grouping described above, we have $O(n^3m^2)$ consecutive groups of failed calls.
Within each group, all calls first reach tail filling with the same active agents and remaining goods.
By the proof of Lemma~\ref{lem:exact-search-call-groups}, at most $O(n^3m^2)$ failed calls have $b_i=9v_i(T)/7$ for some agent $i\in F$ and some bundle $T$ tested before that call first reaches tail filling.
We next count the remaining failed calls within each group.

Consider one group, and let $K\le n$ be the number of active agents when tail filling first begins.
We now count only the calls in this group that have not been counted separately above.
In each such call, we can compute every updated value $b_i$ using only the unsatisfactory bundles tested from the first tail-filling step onward.
By Lemma~\ref{lem:exact-search-bag-candidates}, we have $O(Km^2)$ possible values of $b_i$ for each agent $i$.
Summing over the agents, we obtain $O(nKm^2)$ possible values, counted separately for different agents.
By Lemma~\ref{lem:exact-search-safe-update}, we strictly decrease every updated threshold, so each agent can reach each counted value at most once.
Since every failed call updates at least one agent, we obtain at most $O(nKm^2)=O(n^2m^2)$ such calls in this group.

We divide the failed calls into two types.
A call is of the first type if some agent $i\in F$ has $b_i=9v_i(T)/7$ for a bundle $T$ tested before that call first reaches tail filling.
All other failed calls are of the second type.
As shown in the proof of Lemma~\ref{lem:exact-search-call-groups}, we have at most $O(n^3m^2)$ calls of the first type over the entire search.
By Lemma~\ref{lem:exact-search-call-groups}, we partition all failed calls into $O(n^3m^2)$ consecutive groups.
Within each group, all calls first reach tail filling with the same active agents and remaining goods.
Consider one group, and let $K\le n$ be the number of active agents at that step.
For calls of the second type, every updated value $b_i$ is the maximum over unsatisfactory bundles tested from that step onward.
By Lemma~\ref{lem:exact-search-bag-candidates}, we have $O(Km^2)$ possible values per agent.
By Lemma~\ref{lem:exact-search-safe-update}, each agent can reach each value at most once.
Since every failed call updates at least one agent, we have at most $O(nKm^2)=O(n^2m^2)$ calls of the second type in this group.

We thus have $O(n^3m^2)$ calls of the first type in total and at most $O(n^2m^2)$ calls of the second type in each of the $O(n^3m^2)$ groups.
Including at most one successful call, we obtain
\[
O(n^3m^2)+O(n^3m^2)\,O(n^2m^2)+1
=O(n^5m^4)
\]
calls in total.

We next bound the time required for each call.
Each call has at most $n$ successful rounds and one possible failed round.
In each round, we test $O(m)$ bundles for at most $n$ active agents.
During tail filling, we compute bundle values incrementally by adding the next item's value.
We update each recorded maximum with a single comparison.
Thus each test and record update takes $O(1)$ time, giving $O(n^2m)$ time per call and $O(n^7m^5)$ time over all calls.

Note that the threshold initialization takes $O(nm)$ time.
The sorting and greedy item selection required for the ordered-instance construction and allocation conversion in Lemma~\ref{recon-comp:lem:ordered} also fit within the stated time bound.
We therefore obtain a complete $7/9$-MMS allocation in $O(n^7m^5)$ arithmetic operations and comparisons.

For rational input valuations, this operation count is independent of the encoding lengths of the numerical input values. Every threshold is either its initial proportional share or $9v_i(S)/7$ for some subset $S$ of the original goods, since each update selects a maximum attained by a tested bundle. Bundle values and their partial sums involve at most $m$ input values. Using a common denominator for each agent, these sums, the thresholds, and the cross-products used in comparisons all have encoding length polynomial in the input length. Thus the algorithm also satisfies the polynomial-encoding requirement for strong polynomiality.
\end{proof}

\section{Tightness and Discussion}\label{sec:tightness}

We give a tight example for the threshold algorithm and then discuss related allocation rules and possible
extensions of witness maintenance.

\subsection{A tight example}
We illustrate the tightness of our $7/9$ approximation guarantee through a simple instance in which all agents share identical additive valuations.

\begin{example}
Consider an identical instance with three agents.  
We construct the following valuation profile that serves as a hardness instance for any $\alpha > 7/9$.

\begin{center}
\begin{tabular}{c|ccccc|c}
    & $e_1$ & $e_2$ & $e_3$ & $e_4$ & $e_5$ & $e_6, e_7, \ldots$ \\ \hline
    $\mathbf{v}$ & $7/9$ & $7/9$ & $1/3$ & $1/3$ & $1/3$ & $\epsilon$ \\ 
\end{tabular}
\end{center}

We normalize the maximin share value to $\MMS_i = 1$ by appending additional items of total value $4/9$, consisting of vanishingly small ``water'' items of value $\epsilon \to 0$ that satisfies $\epsilon < \alpha - 7/9$.
Under this normalization, an MMS partition consists of two bundles, each containing one $7/9$-valued item and an additional $2/9$ of water, and a third bundle containing the three $1/3$-valued items.

Following our algorithm, $\reduction{0}$ is applied twice: the first
two agents receive the singleton bundles $\{e_1\}$ and $\{e_2\}$,
each of value exactly $7/9$.  At the beginning of the last round, the
remaining items and their values are as follows:
\begin{center}
\begin{tabular}{c|c|c}
    & $u_1,u_2,u_3$ & $u_4,u_5,\ldots$ \\ \hline
    $\mathbf{v}$ & $1/3$ each & $\epsilon$ 
\end{tabular}
\end{center}
The final agent is then satisfied, for example by $\reduction{2}$.
The first two allocated bundles nevertheless have value exactly
$7/9$.  Hence the algorithm cannot guarantee a value strictly greater
than $7/9$ on this family of instances.
\end{example}

\paragraph{Relation to the $10/13$ algorithm.}
This example should not be interpreted as entering directly into the
later bag-construction phase of the algorithm of
\cite{conf/soda/HeidariKSS26}.  Since $7/9>10/13$, that algorithm can
first apply its singleton reduction to the two largest items.  We use
the example only to show tightness of the threshold algorithm analyzed
here.

\subsection{Discussion}

We conclude by relating our allocation rule to chores allocation~\cite{conf/sigecom/HuangL21,conf/sigecom/HuangS23} and discussing how witness maintenance may support stronger guarantees.
The Heterogeneous First Fit Decreasing (\textsf{HFFD}) algorithm scans chores from largest to smallest, adding a chore to the current bundle whenever the resulting cost is within the threshold of at least one remaining agent~\cite{conf/sigecom/HuangS23}.
It then allocates the bundle to an agent whose threshold is respected.
We adapt this philosophy to the goods setting through the threshold algorithm of Section~\ref{sec:overview}.
The algorithm is given arbitrary nonnegative targets $\alpha_i$ and allocates each selected bundle to an active agent $i$ who values it at least $7\alpha_i/9$.
For each fixed agent whose target is at most her original MMS, we maintain a private witness allocation throughout the real execution.
This witness allocation is used only in the proof and is not computed by the algorithm.

The pair and triple branches first test their listed reductions and then choose the largest satisfactory candidate index.
Otherwise, the algorithm tests $\reduction{3}$ and, if it fails, constructs a bundle by appending tail goods.
A real bundle may exceed a remaining agent's target; we account for the resulting loss of value and preserve the one-sided guarantee through witness maintenance.
The algorithm therefore does not require an upper bound on every remaining agent's value for each allocated bundle, or access to exact MMS values.

For future work, it would be interesting to explore alternative relaxation conditions under which a bundle may exceed the MMS threshold, potentially leading to further improvements in the approximation ratio.
A useful next step is to identify witness conditions that support these relaxations with fewer case distinctions.

\clearpage
\appendix
\renewcommand{\theHsection}{appendix.\arabic{section}}
\section{Stage 1: Deferred Proofs and Procedures}
\label{apdx:stage1}

\subsection{The basic pair update}
Algorithm~\ref{alg:stage1-pair} deletes exactly $w_1,w_\ell$, which dominate the allocated real pair. Exchanges and redistribution do not change any ordered witness-item value, so Proposition~\ref{prop:remove-dominance} preserves validity. An internal deletion gives only gains to survivors; an external deletion replaces $Y$ by $(X\setminus\{w_1\})\cup(Y\setminus\{w_\ell\})$.

We will use one consequence of maximality in the proofs below. Read real and witness items by their ranks at initialization. Stage~1 deletes the same current indexed positions in both sequences and never changes item values, so this rank correspondence persists. After an ordinary pair, no real item originally later than its partner can ever become an eligible partner: the largest remaining item cannot increase, and the active-agent set only shrinks. Thus subsequent ordinary partners have earlier original ranks. In particular, once an ordinary witness partner is a pebble, every subsequent ordinary witness partner is a pebble as well.

\begin{proof}[Proof of Lemma~\ref{lem:stage1-clean-ice}]
Standard reductions and internal pairs only give gains to survivors, so a first deficit must arise in an external ordinary pair. Write $h=v_i(w_1)$ and $r=v_i(w_\ell)$. Its clean inputs give $h+r>1$. Failed $\reduction{0}$ implies $h<7/9$, so $r>2/9$ and the partner is a pebble. Failed $\reduction{1}$ gives $r<7/18$, hence $h>5/9$. Every pebble is then an eligible partner for agent $i$, so maximality makes $w_\ell$ the last pebble.

From the first pebble partner onward, ordinary pairs delete two pebbles and $\reduction{0}$ cannot recur. Neither $\reduction{1}$ nor $\reduction{2}$ can take $P(\bW)\le2k$ to $P'>2(k-1)$: when $P\le2k-2$ this is immediate, and when $P\in\{2k-1,2k\}$ their targets contain enough pebbles. Thus the assumed exit condition forces $P(\bW)>2k$ at the first loss. Failed $\reduction{2}$ now gives $r<7/27$, so $h>20/27$. Maximality also gives $h+\Delta<7/9$, whence $\Delta<1/27$. Since item values never change, this contradicts an ice item of value at least $4/27$ remaining at the exit. No deficit can therefore arise. Validity follows from the basic pair update and the standard reduction argument.
\end{proof}

\subsection{Canonical entry with at most $2k$ pebbles}
\label{apdx:stage1-canonical-proof}
\begin{proof}[Proof of Lemma~\ref{lem:stage1-canonical}]
Write $B=2/9+\Delta$ for this proof. Along with \textnormal{\textsc{WLC}}, we maintain the following auxiliary bounds:
\begin{equation}
\label{eq:stage1-canonical-water}
\begin{aligned}
 D(W)>0&\ \Longrightarrow\ F(W)>B;\\
 D(W)>0,\ P(W)=2&\ \Longrightarrow\ v_i(W)+F(W)>10/9+2\Delta.
\end{aligned}
\end{equation}
Initially every bundle is worth $1$, and $\Delta<2/9$, so all assertions hold. Since $\Delta$ cannot increase, a later decrease only weakens the bounds to be preserved. Before the first pebble partner, ordinary external pairs give positive gains, since $h=v_i(w_1)<7/9$ and their partner has value less than $2/9$. Internal deletions and standard reductions also leave every survivor clean. We henceforth consider pebble partners; the preceding rank argument shows that every later ordinary partner is also a pebble. Failed $\reduction{1}$ gives $r:=v_i(w_\ell)<7/18$.

\paragraph{Whole items for reductions.}
We first justify the standard reductions after non-pebble items are treated as water. At the first ordinary pair with a pebble partner, all non-pebbles have indices greater than $\ell$. Maximality and validity therefore give $h+\Delta<7/9$. Since neither $h$ nor $\Delta$ increases, this bound persists. Assuming the inductive bounds above, every singleton has water greater than $2/9+\Delta>2\Delta$: for a clean singleton use $F(W)\ge1-h$, and for a deficient one use~\eqref{eq:stage1-canonical-water}. A pebble-free bundle has $F(W)>L>3\Delta$.

If a bundle has enough total water to hold a specified set of water items, those items can be collected whole there while keeping every bundle's water value unchanged: replace their shares in other bundles by equally valued water from the collecting bundle. This changes neither bundle values nor ordered item values. Rule $\reduction{0}$ cannot recur after an ordinary pair, so only $\reduction{1}$ and $\reduction{2}$ need consideration.

For $\reduction{1}$, all targets are pebbles when $P(\bW)\ge k+1$. If $P(\bW)<k$, a pebble-free bundle can collect both non-pebble targets. If $P(\bW)=k$ and every bundle is a singleton, collect $w_{k+1}$ in the holder of $w_k$. Otherwise a pebble-free bundle can collect $w_{k+1}$, and some other bundle contains at least two pebbles. In each case the source required by Algorithm~\ref{alg:standard-reduction} and all its target items are available whole.

For $\reduction{2}$, all targets are pebbles when $P(\bW)\ge2k+1$. If $P(\bW)\in\{2k-1,2k\}$ and a bundle has at most one pebble, collect all non-pebbles among the first $2k+1$ items there. There are at most two, so its water suffices. The entire prefix is then available whole and the usual Pigeonhole Principle applies. The remaining case with $P(\bW)=2k$ has exactly two pebbles per bundle. Some bundle contains two of $w_k,\ldots,w_{2k}$; their sum is at most $v_i(w_k+w_{k+1})<7/9$, since $\reduction{1}$ has failed. By \textnormal{\textsc{WLC}}, this bundle has water greater than $\Delta$ and can collect $w_{2k+1}$, giving a source with three whole prefix items.

Finally, if $P(\bW)\le2k-2$, all three $\reduction{2}$ targets are non-pebbles. A pebble-free bundle can collect all three. If none exists, there are at least two singletons; collect two targets in one and the third in the other. The first singleton, together with its pebble, supplies three whole items among the first $2k+1$. Thus the standard exchanges apply in every case. These preparatory water exchanges preserve all inductive bounds; the effects of the standard exchanges are verified below.

\paragraph{Pair updates.}
Put $R=v_i(X\setminus\{w_1\})$. A loss $R<r$ cannot have a deficient anchor with two or more pebbles, since~\eqref{eq:stage1-canonical-water} would give $R>4/9+\Delta$. If the anchor is clean, $h+r>1$ gives $h>5/9$, so maximality of $\ell$ makes $r$ the last pebble. The anchor must consequently have only one pebble, and maximality also gives $h+\Delta<7/9$. Hence $R\ge1-h>B$. A deficient singleton anchor has $R=F(X)>B$.

We next show that every loss removes a pebble no larger than any pebble in a bundle with at least two pebbles. This ensures in particular that the special payment is nonnegative. First consider the execution before any special exchange imports a pebble of value greater than $7/18$. A special recipient then retains water
\[
 F(Y')=v_i(Y)-v_i(q)>L-7/18=7/18+\Delta,
\]
so it cannot later be a singleton anchor with $R<r$. A singleton obtained from a deficient two-pebble bundle also retains more than $B+2/9>7/18$ water: it already had more than $B$, and receives at least $B$ in a loss or at least $r\ge2/9$ without loss. The only remaining deficient singleton that can incur a loss originates from a clean two-pebble bundle. If $\rho$ was the partner in that update, induction makes $\rho$ the last pebble and gives the resulting value bound
\[
 v_i(W)>1-\rho+B.
\]
While this bundle remains a singleton and has no special exchange, its value does not decrease, and every remaining pebble is worth at least $\rho$. If it is the current anchor and $F(W)<r$, then
\[
 h+\rho=v_i(W)-F(W)+\rho
 >1+B-7/18>7/9.
\]
Every remaining pebble is therefore an eligible partner for agent $i$, forcing $r$ to be the last pebble. A whole-pebble gain or a standard exchange increasing its pebble count clears its deficit. If an ordinary merge increases its count while leaving it deficient, its retained water already exceeds $L-r>7/18+\Delta$; any later return to one pebble also has enough water to preclude a loss. This proves the induction up to the first larger special import.

Suppose that such an import first selects $q$ with $s=v_i(q)>7/18$. Every pebble in a multi-pebble bundle is at least $s$, by the global minimum rule. Failed $\reduction{1}$ implies that the number $H$ of pebbles of value at least $s$ is at most $k$. The anchor and donor are two distinct bundles containing such pebbles, whereas every smaller pebble is in a singleton. Consequently
\[
 P(\bW)\le H+(k-2)\le2k-2.
\]
The update deletes two pebbles. Thereafter every ordinary partner is still a pebble, and hence its index is at most $P(\bW)\le2k-2$. Deleting positions $1,\ell$ leaves the three indexed real items tested by $\reduction{2}$ exactly unchanged when $k$ decreases by one. The tests for $\reduction{0}$ and $\reduction{1}$ cannot increase either. Thus no later reduction occurs in Stage~1. Every multi-pebble bundle continues to have all its pebbles worth at least $s$: partners are worth less than $7/18<s$ and therefore lie in singletons, and ordinary exchanges introduce no smaller pebble into a multi-pebble bundle. Every subsequent special donor minimum is consequently at least $s>r$. This completes the history argument.

In the special branch a donor exists because $\ell\ge k+1$ and $w_\ell$ is a pebble. The anchor and partner holder are singletons, so the donor is distinct from both. With $s=v_i(q)\ge r$, the payment $t=\min\{s,1/2\}-r$ is nonnegative. It is affordable because
\[
 F(Y)-t=v_i(Y)-\min\{s,1/2\}>L-1/2=5/18+\Delta>B.
\]
The recipient's value does not decrease. The subsequent ordinary exchange uses the current holder of $w_\ell$, so the donor ultimately loses $q$ and receives $R+t$ water.

We verify the value bounds for every loss. Except when the special donor has $s>1/2$, the affected donor loses its least pebble, of value $a$, and receives water of value $T>B$. An ordinary loss has $a=r,T=R$; a special loss has $a=s,T=R+s-r$. If the donor has $p\ge3$ pebbles and was clean, then
\[
 v_i(W')\ge\frac{p-1}{p}v_i(W)+\frac{F(W)}p+T
 >\frac{p-1}{p}+B>L.
\]
For a deficient donor use $v_i(W)>L$ and $F(W)>B$ in the same expression; even at $p=3$ the result exceeds $22/27+2\Delta>L$. A deficient two-pebble donor gives
\[
 v_i(W')>\frac{v_i(W)+F(W)}2+B>L
\]
by~\eqref{eq:stage1-canonical-water}. A clean two-pebble donor loses only $r-R$ in value, and hence retains value greater than $1-7/18+B=5/6+\Delta>L$. If the special donor has $s>1/2$, it was clean; an output with two remaining pebbles is clean, and a singleton retains one pebble greater than $1/2$ and water greater than $B+1/2-r$, again giving value greater than $5/6+\Delta$.

All newly deficient bundles receive more than $B$ water. A new deficient two-pebble donor comes from $p=3$. In the uncapped case,
\[
 v_i(W')+F(W')
 \ge\frac23v_i(W)+\frac43F(W)+2T.
\]
A clean input gives a bound greater than $2/3+2B=10/9+2\Delta$, and a deficient input gives a still larger bound. The capped case stays clean. Thus~\eqref{eq:stage1-canonical-water} persists in loss rounds.

Without a loss, value and water do not decrease. If a deficient three-pebble holder becomes a two-pebble bundle, its old water exceeds $B$ and it receives at least $r\ge2/9$, which proves the second bound in~\eqref{eq:stage1-canonical-water}. The only other count-changing merge that can yield a deficient two-pebble bundle has $P(X)=3,P(Y)=1$; here the retained water is at least $F(Y)=v_i(Y)-r>L-7/18$, so its new value plus water exceeds $7/6+2\Delta>10/9+2\Delta$. Count-preserving exchanges preserve both auxiliary bounds. Internal deletions and reduction remainders distribute gains; a whole pebble clears every old deficit. In a standard exchange, replacing a non-pebble by a pebble gains at least $2/9-\Delta$, which also clears every deficit permitted by \textnormal{\textsc{WLC}}. Other standard exchanges do not decrease value or water. These observations complete the simultaneous induction.

At the end of Stage~1, pair failure and $P(\bW)\le2k$ give
\[
 v_i(C_1)\le v_i(w_1)+v_i(w_{k+1})+\Delta<L.
\]
Lemma~\ref{lem:canonical-entry} now supplies the valid canonical reorganization.
\end{proof}

\subsection{Deficit bounds without ice}
\label{apdx:stage1-repair-proof}
\begin{proof}[Proof of Lemma~\ref{lem:stage1-repair}]
In addition to~\eqref{eq:no-ice-bundle-bounds}, we prove the following assertions whenever $D(W)>0$. For a two-pebble bundle, $a$ denotes its smaller pebble; for a three-pebble bundle, $b$ denotes its largest pebble.
\begin{equation}
\label{eq:stage1-noice-water}
\begin{aligned}
 P(W)=1&\ \Longrightarrow\ F(W)\ge2/9;\\
 P(W)=1,\ D(W)>1/27&\ \Longrightarrow\ F(W)\ge4/9;\\
 P(W)=2&\ \Longrightarrow\ F(W)+(v_i(a)-7/27)_+\ge2/9;\\
 P(W)=3&\ \Longrightarrow\ F(W)+(v_i(b)-7/27)_+\ge2/9.
\end{aligned}
\end{equation}
Here $(x)_+=\max\{x,0\}$. All these assertions hold initially because every bundle has value $1$.

By the rank argument above, after the first ordinary pair with a pebble partner, every ordinary pair has a pebble partner. Before that round all bundles remain nondeficient: a non-pebble partner has value less than $2/9$, while a clean anchor has remainder greater than $2/9$ because $\reduction{0}$ has failed.

From the first pebble-partner round onward, $\reduction{0}$ never recurs and ordinary pairs delete two pebbles. Neither $\reduction{1}$ nor $\reduction{2}$ can take $P(\bW)\le2k$ to $P'>2(k-1)$: for $P\le2k-2$ this is immediate; for $P=2k-1$ their targets contain at least one pebble, and for $P=2k$ at least two. Hence the assumed final inequality $P(\bW)>2k$ implies this inequality throughout these rounds. All target items of the intervening $\reduction{1}$ and $\reduction{2}$ updates are therefore solid pebbles, so making non-pebbles fluid does not affect their standard maintenance. At an ordinary pair, failed $\reduction{1}$ gives $r:=v_i(w_\ell)<7/18$, and failed $\reduction{2}$ gives $v_i(w_{P(\bW)})<7/27$.

Write $h=v_i(w_1)$ and $R=v_i(X\setminus\{w_1\})$. We claim that every loss, meaning $R<r$, has a one-pebble anchor, $h>5/9$, $R\ge2/9$, and $r$ is the last pebble and is less than $7/27$. Three pebbles in $X$ would give $R\ge4/9>r$. A deficient two-pebble anchor also has $R\ge4/9$ by~\eqref{eq:stage1-noice-water}. For a clean anchor, $h+r>1$ implies $h>5/9$. For a deficient one-pebble anchor, deficit greater than $1/27$ would give $R\ge4/9$; otherwise $h>26/27-r>5/9$. Maximality of $\ell$ now puts every later witness item below $7/9-h<2/9$, so $r$ is the last pebble. A clean anchor with another pebble would consequently have $R\ge r$, a contradiction. The asserted bound on $R$ follows from cleanliness or~\eqref{eq:stage1-noice-water}.

Consider an external update without the special exchange. If $P(X)=1$, the holder of $r$ loses one pebble and receives at least $2/9$ water. A clean holder can acquire deficit only in a loss round, and that deficit is less than $1/27$. A deficient three-pebble holder leaves two with deficit at most $2/27$. A deficient holder with at most two pebbles is processed this way only without loss, so its deficit does not increase. When such a two-pebble holder becomes a singleton, its new water is at least $r+F(Y)\ge4/9$, using~\eqref{eq:stage1-noice-water}. Every other newly deficient output receives at least $2/9$ water. Outputs with at least four pebbles are clean, since their pebbles and the received water alone are worth at least $10/9$.

If $P(X)=2$, no loss is possible, and the holder's pebble count is unchanged. Replacing $r$ by the other pebble of $X$ either increases the relevant smallest or largest pebble, or the water of $X$ pays for the entire decrease; thus~\eqref{eq:stage1-noice-water} persists. Singleton water only increases. If $P(X)\ge3$, the gain exceeds $4/9-7/18=1/18$. The only potentially deficient outputs with an increased count are obtained from a three-pebble anchor and a holder with one or two pebbles; their deficits are at most $1/54$. In the former case the old singleton water bound suffices; in the latter the old smallest pebble is no larger than the new largest pebble, so the old two-pebble water inequality suffices. Every output with at least four pebbles is clean: a three-pebble holder's old deficit is cleared by the gain, and an anchor with at least four pebbles gives gain greater than $5/18$, clearing every allowed deficit.

In the special branch the loss claim gives $P(X)=1$, $R\ge2/9$, and $r<7/27$. A donor with at least three pebbles exists because $P(\bW)>2k$, and it is distinct from $X,Y$. Put $s=v_i(q)$ and $t=(\min\{s,1/3\}-7/27)_+$. Since $r$ is the last pebble, $s\ge r$; hence $0\le t\le2/27$ and $s-r-t\ge0$. The deficient holder has at least $2/9$ water, so it can pay $t$ without losing value. If it has one pebble, its new water is at least $25/27-r-t\ge16/27$. If it has two, let $b$ be its other pebble. For $s\le v_i(b)$, its new water expression is
\[
 F(Y)-t+(s-7/27)_+\ge F(Y)\ge2/9.
\]
For $v_i(b)<s$, the old value bound instead gives this expression, with $s$ replaced by $v_i(b)$, at least
$25/27-r-t-7/27\ge1/3$.
The donor loses $s$ and receives $R+t$ water. For $s\le1/3$ its net loss is at most $1/27$, giving precisely the required bounds for a three- or four-pebble donor. For $s>1/3$, even a three-pebble donor retains value greater than $2/3+8/27=26/27$. A larger donor retaining at least four pebbles is clean. All possibly deficient donor outputs receive at least $2/9$ water.

Internal pairs distribute only gains. In standard reductions, outside holders keep their pebble count and water, and replace target pebbles by weakly larger pebbles. Redistribute any remaining pebbles before water: each pebble gain clears every old deficit, and water gains preserve~\eqref{eq:stage1-noice-water}. This proves the induction and the lemma.

If a deficit arises, its first creation has clean inputs and therefore $h+r>1$. The loss claim gives $r<7/27$, so $h>20/27$. Maximality of the partner then gives $\Delta<7/9-h<1/27$; this bound persists afterward. In particular, melting ice does not have to be used to infer the required bound on $\Delta$ in executions with a deficit.

We record the stronger singleton bound needed by the repair. At any non-pair checkpoint before repair, every deficient singleton has $F(W)\ge4/9$. The preceding branches already give this bound except when a clean two-pebble holder loses the last pebble $r$. Its resulting value exceeds $11/9-r$. Until it ceases to be a singleton, its value cannot decrease; a special exchange instead gives at least $16/27$ water. The recorded $r$ is no larger than every subsequent pebble. At the checkpoint let $b$ be the singleton's current pebble and $\sigma=\min\{v_i(w_{k+1}),7/27\}$. Then $r\le\sigma$ and pair failure gives $v_i(b)+\sigma<7/9$, whence
\[
 F(W)>11/9-r-v_i(b)>4/9.
\]
The existing two- and three-pebble inequalities in~\eqref{eq:stage1-noice-water} remain sufficient when $7/27$ is replaced by $\sigma$.

After the first non-pair checkpoint, $h<5/9$ persists. The stronger singleton water bound also persists through the initial pair and standard reduction updates of Stage~2 while $P(\bW)>2k$. Indeed, every anchor remainder is at least $4/9$: this follows from cleanliness, the singleton bound, the two-pebble water inequality, or the presence of two remaining pebbles. An external ordinary pair therefore gives its holder a gain greater than $1/18$, and any newly formed singleton receives at least $4/9$ water. Standard reductions only increase outside pebble values and distribute gains, preserving this bound. Thus it is available even if an equality-producing $\reduction{2}$ occurs in a later pair round.
\end{proof}

\section{Stage 2: Proofs and Procedures}
\label{apdx:stage2}
\subsection{Ordinary Rounds Without Ice}
\label{apdx:stage2-noice}
We give the complete maintenance underlying Subsection~\ref{subsec:stage2-noice} and prove Lemmas~\ref{lem:noice-high} and~\ref{lem:noice-ordinary-triple}. The ordinary-triple procedures assume the full entrance inequalities following the repair in Subsection~\ref{subsec:stage2-repair}; verification of that entrance is separated below. Intervening pair and reduction rounds remain the subject of Appendix~\ref{apdx:stage2-other}.

All updates below use the divisibility of water permitted in Section~\ref{sec:analysis}. A water transfer changes its allocation among bundles without creating an additional item in the ordered witness sequence. Pebbles remain whole. Thus exchanges do not alter the witness order, and validity after deletion follows from the dominance of the three deleted pebbles. Removing a container includes distributing its unused material as specified in the complete procedures.

\subsubsection{Repair before no-ice maintenance}
\label{apdx:stage2-repair}
\begin{proof}[Proof of Lemma~\ref{lem:stage2-repair}]
The input is either an allocation produced by the Stage~1 maintenance, possibly followed by the initial pair and standard reduction updates of Stage~2, or a clean allocation. Write $h=v_i(w_1)<5/9$. At an ordinary-triple checkpoint put $\sigma=\min\{v_i(w_{k+1}),7/27\}$; pair failure gives $h+\sigma<7/9$. For an earlier equality-producing reduction, use $\sigma=7/27$ only in the water estimates below. A clean one-pebble bundle has water greater than $1-h>4/9$; a deficient one can arise only in the first type of input and has water at least $4/9$ by Appendix~\ref{apdx:stage1-repair-proof}. A zero-pebble bundle has water at least $25/27$ in either case. None of the following estimates requires $\reduction{2}$ to have failed.

Let $s$ be the value of the imported pebble and
\[
 t=\max\{2/9,\min\{s,1/3\}-1/27\}.
\]
Then $2/9\le t\le8/27$ and $s-t\ge0$. If $s\le1/3$, the donor loses at most $1/27$ in value. Thus a three-pebble donor leaves two with deficit at most $2/27$, and a four-pebble donor leaves three with deficit at most $1/27$. If $s>1/3$, a three-pebble donor retains value greater than $2/3+8/27=26/27$, and a larger donor remains clean. A donor currently having at least five pebbles retains at least four pebbles and at least $2/9$ water and is clean. Apply these estimates inductively: a donor may be selected again while it has at least three pebbles, using its current count-dependent deficit bound. Donors never become targets, so their water only increases, by at least $2/9$ at every donation. Every donor therefore satisfies the required deficit and water bounds.

The target gains $s-t\ge0$, so its deficit does not increase. A zero-pebble target receives two pebbles and retains water at least $25/27-16/27=1/3$. For a one-pebble target, let its original water be $F\ge4/9$ and its pebble value be $b$. If $s\le\sigma$, then $t=2/9$. If $\sigma<s\le b$ and $s\le7/27$, then
\[
 F-t+s-\sigma>F-2/9\ge2/9.
\]
For $\sigma<s\le b$ and $s\ge7/27$, use $s-t\ge1/27$ and $\sigma\le7/27$ to obtain the same bound. Finally, if $s>b$, then $F+b\ge25/27$ and $t\le8/27$ give
\[
 F-t+(b-\sigma)_+\ge25/27-8/27-7/27=10/27.
\]
Thus every repaired target satisfies the two-pebble water inequality. These estimates also prove affordability at each step, including both imports into a zero-pebble target.

In the Stage~1 case, all untouched bundles retain~\eqref{eq:stage1-noice-water}, which is sufficient because $\sigma\le7/27$. In the clean case, untouched bundles have no deficit, while every newly deficient donor receives at least $2/9$ water. A donor exists whenever a bundle has fewer than two pebbles, by $P(\bW)>2k$. Each step fills one missing pebble and leaves at least two in the donor, so the procedure terminates. Item values and their order do not change, preserving validity and, at an ordinary-triple checkpoint, $h+\sigma<7/9$. This proves~\eqref{eq:noice-high-invariants} at the high-anchor entrance. For any deficient two-pebble output with smaller pebble $a$, its water inequality gives $v_i(a)+F(W)\ge4/9$ because $v_i(a),\sigma\ge2/9$; a clean output has the same remainder bound from $1-h>4/9$. Thus the earlier equality-producing reduction has the required entrance as well. Even for a clean input, the repair may create deficits; these satisfy the count-dependent bounds in~\eqref{eq:no-ice-bundle-bounds}.
\end{proof}

\subsubsection{High-anchor maintenance when $P(\bW)\ge2k+2$}
\label{apdx:noice-high-general}
Write $h=v_i(w_1)$, $P=P(\bW)$, $\tau=\sigma+2/9$, and $s_W,m_W$ for the smallest and largest pebble values in $W$.
Thus $2/9\le\sigma\le7/27$, $4/9\le\tau\le13/27$, and every bundle has value at least $25/27$.
All ranks are fixed during an update; exchanges move whole pebbles and existing water without changing item values. Ties use the fixed witness order, and one largest donor pebble is retained throughout preparation.

The failed $\reduction{2}$ condition gives
\begin{equation}
\label{eq:app-high-call-bounds}
\begin{gathered}
 v_i(w_{2k+1})<7/27,\qquad
 2v_i(w_{2k})+v_i(w_{2k+1})<7/9,\\
 v_i(w_{2k}+w_{2k+1})<14/27,\qquad v_i(w_{2k})<5/18.
\end{gathered}
\end{equation}
The last two inequalities follow from the preceding one, using $v_i(w_{2k})\ge v_i(w_{2k+1})\ge2/9$.
Also $h+v_i(w_{P-1})+v_i(w_P)>7/9$, so maximality gives $\ell\ge P-1$ and $t\ge P$.
Consequently every triple of pebbles containing $w_1$ dominates the real triple.

\paragraph{One exchange for the preparation.}
The following exchange covers both pebbles worth at most $\sigma$ and the exceptional pebble $b=w_{2k}>\sigma$.

\begin{algorithm}[H]
\caption{Exchange used in the high-anchor maintenance}
\label{alg:noice-exchange}
\begin{algorithmic}[1]
\Require A two-pebble bundle $B$, a pebble $q\in B$, and a donor pebble $a\in Y$, where $B\ne Y$.
\Require If $v_i(q)\le\sigma$, then $v_i(a)>\sigma$; otherwise $q=b=w_{2k}$, $\sigma=7/27$, $q$ is smallest in $B$, $s_B+F(B)\ge13/27$, and $v_i(a)\ge v_i(b)$.
\Require The chosen $a$ is distinct from the retained largest pebble of $Y$.
\State Set $\lambda=\max\{\sigma,v_i(q)\}$ and $t=\min\{v_i(a)-\lambda,4/9-v_i(q)\}$.
\State Exchange $a,q$ and send water of value $t$ from $B$ to $Y$.
\end{algorithmic}
\end{algorithm}

For $v_i(q)\le\sigma$, the invariant gives $F(B)\ge2/9$ if $B$ is deficient; otherwise $F(B)\ge1-h-\sigma>2/9$.
Hence $0\le t\le4/9-v_i(q)\le2/9$ is affordable.
For $q=b$, the formula is exactly
\begin{equation}
\label{eq:app-high-scarce-payment}
 t=\min\{v_i(a),4/9\}-v_i(b),
\end{equation}
and $F(B)\ge13/27-v_i(b)>4/9-v_i(b)\ge t\ge0$.
In both cases $B$ gains value, since $v_i(a)-v_i(q)-t\ge\lambda-v_i(q)\ge0$.
If its unchanged pebble remains largest, its new smaller-pebble-plus-water value is at least $\lambda+F(B)\ge\tau$.
Otherwise it is at least
\[
 v_i(B)-v_i(q)-t\ge25/27-4/9=13/27\ge\tau.
\]
Thus every affected two-pebble holder retains its deficit bound and the stronger inequality $s_B+F(B)\ge\tau$.

When $t=v_i(a)-\lambda$, the effective cost to $Y$ of importing and then deleting $q$ is $\lambda$.
In the other case, its retained largest value $m\ge v_i(a)$ satisfies
\begin{equation}
\label{eq:app-high-large-exchange}
 m+t>8/9+\lambda-2v_i(q)\ge8/9-\lambda>11/18,
\end{equation}
because $\lambda\le7/27$ for $q\le\sigma$ and $\lambda=v_i(b)<5/18$ otherwise.
Later preparation exchanges only add water and retain $m$.
Thus a two-pebble deletion from this donor, followed by an anchor remainder worth at least $4/9$, leaves value greater than $1$.

\paragraph{Choosing the donor pebbles.}
Let $E=P-2k$, let $M$ count the pebbles worth at most $\sigma$, and let $R$ count bundles with at least three pebbles.
Then
\[
 M\ge E\ge R;
 \qquad \sigma<7/27\ \Longrightarrow\ M\ge E+1.
\]
Indeed, $E=\sum_W(P(W)-2)\ge R$; the first bound on $M$ follows from $w_{2k+1}<7/27$ when $\sigma=7/27$.
Otherwise~\eqref{eq:noice-indices} and $w_{2k}<5/18$ give $w_{2k}\le\sigma$.
The following algorithm gives the complete two-pebble preparation.

\begin{algorithm}[H]
\caption{Preparing two pebbles for the high-anchor exchange}
\label{alg:noice-high-full-pair}
\begin{algorithmic}[1]
\Require The other pebble of $X$ exceeds $\sigma$, and either $E\ge2$, $M\ge2$, or the unique pebble worth at most $\sigma$ is in a three-pebble bundle.
\State Choose $Y$ with at least three pebbles maximizing the number worth at most $\sigma$; retain one largest pebble.
\If{$Y$ contains at least two pebbles worth at most $\sigma$}
    \State Choose its two least pebbles other than the retained maximum.
\ElsIf{$Y$ contains one and a two-pebble holder contains another}
    \State Import the latter by Algorithm~\ref{alg:noice-exchange}, using an original nonmaximum pebble of $Y$ worth more than $\sigma$.
\ElsIf{$Y$ contains none}
    \State Import two from two-pebble holders by Algorithm~\ref{alg:noice-exchange}, using two distinct original nonmaximum pebbles of $Y$.
\Else
    \State Set $b=w_{2k}$.
    \If{$b$ is in a three-pebble bundle}
        \State Choose that bundle as $Y$, retaining a largest pebble distinct from $b$.
    \Else
        \State Import $b$ into $Y$ by Algorithm~\ref{alg:noice-exchange}, using its smaller pebble worth more than $\sigma$; the holder of $b$ may be $X$.
    \EndIf
    \State Use $b$ and the pebble of $Y$ worth at most $\sigma$.
\EndIf
\State \Return $Y$ and the chosen or prepared pair.
\end{algorithmic}
\end{algorithm}

The first three cases follow from $M\ge E$; two imports may use the same holder, since the first exchange preserves its required inequality.
In the last case no two-pebble holder has a pebble worth at most $\sigma$, and each other bundle has at most one.
Thus $M=E=R$, every such bundle has exactly three pebbles, and $\sigma=7/27$.
Exactly $2k$ pebbles exceed $\sigma$, so $b$ is their smallest and
\begin{equation}
\label{eq:app-high-scarce-pair}
 7/27<v_i(b)<5/18,\qquad v_i(b+q)<14/27
 \quad\text{whenever }v_i(q)\le\sigma.
\end{equation}
If $b$ is in a two-pebble holder, both pebbles exceed $\sigma$.
Its invariant gives $s_B+F(B)\ge13/27$ when deficient; when nondeficient this follows from $s_B+F(B)\ge1-h>13/27$.
Therefore the last import satisfies Algorithm~\ref{alg:noice-exchange}, including when its holder is $X$.
All donor maxima are retained, and the prepared pair has effective cost at most $2\sigma$, unless~\eqref{eq:app-high-large-exchange} applies.

\paragraph{The update and its bounds.}
The complete update for $E\ge2$ is as follows.

\begin{algorithm}[H]
\caption{High-anchor maintenance without ice when $P(\bW)\ge2k+2$}
\label{alg:noice-high-full-general}
\begin{algorithmic}[1]
\Require The bundle bounds~\eqref{eq:no-ice-bundle-bounds}, the high-anchor invariants, $h>1/3$, and $E\ge2$.
\State Let $X$ contain $w_1$.
\If{$P(X)\ge3$}
    \State Delete $w_1$ and two other pebbles; distribute the remainder, remove $X$, re-index, and return.
\EndIf
\State Let $s$ be the other pebble of $X$.
\If{$v_i(s)\le\sigma$}
    \State Choose $Y$ with at least three pebbles containing a pebble $q$ worth at most $\sigma$, if one exists.
    \State Otherwise choose any $Y$ with at least three pebbles and import a pebble $q$ with $v_i(q)\le\sigma$ from a two-pebble bundle $B\ne X,Y$ by Algorithm~\ref{alg:noice-exchange}, using the minimum pebble of $Y$.
    \State Exchange $q$ for all water of $X$ and remove $X$.
\Else
    \State Prepare $Y$ and its pair by Algorithm~\ref{alg:noice-high-full-pair}.
    \State Exchange that pair for $X\setminus\{w_1\}$ and remove $X$.
\EndIf
\State Re-index the surviving items.
\end{algorithmic}
\end{algorithm}

All net costs and donor deficits in the following estimates refer to the donor before preparation.
If $P(X)\ge3$, other bundles receive only gains. A received pebble clears any deficit, since $2/9>2/27$; water gains preserve the inequalities.
Suppose next that $v_i(s)\le\sigma$. Then $F(X)\ge2/9$, by its invariant or $F(X)\ge1-h-\sigma>2/9$.
If no bundle with at least three pebbles contains a pebble worth at most $\sigma$, every such pebble lies in a two-pebble bundle. Since $M\ge E\ge2$ and at most one lies in $X$, a holder $B\ne X$ exists; also $B\ne Y$ because $P(Y)\ge3$. Thus the preparation leaves $F(X)$ unchanged.
The effective donor cost is at most $\sigma$, so its final loss is at most $\sigma-F(X)\le1/27$.
A donor with three pebbles therefore leaves two with deficit at most $2/27$; a donor with four leaves three with deficit at most $1/27$; a larger donor leaves four pebbles and at least $2/9$ water and is nondeficient.
Every deficient output receives at least $2/9$ water.
If the payment instead uses the $4/9-v_i(q)$ cap, the removed donor minimum exceeds $4/9$; two remaining pebbles and $F(X)\ge2/9$ already exceed value $1$.

Finally suppose $v_i(s)>\sigma$.
The anchor remainder has value $H=v_i(s)+F(X)\ge\tau$, by its invariant or $1-h>\tau$.
Preparation changes it only if $b=s$ is imported into $Y$; in that case the exchange does not decrease $X$, so still $H\ge\tau$ after preparation.
If a cap is used,~\eqref{eq:app-high-large-exchange} makes the final merged survivor nondeficient.
Otherwise the prepared pair has effective cost $Q\le2\sigma$, giving
\[
 D(\text{survivor})\le D(Y)+Q-H
       \le D(Y)+\sigma-2/9\le D(Y)+1/27.
\]
This proves the required two- and three-pebble deficit bounds.
A donor with at least five pebbles leaves three pebbles in its own remainder, whose value together with $H\ge4/9$ exceeds $1$.
For a three-pebble donor that was deficient, either its old water is at least $2/9$, or its retained largest pebble and old water have value at least $\tau$.
A nondeficient donor instead has a post-preparation remainder worth at least $1-Q\ge1-2\sigma\ge\tau$.
Together with $H\ge\tau$, these one-pebble remainders yield either $2/9$ water or smaller-pebble-plus-water value at least $\tau$.
For a four-pebble donor, the largest survivor pebble and water have value at least $H\ge\tau$.
All outside holders retain their bounds by the exchange proof.

Every branch deletes three pebbles including $w_1$ and one bundle, so $P'-2k'=E-1>0$.
Validity follows from dominance; no item value increases.
The fixed-$\sigma$ inequality persists, and~\eqref{eq:noice-indices} persists because deleting old $w_1$ leaves every old top-$k$ survivor among the new first $k-1$ items.
The value bound $25/27>7/9+\Delta'$ gives \textnormal{\textsc{WLC}}.
If the new largest value is at most $1/3$, either water bound for a deficient three-pebble bundle gives largest-pebble-plus-water value at least $4/9$, supplying~\eqref{eq:noice-low-invariants}.

This proves Lemma~\ref{lem:noice-high}.

\subsubsection{High-anchor maintenance when $P(\bW)=2k+1$}
\label{apdx:noice-high-boundary}
Here we need only validity and a final deficit of at most $2/27$ in every survivor; no water or index inequality is required after the update. There is a unique three-pebble bundle, and all other bundles have two pebbles. If it contains $w_1$, delete its three pebbles and distribute its water. Otherwise let $X$ contain $w_1$, let $s$ be its other pebble, and let $Y$ be the three-pebble bundle. When $M\ge2$ or the unique pebble worth at most $\sigma$ lies in $Y$, the preceding construction already gives the required deficit bound.

It remains to handle $M=1$ with the unique such pebble $q$ outside $Y$. Then $\sigma=7/27$, and $b=w_{2k}$ satisfies
\begin{equation}
\label{eq:app-boundary-values}
 7/27<v_i(b)<5/18,\qquad 2v_i(b)+v_i(q)<7/9.
\end{equation}
The bundle $Y$ is nondeficient: all its pebbles exceed $7/27$, so either water alternative for a deficient three-pebble bundle would give value greater than $1$.

\begin{algorithm}[H]
\caption{The high-anchor boundary with one pebble worth at most $\sigma$}
\label{alg:noice-high-full-boundary}
\begin{algorithmic}[1]
\Require $E=1$, $P(X)=2$, $\sigma=7/27$, and the unique $q$ worth at most $\sigma$ is outside $Y$.
\State Set $b=w_{2k}$; retain a largest pebble of $Y$ distinct from $b$ if $b\in Y$.
\If{$q=s$}
    \State If needed, import $b$ by Algorithm~\ref{alg:noice-exchange}, using the minimum pebble of $Y$.
    \State Exchange $b$ for all water of $X$ and remove $X$.
\Else
    \State Import $q$ by Algorithm~\ref{alg:noice-exchange}, using an original nonmaximum pebble of $Y$ distinct from $b$.
    \State If needed, import $b$ by the same exchange, using the other original nonmaximum pebble of $Y$.
    \State Exchange $b,q$ for $X\setminus\{w_1\}$ and remove $X$.
\EndIf
\State Re-index the surviving witness items.
\end{algorithmic}
\end{algorithm}

The preceding exchange calculation proves feasibility and shows that outside holders only gain value. For an import of $b$, its holder has smaller-pebble-plus-water value at least $13/27$, including after importing $q$ from that holder and when $b=s$.

A capped import leaves a nondeficient output by the preceding cap estimate; in the one-pebble branch, the removed donor minimum exceeds $4/9$, so the two retained pebbles and $F(X)\ge2/9$ suffice. Without a cap, if $q=s$, the donor loses less than $5/18-2/9=1/18$. If $b=s$, its loss after merging is at most $7/27-F(X)<1/18$, since $F(X)\ge13/27-v_i(b)>11/54$. Otherwise the retained donor remainder exceeds $1-7/27-5/18=25/54$ and receives an anchor remainder worth at least $13/27$. In every case the final value exceeds $17/18>25/27$. Thus every surviving deficit is at most $2/27$.

The deleted triple contains $w_1$ and dominates the real triple, so validity is preserved. Put $k'=k-1$. We have $P'=2k'$ and \textnormal{\textsc{WLC}}, since every survivor has value at least $25/27$ and $\Delta'<4/27$. Deleting $w_1$ and the old failed pair test give
\[
 v_i(w'_1+w'_{k'+1})\le v_i(w_1+w_{k+1})<7/9.
\]
The item $w'_{2k'+1}$ is a non-pebble, so Lemma~\ref{lem:canonical-entry} gives immediate canonical entry. This proves Lemma~\ref{lem:noice-high-boundary} using only the final $2/27$ deficit bound.

\subsubsection{Low-anchor ordinary triples}
\label{apdx:noice-low}

We give the low-anchor part of Algorithm~\ref{alg:noice-low-overview} and Lemma~\ref{lem:noice-ordinary-triple}. Throughout this construction, the real ordinary triple is $\{u_1,u_\ell,u_t\}$, where $t=\max\{\ell+1,2k+1\}$, and it is allocated to an agent other than the fixed agent $i$. Thus $k\ge2$. Assume that no ice remains, $P(\bW)>2k$, $v_i(w_1)\le1/3$, and the count, deficit, and water inequalities in~\eqref{eq:no-ice-bundle-bounds} and~\eqref{eq:noice-low-invariants} hold. Item indices below are the indices at the start of the round. Moving items between bundles does not change their values or these indices.

The complete update has three consecutive parts. The first gives the bundle containing $w_1$ two or three pebbles. The second is used only when it has two, and records the donor from which a third pebble is obtained. The final part corrects the two remaining indices, deletes the resulting dominance triple, and returns the remaining water. All water transfers use the divisibility permitted in Section~\ref{sec:analysis}.

\begin{algorithm}[H]
\caption{Low-anchor maintenance: preparing a two- or three-pebble bundle}
\label{alg:noice-low-full-anchor}
\begin{algorithmic}[1]
\Require The low-anchor assumptions above.
\State Let $X$ be the bundle containing $w_1$.
\If{every bundle has at least four pebbles}
    \State Enter the trivial dense case and \Return.
\EndIf
\If{$P(X)\ge4$}
    \State Choose a bundle $B$ with two or three pebbles, and let $y$ be its largest pebble.
    \State Exchange $w_1$ and $y$.
    \State Transfer water of value $v_i(w_1)-v_i(y)$ from $B$ to the old bundle $X$.
    \State Let $X$ now denote $B$, the bundle containing $w_1$.
\EndIf
\State \Return $X$, which has two or three pebbles.
\end{algorithmic}
\end{algorithm}

\begin{algorithm}[H]
\caption{Low-anchor maintenance: obtaining a third pebble}
\label{alg:noice-low-full-third}
\begin{algorithmic}[1]
\Require The prepared bundle $X$ has two pebbles.
\State Let $e=w_{2k+1}$.
\If{$e\in X$}
    \State Choose a bundle $Y$ with at least three pebbles, and let $q$ be its least-valued pebble.
    \State Move $q$ from $Y$ to $X$, and transfer water of value $v_i(q)$ from $X$ to $Y$.
\Else
    \State Let $B$ be the bundle containing $e$.
    \If{$P(B)\ge3$}
        \State Set $Y=B$.
    \Else
        \State Choose a bundle $Y$ with at least three pebbles containing a pebble $a$ whose index is at most $2k$.
        \State Exchange $a$ and $e$, and transfer water of value $v_i(a)-v_i(e)$ from $B$ to $Y$.
    \EndIf
    \State Move $e$ from $Y$ to $X$, and transfer water of value $v_i(e)$ from $X$ to $Y$.
\EndIf
\State \Return $X$ and the donor $Y$; every transfer in this part preserves each participating bundle's value.
\end{algorithmic}
\end{algorithm}

\begin{algorithm}[H]
\caption{Low-anchor maintenance: correcting the indices and deleting the triple}
\label{alg:noice-low-full-delete}
\begin{algorithmic}[1]
\Require The prepared bundle $X$ has three pebbles; remember $Y$ if a third pebble was acquired.
\State Write the two pebbles of $X\setminus\{w_1\}$ as $q_1,q_2$, in increasing order of index.
\If{the index of $q_1$ exceeds $\ell$}
    \State Let $B$ contain $w_\ell$; exchange $q_1$ and $w_\ell$.
    \State Transfer water of value $v_i(w_\ell)-v_i(q_1)$ from $X$ to $B$, and set $q_1=w_\ell$.
\EndIf
\If{the index of $q_2$ exceeds $t$}
    \State Let $B$ contain $w_t$; exchange $q_2$ and $w_t$.
    \State Transfer water of value $v_i(w_t)-v_i(q_2)$ from $X$ to $B$, and set $q_2=w_t$.
\EndIf
\State Delete $w_1,q_1,q_2$ from $X$.
\If{a third pebble was acquired}
    \State Transfer all remaining water of $X$ to the remembered donor $Y$.
\Else
    \State Distribute all remaining water of $X$ among the other bundles.
\EndIf
\State Remove $X$ and re-index the remaining witness items.
\end{algorithmic}
\end{algorithm}

\begin{proof}[Low-anchor part of Lemma~\ref{lem:noice-ordinary-triple}]
Write $h=v_i(w_1)\le1/3$. The preparation gives every bundle at least two pebbles. By~\eqref{eq:no-ice-bundle-bounds}, every bundle has value at least $25/27$, and every bundle with at least four pebbles has value at least $1$. The positive part $\max\{D(W),0\}$ is the shortfall used in the proof; a transfer of surplus from a bundle that remains worth at least $1$ does not create a positive deficit.

\paragraph{Preparing the anchor.}
If every bundle has at least four pebbles, then $P(\bW)\ge4k$ and $S(\bW)\ge4k$, so the trivial dense case applies. Otherwise, suppose that the current bundle $X$ containing $w_1$ has at least four pebbles. A bundle $B$ with two or three pebbles exists. Let $y$ be its largest pebble. We first show
\[
    F(B)\ge h-v_i(y).
\]
For a two-pebble bundle,
\[
    F(B)\ge25/27-2v_i(y)\ge h-v_i(y),
\]
because $h+v_i(y)\le2/3$. For a three-pebble bundle of value at least $1$,
\[
    F(B)\ge1-3v_i(y)\ge h-v_i(y),
\]
because $h+2v_i(y)\le1$. For a deficient three-pebble bundle, the required inequality is exactly its low-anchor water inequality in~\eqref{eq:noice-low-invariants}.

After exchanging $w_1$ and $y$, transfer exactly $h-v_i(y)$ water from $B$ to the old bundle $X$. The preceding inequality makes this payment feasible. The old bundle $X$ recovers exactly the pebble value it lost; the new bundle containing $w_1$ pays exactly the value it gained. Both total bundle values are therefore unchanged.

The new anchor has the old pebble count of $B$, so it satisfies the appropriate two- or three-pebble deficit bound. The other bundle still has at least four pebbles and value at least $1$. If the new anchor has three pebbles and positive deficit, its largest pebble is $w_1$, so its low-anchor water inequality holds automatically. No other bundle changes. Thus this preparation preserves all required inequalities and creates no positive deficit.

\paragraph{Obtaining a third pebble.}
Suppose now that the prepared anchor $X$ has two pebbles. The item $e=w_{2k+1}$ is a pebble because $P(\bW)>2k$, and the failed $\reduction{2}$ test gives
\[
    2/9\le v_i(e)<7/27.
\]
Since both anchor pebbles are worth at most $h$, its water value is at least
\[
    25/27-2h\ge7/27.
\]
If $e$ belongs to another bundle with at least three pebbles, buying it with exactly $v_i(e)$ water is therefore feasible. Its donor keeps at least two pebbles and the same total value.

If $e$ belongs to a two-pebble bundle $B$, a donor $Y$ with at least three pebbles and a pebble $a$ among the first $2k$ exists. Indeed, if no bundle with at least three pebbles contained such an item, the first $2k$ pebbles would all lie in the two-pebble bundles. As at least one bundle has more than two pebbles, those bundles have capacity at most $2(k-1)<2k$, a contradiction. The index of $a$ gives $v_i(a)\ge v_i(e)$, and
\[
    0\le v_i(a)-v_i(e)\le1/3-2/9=1/9.
\]
The water of $B$ is at least $25/27-2h\ge7/27$, so it can pay this difference while exchanging $e$ for $a$. Both bundles retain their values. The pebble $e$ can then be bought from $Y$ with $v_i(e)$ water from the anchor, as above.

If $e$ is already in the anchor, its water value is strictly greater than
\[
    25/27-1/3-7/27=1/3.
\]
A donor with at least three pebbles exists because $P(\bW)>2k$, and its least pebble has value at most $h\le1/3$. Buying that pebble with equal-value water is feasible. In every case the anchor now has three pebbles, retains its value, and contains a pebble of value at most $v_i(e)<7/27$.

The donor loses exactly one pebble while keeping its value. A three-pebble donor becomes a two-pebble bundle with deficit at most $1/27$, which is within the allowed $2/27$. A donor with four pebbles was worth at least $1$ and remains worth at least $1$ after becoming a three-pebble bundle. A larger donor retains at least four pebbles and value at least $1$. The optional two-pebble holder keeps both its count and value.

The routing exchange also preserves the water inequality of any deficient three-pebble donor before the purchase. More generally, when a bundle gives a pebble of value $b$ in exchange for a pebble of value $a\le b$ and water $b-a$, the value of its largest pebble together with its water does not decrease. If the removed pebble was not largest, the old largest remains and water increases. If it was largest, the new maximum is at least $a$, and adding the received water gives at least $a+(b-a)=b$, in addition to the old water. This observation will also cover the exchanges correcting the two indices below.

\paragraph{Correcting the indices.}
At this point the anchor has exactly three pebbles and value at least $25/27$. Order its two nonleading pebbles as $q_1,q_2$ by their indices. If the index of $q_1$ exceeds $\ell$, the item $w_\ell$ is external to the anchor: both $q_1$ and $q_2$ have larger indices, and $w_1$ has index $1$. It is also a pebble, because its index is earlier than that of a pebble. We show that exchanging $q_1$ for $w_\ell$ and paying the exact increase from anchor water is feasible.

The maximal choice of $\ell$, together with validity, gives
\[
    v_i(w_1)+v_i(w_{\ell+1})
       +v_i\bigl(w_{\max\{\ell+2,2k+1\}}\bigr)<7/9.
\]
If $\ell\ge2k$, the index of $q_2$ is at least $\ell+2$, and all pebbles have values between $2/9$ and $1/3$. The new three-pebble value is therefore strictly less than
\[
    7/9+\bigl(v_i(w_\ell)-v_i(w_{\ell+1})\bigr)
    \le7/9+1/9=8/9.
\]
If $\ell<2k$, the maximality inequality instead uses $e=w_{2k+1}$. We have $v_i(q_2)\le v_i(w_{\ell+1})$, so the new three-pebble value is strictly less than
\[
    7/9+\bigl(v_i(w_\ell)-v_i(e)\bigr)
    \le8/9.
\]
Since the anchor's total value is at least $25/27>8/9$ and the exchange preserves that value, its water can pay the required increase.

After this step, the middle index is at most $\ell$. If the last index exceeds $t$, then $w_t$ is external to the anchor and is a pebble. Moreover,
\[
    v_i(w_t)\le v_i(e)<7/27.
\]
The other two pebbles have values at most $1/3$, so the new three-pebble value is strictly less than $25/27$. This exchange is again payable from anchor water. The outside holder receives equal total value and retains its pebble count. The preceding observation shows that a deficient three-pebble holder also retains its low-anchor water inequality.

\paragraph{Deletion and the water return.}
The final pebble indices are $1$, at most $\ell$, and at most $t$. They therefore form a dominance bundle for the real triple. Deleting them preserves validity by Proposition~\ref{prop:remove-dominance}. All earlier operations merely changed their containers and the distribution of water, so they also preserved validity.

If no third pebble was acquired, all remaining material in the deleted anchor is water and can be distributed as gains. If a third pebble was acquired, the prepared anchor contained a pebble worth less than $7/27$. Correcting the middle index leaves that last pebble in place; correcting the last index replaces it by a pebble no larger than $e$. Thus the final triple has value strictly below $25/27$. Throughout acquisition and index correction the anchor retained its pre-acquisition value, which was at least $25/27$. Returning all remaining water to the original donor gives that donor a strictly positive gain equal to the pre-acquisition anchor value minus the deleted triple value. All other outside exchanges preserved value. In particular, no surviving bundle acquires a new positive deficit or an increase in its positive deficit.

Exactly three pebbles and one bundle have been removed. Every surviving bundle has at least two pebbles and satisfies~\eqref{eq:no-ice-bundle-bounds}. A newly three-pebble donor came from a bundle of at least four pebbles and still has value at least $1$; every other deficient three-pebble bundle retained the value of its largest pebble together with its water. Since the new largest item has value at most $h$,~\eqref{eq:noice-low-invariants} remains true for the new round whenever $P(\bW)>2k$ continues to hold. Non-pebble values remain below $4/27$, and every bundle has value at least $25/27$, so \textnormal{\textsc{WLC}} holds as well.

Finally, let $k'=k-1$. If the updated pebble count is $2k'$, each bundle has exactly two pebbles. Write the new order and non-pebble bound as $w'_1,w'_2,\ldots$ and $\Delta'$. The new largest pebble is still worth at most $1/3$, while $w'_{2k'+1}$ is a non-pebble, so
\[
    v_i\bigl(w'_1+w'_{k'+1}+w'_{2k'+1}\bigr)
    \le2/3+\Delta'
    <7/9+\Delta'.
\]
Validity, \textnormal{\textsc{WLC}}, and $P(\bW)=2k'\le3k'$ now permit canonical reorganization by Lemma~\ref{lem:canonical-entry}. This proves the low-anchor ordinary-triple claim. Pair or reduction rounds occurring between ordinary triples still require their separate maintenance interfaces.
\end{proof}

\subsection{Ordinary Rounds With Ice}
\label{apdx:stage2-ice}
\subsubsection{Clean maintenance with solid ice}
\label{apdx:clean-ice}
We complete the branches of Algorithm~\ref{alg:stage2-clean-ice}. The input is an ordinary real triple allocated to another agent, a valid witness allocation with $P(\bW)>2k$, and $v_i(W)\ge1$ for every bundle. Thus $k\ge2$, and the pair and $\reduction{2}$ tests have failed. Write $h=v_i(w_1)$ and let $X$ contain $w_1$. The case $h<13/27$ is given in Appendix~\ref{apdx:clean-divisible}; the following procedures apply when $h\ge13/27$ and retain every remaining ice item whole.

\paragraph{A dominance triple and its deletion.}
Let $a$ be an ice item with index $s$, and let $q\ne w_1$ be a pebble or an ice item of smaller index. Since $s>P(\bW)>2k$,
\[
    v_i(w_1+w_{s-1}+w_s)\ge13/27+2(4/27)=7/9.
\]
Validity and maximality of the real second index imply $\ell\ge s-1$ and $t\ge s$. Hence $\{w_1,q,a\}$ dominates the real triple. For two ice items, take $a$ to be the later one. Their triple has value less than $5/9+2(2/9)=1$. For a pebble $q$ of index greater than $k$, the failed pair condition gives $h+v_i(q)<7/9$, so the triple again has value less than $1$.

The next subroutine is used whenever the triple lies in $X$ and one further bundle $Y$.

\begin{algorithm}[H]
\caption{Deleting a triple from at most two witness bundles}
\label{alg:clean-ice-delete}
\begin{algorithmic}[1]
\Require A dominance triple $T\subseteq X\cup Y$, with $w_1\in X$, $v_i(T)<1$, and every bundle worth at least $1$.
\If{$X=Y$}
    \State Distribute $X\setminus T$ among the other bundles, keeping ice and pebbles whole.
\Else
    \State Replace $Y$ by $(X\cup Y)\setminus T$.
\EndIf
\State Delete $T$, remove $X$, and re-index.
\end{algorithmic}
\end{algorithm}

If $X\ne Y$, the merged survivor has value at least $2-v_i(T)>1$. If $X=Y$, all survivors only receive gains. In either case exactly the triple and one container are removed. Regrouping keeps the item values unchanged, and the dominance deletion preserves validity.

\paragraph{Preparing a pebble and an ice item.}
Assume that no bundle contains two ice items or both ice and a pebble of index greater than $k$. The following subroutine handles either a bundle with fewer than two pebbles or a two-pebble bundle with a pebble of index greater than $k$.

\begin{algorithm}[H]
\caption{Preparing a pebble and an ice item for deletion}
\label{alg:clean-ice-prepare}
\begin{algorithmic}[1]
\Require The preceding failed tests, $h\ge13/27$, and an indicated bundle $Y$.
\State Choose the ice item $a$ of $Y$ if present; otherwise choose the largest current ice item.
\If{$P(Y)<2$}
    \State Let $q$ be the smallest current pebble.
\Else
    \State Let $q$ be the least-valued pebble of $Y$ whose index exceeds $k$.
\EndIf
\For{each of $q,a$ not already in $Y$}
    \State Move that item into $Y$ and pay its former holder the same value in water from $Y$.
\EndFor
\State Complete the preceding two-bundle deletion with target $\{w_1,q,a\}$ in $X,Y$.
\end{algorithmic}
\end{algorithm}

Every selected $q$ has index greater than $k$. If $Y$ has fewer than two pebbles, denote its possible pebble by $z$, reading its value as zero when absent. If $Y$ already contains $a$, then its pebble, if present, has index at most $k$, so $q$ must be imported. Since $v_i(z)\le h$ and $v_i(q)\le v_i(w_{k+1})$,
\[
    v_i(z+q+a)\le h+v_i(w_{k+1})+v_i(a)<1.
\]
Thus $F(Y)>v_i(q)$ when $a$ is already there, and $F(Y)>v_i(q)+v_i(a)$ when it is not. These bounds also fund the payments when $q$ is already present. If $Y$ has two pebbles, it is ice-free by the preceding failed test, and $q$ is already present. Its other pebble $z$ satisfies
\[
    F(Y)\ge1-v_i(z+q)>2/9>v_i(a).
\]
All payments are therefore available. The exchanges preserve every bundle value. The resulting triple is dominant and worth less than $1$ by the preceding argument, so the two-bundle deletion keeps all survivors worth at least $1$.

\paragraph{Reorganizing three bundles.}
The remaining exceptional bundle $B$ has three pebbles, all of index greater than $k$. The preceding failed tests make $B$ ice-free, and every ice-containing bundle has at least two pebbles, all among the first $k$.

\begin{algorithm}[H]
\caption{Three-bundle reorganization before the critical state}
\label{alg:clean-ice-three}
\begin{algorithmic}[1]
\Require The preceding failed tests and a three-pebble bundle $B$ with no pebble among the first $k$.
\State Order the pebbles of $B$ by value as $x_1,x_2,x_3$.
\State Choose a bundle $Y$ with ice item $a$, and set $T=\{w_1,x_1,a\}$.
\If{$Y=X$}
    \State Complete the two-bundle deletion with $X,B,T$, and return.
\EndIf
\State Let $y$ be the least-valued pebble of $Y$.
\State Form $Q_0$ from $x_2,x_3,y$, adding water of value $\max\{0,1-v_i(x_2+x_3+y)\}$ from $B$.
\State Form $Q_1$ from all material of $X\cup B\cup Y$ outside $T\cup Q_0$.
\State Delete $T$, replace $X,B,Y$ by $Q_0,Q_1$, and re-index.
\end{algorithmic}
\end{algorithm}

The three pebbles of $B$ are later than all pebbles of $Y$, so $v_i(y)\ge v_i(x_1)$. Consequently
\[
    F(B)\ge\max\{0,1-v_i(x_1+x_2+x_3)\}
       \ge\max\{0,1-v_i(x_2+x_3+y)\},
\]
which funds $Q_0$. The target $T$ is dominant and worth less than $1$. Also $B\ne X,Y$. If $Y=X$, the previous deletion proof applies. Otherwise the three input bundles are distinct. If the bare triple $x_2,x_3,y$ is worth less than $1$, then $v_i(Q_0)=1$ and $v_i(Q_1)\ge3-v_i(T)-1>1$.

If that bare triple is worth at least $1$, no water is used in $Q_0$. The failed pair test, $h\ge13/27$, and the late index of $x_1$ give
\[
    v_i(X\setminus\{w_1\})>2/9+v_i(x_1),
    \qquad v_i(x_1)<8/27,
    \qquad F(B)>1/9.
\]
If $P(Y)=2$, its remaining pebble is worth at least $v_i(y)$; hence
\[
    v_i(Q_1)>v_i(y)+v_i(x_1)+1/3
       >v_i(y+x_2+x_3)\ge1.
\]
The second strict inequality follows from $v_i(x_2)\le v_i(x_1)$ and $v_i(x_3)<1/3$. If $P(Y)\ge3$, two remaining pebbles of $Y$ instead give
\[
    v_i(Q_1)>2/9+v_i(x_1)+1/9+4/9
       =7/9+v_i(x_1)\ge1.
\]
Thus both outputs have value at least $1$. All retained ice stays whole, and exactly the dominance triple is deleted, proving validity.

\paragraph{Proof of the clean-maintenance lemma and the critical entrance.}
\begin{proof}[Proof of Lemma~\ref{lem:stage2-clean-ice}]
For $h<13/27$, Appendix~\ref{apdx:clean-divisible} gives the full procedure and proof.

Otherwise the first ice branch chooses either two ice items in one bundle or an ice item together with a pebble of index greater than $k$. The dominance and deletion arguments above handle both choices. The branch for a bundle with fewer than two pebbles, or with two pebbles including a late one, uses the preparation subroutine. The last exceptional branch uses the three-bundle subroutine. Therefore each completed update preserves validity and keeps every survivor worth at least $1$. All retained ice items in these branches remain whole. If a completed update has $\largestwater<4/27$, the transition in Appendix~\ref{apdx:clean-noice-handoff} supplies the no-ice entrance. An equality output instead uses Lemma~\ref{lem:stage2-equality}.

If none of the tests applies, the counting argument in Subsection~\ref{subsec:clean-ice} proves~\eqref{eq:critical-bounds}. No item is changed at this handoff, so every bundle is still worth at least $1$.
\end{proof}

\subsubsection{Clean maintenance when $v_i(w_1)<13/27$}
\label{apdx:clean-divisible}
Fix an ordinary triple allocated to another agent, so $k\ge2$. Suppose that every bundle has value at least $1$, $P(\bW)>2k$, and $h=v_i(w_1)<13/27$. Melt all remaining ice. We use the bound $\largestwater<2/9$ in this procedure.

First give every bundle at least two pebbles. While $Z$ has fewer than two, choose $Y$ with at least three pebbles, move its least-valued pebble $q$ to $Z$, and pay $Y$ exactly $v_i(q)$ in water from $Z$. A donor exists because $P(\bW)>2k$. If $Z$ currently has one pebble $z$, its water value is at least
\[
    1-v_i(z)\ge1-h>h\ge v_i(q).
\]
A zero-pebble bundle can make its first purchase as well; afterwards the same one-pebble inequality applies. Each purchase preserves both bundle values and reduces the number of missing pebbles. Thus this preparation terminates, changes no item values, and leaves every bundle with at least two pebbles and value at least $1$.

\begin{algorithm}[H]
\caption{Clean ordinary-triple maintenance using water}
\label{alg:clean-divisible-triple}
\begin{algorithmic}[1]
\Require An ordinary triple, $h<13/27$, $P(\bW)>2k$, and every bundle worth at least $1$.
\State Melt all remaining ice.
\While{a bundle $Z$ has fewer than two pebbles}
    \State Choose $Y$ with at least three pebbles and its least pebble $q$.
    \State Move $q$ to $Z$ and pay $Y$ exactly $v_i(q)$ in water from $Z$.
\EndWhile
\If{$h\le1/3$}
    \State Use Algorithms~\ref{alg:noice-low-full-anchor}--\ref{alg:noice-low-full-delete}, paying with water.
    \State Enter the canonical case if the new count is $P(\bW)=2k$, and return.
\EndIf
\State Let $X$ contain $w_1$.
\If{$P(X)\ge3$}
    \State Delete $w_1$ and two other pebbles of $X$; distribute the remainder and remove $X$.
\Else
    \State Choose $Y$ with $P(Y)\ge3$ and distinct pebbles $q,r\ne w_1$ minimizing $v_i(q+r)$.
    \State Require each of $q,r$ to lie in $Y$ or in a two-pebble bundle.
    \State Import each missing pebble into $Y$, exchanging an unchosen pebble $a$ of $Y$ and receiving $v_i(a)-v_i(q)$ or $v_i(a)-v_i(r)$ from its old holder.
    \State Delete $w_1,q,r$, merge the remainders of $X,Y$, and remove $X$.
\EndIf
\State Re-index; if $P(\bW)=2k$ for the new $k$, enter the canonical case immediately.
\end{algorithmic}
\end{algorithm}

\paragraph{The case $1/3<h<13/27$.}
Any three pebbles including $w_1$ dominate the real triple, by the maximality argument used in the high-anchor maintenance. If $P(X)\ge3$, deleting three such pebbles and distributing the remainder gives only gains to survivors.

When $P(X)=2$, the minimizing choice in the algorithm has
\begin{equation}
\label{eq:clean-divisible-pair}
    v_i(q+r)<14/27.
\end{equation}
To verify existence, write $E=P(\bW)-2k$, let $M$ count pebbles worth less than $7/27$, and let $R$ count bundles with at least three pebbles. The failed $\reduction{2}$ test and the prepared pebble counts give $M\ge E\ge R\ge1$. If a bundle with at least three pebbles contains two of these $M$ pebbles, choose it. If a two-pebble holder contains one, two such pebbles can be chosen from the allowed holders whenever $M\ge2$: either another lies in a bundle with at least three pebbles, which is chosen as $Y$, or both lie in two-pebble holders. Their sum is strictly smaller than $14/27$.

Otherwise, when $M\ge2$, all these pebbles lie in bundles with at least three pebbles, at most one per bundle. Then $M=E=R$, every such bundle has exactly three pebbles, and exactly $2k$ pebbles have value at least $7/27$. Put $b=w_{2k}$. If $b$ lies in a three-pebble bundle, choose that bundle and its pebble worth less than $7/27$; otherwise choose any three-pebble bundle and take $b$ from its two-pebble holder. If $M=1$, then $E=R=1$: the unique three-pebble bundle can similarly supply or import both $b=w_{2k}$ and the unique pebble worth less than $7/27$. In all these cases, the failed reduction gives
\[
    v_i\bigl(b+w_{2k+1}\bigr)<14/27,
\]
and every chosen pebble worth less than $7/27$ is no larger than $w_{2k+1}$. This proves~\eqref{eq:clean-divisible-pair} for an allowed choice, and hence for the minimizing choice.

Every import is feasible and preserves values. Minimality ensures that an unchosen pebble $a$ of $Y$ has value at least that of a selected external pebble; otherwise it would give a smaller allowed pair. Such an unchosen pebble exists for each import because $P(Y)\ge3$. The old two-pebble holder $B$ of an imported pebble $q$, with other pebble $z$, has water of total value at least
\[
    1-v_i(z)-v_i(q)\ge1-h-v_i(q)
       >h-v_i(q)\ge v_i(a)-v_i(q).
\]
Thus it can pay the exact difference. This argument also applies to two successive imports from the same holder, and to an import from $X$; both its pebble count and value remain unchanged. After preparation, $X$ and $Y$ still have value at least $1$. The deleted dominance triple has value
\[
    h+v_i(q+r)<13/27+14/27=1.
\]
Their merged remainder therefore has value greater than $1$ and at least two pebbles. All other bundles retain their values and counts.

\paragraph{The case $h\le1/3$.}
The cited low-anchor algorithms specify the whole-item exchanges; the same operations work here with payments in water. We verify the needed bounds. To move $w_1$ from a bundle with at least four pebbles into a bundle with $p\in\{2,3\}$ pebbles and largest pebble $y$, the latter has water of total value at least $1-pv_i(y)\ge h-v_i(y)$, so the exchange is affordable. If no such bundle exists, dense closure applies.

For a two-pebble $X$, its water value is at least $1-2h\ge1/3$. The pebble $e=w_{2k+1}$ is worth less than $7/27$. If it lies in an external bundle with at least three pebbles, buy it directly. If its holder has two pebbles, route $e$ through a donor with at least three pebbles by the specified equal-value exchange; its price difference is at most $1/9$, which the holder can pay. Such a donor with a pebble among the first $2k$ exists by the same counting argument in Algorithm~\ref{alg:noice-low-full-third}. If $e$ is already in $X$, its water value exceeds $1-h-v_i(e)>1/3$, so it can instead buy any donor's least pebble. All these exchanges preserve values and leave each donor at least two pebbles.

The prepared $X$ has three pebbles, includes $w_1$, and has value at least $1$. The index-correction calculation in Appendix~\ref{apdx:noice-low} uses only this structure, $h\le1/3$, maximality of $\ell$, and the failed-test bounds; it does not require other bundles to contain no ice. It bounds the replacement triples by $8/9$ and $25/27$. Hence both exact payments are available from $X$, and the resulting triple is dominant. Deleting it and distributing the remaining water creates only gains. Thus every survivor remains worth at least $1$ and has at least two pebbles.

\paragraph{Continuation and canonical entry.}
Both cases delete three pebbles, including the old $w_1$, and one container. They preserve validity, all surviving values at least $1$, and all item values. Consequently $\largestwater<2/9$ and \textnormal{\textsc{WLC}} hold. The largest pebble value cannot increase, so the same construction remains applicable at subsequent ordinary triples. If the completed update reaches $P'=2k'$ with $k'=k-1$, deleting the old first item gives
\[
    v_i(w'_1)\le v_i(w_1),\qquad
    v_i(w'_{k'+1})\le v_i(w_{k+1}).
\]
The old pair condition failed and $w'_{2k'+1}$ is non-pebble, so $v_i(C'_1)<7/9+\Delta'$. Lemma~\ref{lem:canonical-entry} therefore gives immediate canonical reorganization.

\paragraph{Preparing a clean equality output.}
The following preparation is used for a clean output with $P(\bW)=2k$ and $h=v_i(w_1)<5/9$. It preserves cleanliness and gives every bundle exactly two pebbles for the continuation in Lemma~\ref{lem:stage2-equality}. Inputs with more than $2k$ pebbles use the unified repair at the no-ice entrance instead.

\begin{algorithm}[H]
\caption{Clean preparation for equality continuation}
\label{alg:clean-equality-preparation}
\begin{algorithmic}[1]
\Require Every bundle is worth at least $1$, $P(\bW)=2k$, and $h<5/9$.
\State Melt all remaining ice.
\While{some bundle $Z$ has fewer than two pebbles}
    \State Choose $Y$ with at least three pebbles and let $q$ be its least pebble.
    \State Move $q$ from $Y$ to $Z$.
    \State Pay water of value $\min\{v_i(q),1/3\}$ from $Z$ to $Y$.
\EndWhile
\State \Return the witness allocation.
\end{algorithmic}
\end{algorithm}

A donor exists whenever a bundle has fewer than two pebbles, since $P(\bW)=2k$. A target has water worth at least $1-h>4/9$, while the payment is at most $1/3$. Thus it can pay and its value cannot decrease. If $v_i(q)\le1/3$, the donor receives exactly the value removed. Otherwise its two remaining pebbles and the payment have value at least $2v_i(q)+1/3>1$. Every bundle therefore remains worth at least $1$. Each iteration supplies one missing pebble and creates no new shortage, so the process terminates. All items keep their values and validity is preserved. Every bundle then has exactly two pebbles; \textnormal{\textsc{WLC}} and the larger-pebble remainder bound follow from $v_i(W)\ge1$ and $h<5/9$, as required by Lemma~\ref{lem:stage2-equality}.

\subsubsection{The clean-to-no-ice handoff}
\label{apdx:clean-noice-handoff}
Suppose that a completed clean update has $P(\bW)>2k$ and $\Delta<4/27$. Continue the clean pair and reduction procedures until an ordinary triple is reached, unless a closure case or equality continuation applies first. These updates preserve validity and all bundle values at least $1$, and $\Delta$ cannot increase.

At that ordinary triple, invoke Algorithm~\ref{alg:stage2-repair} with the clean input covered by Appendix~\ref{apdx:stage2-repair}. The resulting allocation has at least two pebbles per bundle and satisfies the no-ice deficit bounds in~\eqref{eq:no-ice-bundle-bounds} and the required water inequalities. At this checkpoint, pair failure gives
\[
    \sigma=\min\{v_i(w_{k+1}),7/27\},
    \qquad h+\sigma<7/9.
\]
Fix this value when the high-anchor case applies. If $\sigma<7/27$, sortedness gives~\eqref{eq:noice-indices}. When $h\le1/3$, every deficient three-pebble bundle has at least $2/9$ water and hence satisfies~\eqref{eq:noice-low-invariants}. Thus the applicable ordinary-triple entrance is established immediately before it is used.

The separate preparation in Algorithm~\ref{alg:clean-equality-preparation} remains available for a clean equality output. Once critical maintenance has begun, it continues independently of this handoff.

\subsubsection{Critical maintenance}
\label{apdx:critical-operations}
Assume~\eqref{eq:critical-bounds},~\eqref{eq:critical-deficit-bounds}, and \eqref{eq:critical-value-bound}. Write $h=v_i(w_1)$ and consider an ordinary triple allocated to another agent, so $k\ge2$. We use the failed-test bounds~\eqref{eq:ordinary-triple-bounds}. All retained pebbles and ice items stay whole; water exchanges preserve values and the ordered witness items.

Decreasing ice in a bundle with at least four pebbles to $1/9$ and melting it leaves value at least $4(2/9)+1/9=1$. Repeat this whenever needed. Redistribution is safe as well: a deficient three-pebble recipient that gains a pebble gains more than $1/9$ even after decreasing its old ice, clearing its deficit; receiving ice clears a deficient ice-free three-pebble recipient because $4/27>1/18$. Under this redistribution, a survivor that remains deficient receives only water, preserving~\eqref{eq:critical-value-bound}. Route any ice from a deleted bundle to an ice-free survivor. If none exists, the deleted bundle has three pebbles and ice, and all other bundles have at most three pebbles and ice. Since $P(\bW)\ge3k$, all have exactly three pebbles and ice, so dense closure applies before deletion.

\subsubsection{High-anchor critical maintenance}
\label{apdx:critical-high}
\begin{proof}[Proof of Lemma~\ref{lem:stage2-critical-high}]
Since $h>1/3$, the candidate using the last two pebbles with $w_1$ is feasible. Maximality gives $\ell\ge P(\bW)-1$ and $t\ge P(\bW)$; hence any three pebbles including $w_1$ dominate the real triple. The branch $P(X)\ge3$ now follows from the redistribution argument above.

If $P(X)=2$, choose the ice-free $Y$ from the algorithm, put $m=P(Y)\ge4$, and write $H=X\setminus\{w_1\}$ and $C=Y\setminus\{q,r\}$. Both inputs are nondeficient. The failed pair bound and the choice of the two least pebbles give
\[
    v_i(H)\ge1-h>4/9,\qquad
    v_i(C)\ge(1-2/m)v_i(Y)\ge1-2/m.
\]
For $m=4$, the output $C\cup H$ has three pebbles and value greater than $17/18$. If it is deficient and ice-free, its largest pebble together with its water has value at least $v_i(H)>4/9$. For $m\ge5$, the output has at least four pebbles and value greater than $3/5+4/9>1$; a subsequent ice decrease also leaves value at least $4(2/9)+1/9=1$.

Thus only a three-pebble output of two nondeficient inputs can acquire deficit, and that deficit is below $1/18$. No previously deficient survivor loses value. Every survivor has at least two pebbles and at most one ice item. Exactly three pebbles and one bundle disappear, so the new count is at least $3(k-1)$. The permitted decreases and dominance deletion preserve validity, proving the lemma.
\end{proof}

\subsubsection{Low-anchor critical maintenance}
\label{apdx:critical-low}
Here is the complete form of the main-text overview. In the balancing step, read $v_i(a)=0$ when $Z$ has no ice. Re-index after value decreases and keep indices fixed during exchanges.

\begin{algorithm}[H]
\caption{Critical low-anchor maintenance}
\label{alg:critical-low-complete}
\begin{algorithmic}[1]
\Require Conditions~\eqref{eq:critical-bounds}--\eqref{eq:critical-value-bound}, $h\le1/3$, and real indices $1,\ell,t$.
\State Decrease ice in every bundle with at least four pebbles to $1/9$ and melt it.
\While{a bundle $Z$ has two pebbles}
    \State Choose an ice-free $Y$ with at least four pebbles and its least pebble $q$.
    \State Exchange $q$ for any ice $a$ of $Z$ and water of value $v_i(q)-v_i(a)$.
    \State Decrease and melt the received ice if $Y$ still has at least four pebbles.
\EndWhile
\If{no ice-free three-pebble bundle exists}
    \State \Return the trivial dense case.
\EndIf
\State Choose an ice-free three-pebble bundle $X$ and its largest pebble $y$.
\State If $w_1\notin X$, exchange $y$ with $w_1$, paying $h-v_i(y)$ water from $X$ to the old holder of $w_1$.
\State Order the other two pebbles of $X$ by index as $a,b$.
\State If the index of $a$ exceeds $\ell$, exchange it with $w_\ell$, paying its source $v_i(w_\ell)-v_i(a)$ water from $X$.
\State If the index of $b$ exceeds $t$, exchange it with $w_t$, paying its source $v_i(w_t)-v_i(b)$ water from $X$.
\State Delete the three pebbles of $X$, distribute its water, remove $X$, and re-index.
\end{algorithmic}
\end{algorithm}

\begin{proof}[Proof of Lemma~\ref{lem:stage2-critical-low}]
While a two-pebble $Z$ exists, $P(\bW)\ge3k$ supplies a donor with at least four pebbles. Both are nondeficient. If $x,z$ are the pebbles of $Z$, its payment is available because
\[
    F(Z)-\bigl(v_i(q)-v_i(a)\bigr)
       \ge1-v_i(x)-v_i(z)-v_i(q)\ge0.
\]
The exchange preserves values, and any subsequent ice decrease leaves the donor nondeficient. Each iteration eliminates one two-pebble bundle and creates none. Deficient bundles are untouched. If no ice-free three-pebble bundle then exists, every bundle has at least four items counted by $S$, so dense closure applies.

Otherwise, for the chosen $X$ and its largest pebble $y$, the payment for $w_1$ is available: when $D(X)\le0$, $F(X)\ge1-3v_i(y)\ge h-v_i(y)$; when $D(X)>0$, \eqref{eq:critical-value-bound} gives $F(X)\ge4/9-v_i(y)>h-v_i(y)$. This leaves $X$ ice-free with three pebbles, including $w_1$, and value at least $17/18$.

The index-correction calculation in Appendix~\ref{apdx:noice-low} now applies unchanged. Its hypotheses hold: $X$ is ice-free with three pebbles, $h\le1/3$, the same maximal $\ell$ and failed-test bounds apply, and $v_i(X)\ge17/18>25/27$. It bounds the required middle and last replacement triples by $8/9$ and $25/27$, respectively, proving both payments feasible and the final triple dominant.

When placing $w_1$ and correcting indices, every outside exchange replaces a larger pebble by a smaller pebble plus the difference in water. Its largest-pebble value plus water cannot decrease: either its old largest remains, or the received water compensates for its replacement. Hence \eqref{eq:critical-value-bound} is preserved. All exchanges preserve values, ice decreases affect only nondeficient bundles, and final water distribution gives gains. Thus no positive deficit is created or increased. Each survivor has at least three pebbles and at most one ice item. Exactly three pebbles are deleted, leaving $P(\bW)\ge3(k-1)$. These observations prove validity and all the critical inequalities.
\end{proof}

Both arguments remain valid after the last ice disappears. Complete the pending deletion first. Every resulting bundle then has value at least $17/18$ and $\largestwater<4/27$, so
\[
    v_i(W)\ge17/18>25/27=7/9+4/27>7/9+\largestwater.
\]
Thus \textnormal{\textsc{WLC}} holds, and the same critical maintenance can continue.

\subsection{Intervening Rounds and Closure}
\label{apdx:stage2-other}
All constructions below concern a round allocated to another agent, so $k\ge2$. Closures take priority. Standard $\reduction{1}$ and $\reduction{2}$ updates share one deletion argument; the high-anchor index exception and the continuation at $P(\bW)=2k$ are treated separately.

\subsubsection{Pair rounds}
\label{apdx:stage2-pair-complete}

Consider an ordinary pair $\{u_1,u_\ell\}$ allocated to another agent, with $P(\bW)>2k$. We assume the current clean or no-ice inequalities, or the critical conditions~\eqref{eq:critical-bounds}--\eqref{eq:critical-value-bound}. Write $h=v_i(w_1)<5/9$: this holds at the first applicable non-pair checkpoint and persists because witness values never increase. Critical bundles with at least four pebbles are assumed ice-free after the preceding ice decreases.

For an internal pair, redistribute pebbles first. In the critical case, normalize ice in recipients with at least four pebbles, place any residual ice in an ice-free survivor, and normalize again if necessary; distribute water last. Other modes permit unrestricted gains. The only additional operation below repairs a critical external merge containing two ice items.

\begin{algorithm}[H]
\caption{Complete pair maintenance}
\label{alg:stage2-pair-complete}
\begin{algorithmic}[1]
\Require An ordinary real pair, the current maintenance inequalities, and $P(\bW)>2k$.
\State In the critical case, decrease ice in bundles with at least four pebbles to $1/9$ and melt it.
\State Set $q=w_{\min\{\ell,P(\bW)\}}$; let $X,Y$ contain $w_1,q$.
\If{$X=Y$}
    \State Distribute the remaining pebbles of $X$ whole; in the critical case, decrease ice in recipients with at least four pebbles to $1/9$ and melt it.
    \State Distribute retained ice whole; in the critical case, send it to a bundle with at least four pebbles and decrease it to $1/9$.
    \State Distribute the remaining water and delete $X=\{w_1,q\}$.
\Else
    \State Delete $w_1,q$, replace $Y$ by $M=(X\setminus\{w_1\})\cup(Y\setminus\{q\})$, and remove $X$.
    \If{the critical maintenance is in use and $P(M)\ge4$}
        \State Decrease every ice item of $M$ to $1/9$ and melt it.
    \ElsIf{the critical maintenance is in use and $M$ contains two ice items}
        \State Perform the two-ice repair below.
    \EndIf
\EndIf
\State Re-index.
\end{algorithmic}
\end{algorithm}

\begin{algorithm}[H]
\caption{Repairing two ice items after a critical pair merge}
\label{alg:critical-pair-two-ice}
\begin{algorithmic}[1]
\Require A critical merge $M$ with two or three pebbles and two ice items.
\State Let $a,b$ be its ice items, with $v_i(a)\ge v_i(b)$.
\State Choose an ice-free $Z$ with at least four pebbles and its least pebble $z$.
\State Exchange $a,z$; decrease $b$ to $1/9$ and melt it in $M$.
\State Set $t=0$.
\If{$Z$ still has at least four pebbles}
    \State Decrease $a$ to $1/9$ and melt it in $Z$.
\Else
    \State Set $t=\max\{0,17/18-v_i(Z)\}$ and transfer that water value from $M$ to $Z$.
\EndIf
\State \Return the surviving bundles.
\end{algorithmic}
\end{algorithm}

\begin{proof}[Correctness of the pair maintenance]
We have $\ell\ge k+1$, so the chosen $q$ is a pebble of index between $k+1$ and $\ell$. Failed $\reduction{1}$ and validity give $v_i(q)<7/18$. Thus $\{w_1,q\}$ dominates the real pair. Exactly two pebbles and one bundle are deleted; writing $k'=k-1$,
\[
    P'=P(\bW)-2>2k',
    \qquad P'\ge3k'+1\quad\text{in the critical case}.
\]
In particular, this pair update does not reach the equality boundary.

\paragraph{The merge and the no-ice inequalities.}
The anchor remainder $R=v_i(X\setminus\{w_1\})$ is at least $4/9$. For a clean anchor, $R\ge1-h>4/9$. For an anchor with at least three pebbles, two pebbles remain. A critical two-pebble anchor is clean. A deficient no-ice two-pebble high anchor has either $2/9$ water or remainder at least $\sigma+2/9\ge4/9$; a low anchor has $R\ge25/27-h\ge16/27$. Hence an external merge gives $Y$ a gain
\[
    R-v_i(q)>1/18.
\]
It remains clean in the clean mode. In the no-ice mode its new count is $P(X)+P(Y)-2$. If $P(X)=2$, the count is unchanged and value increases. If $P(X)=3,P(Y)=2$, the new three-pebble deficit is less than $2/27-1/18=1/54$. If $P(X)=3,P(Y)\ge3$, the gain clears the possible $1/27$ deficit. If $P(X)\ge4$, then $R\ge2/3$ and the gain exceeds $5/18$, clearing every allowed deficit. Thus~\eqref{eq:no-ice-bundle-bounds} holds.

For a two-pebble output, let $s,r$ be the remaining pebbles of $X,Y$. The latter remainder satisfies
\[
    v_i(r)+F(Y)=v_i(Y)-v_i(q)>29/54>\sigma+2/9.
\]
If $F(X)\ge2/9$, the output has the required water directly. Otherwise the high-anchor inequality, or $v_i(X)\ge1$, gives $v_i(s)+F(X)\ge\sigma+2/9$. Combining these two one-pebble remainders gives the required smaller-pebble-plus-water bound. A deficient three-pebble output is possible only for $P(X)=3,P(Y)=2$; its largest pebble and water are worth at least $29/54$, sufficient for both the high- and low-anchor inequalities.

Internal pairs only distribute gains. Receiving a pebble clears any old no-ice deficit; water gains retain its water inequality. Deletion of $w_1$ keeps every old top-$k$ survivor among the new first $k-1$ items. Thus the fixed-$\sigma$ and index inequalities persist. No-ice bundle values remain at least $25/27$, giving \textnormal{\textsc{WLC}}.

\paragraph{Critical distribution and the two-ice repair.}
An external critical merge has value greater than $1$, because its old holder had value at least $17/18$. If it has two pebbles, both original holders were clean, so
\begin{equation}
\label{eq:critical-pair-two-pebble-value}
    v_i(M)\ge2-h-v_i(q)>19/18.
\end{equation}
An internal critical pair redistributes pebbles, ice, and water in that order. A pebble gain clears an old $1/18$ deficit even after lowering a recipient's ice to $1/9$. After the pebbles are distributed, $P'\ge3k'+1$ supplies a survivor with at least four pebbles, hence an ice-free recipient for residual ice. Water gains preserve~\eqref{eq:critical-value-bound} for any still-deficient ice-free three-pebble bundle.

An external merge with at most one ice item is already valid structurally. If it has at least four pebbles, lowering its ice to $1/9$ leaves it clean. Otherwise the two-ice repair applies. Since $P'\ge3k'+1$ and $P(M)\le3$, the required outside $Z$ with at least four pebbles exists and is clean and ice-free. Put $a_0=v_i(a)$, $b_0=v_i(b)$, and $z_0=v_i(z)$, so $4/27\le b_0\le a_0<2/9$.

If $Z$ had at least five pebbles, its remaining four and the melted $1/9$ guarantee value at least $1$. If it had four, its remainder after giving its minimum is worth at least $\max\{3z_0,1-z_0\}\ge3/4$. Consequently
\begin{equation}
\label{eq:critical-pair-payment}
    0\le t\le\max\{0,7/36-a_0\}\le5/108<1/9.
\end{equation}
The newly melted $b$ supplies this payment. If $Z$ had four pebbles, it finishes with three pebbles, one ice item, and value at least $17/18$; otherwise it remains ice-free and worth at least $1$.

Suppose $M$ originally had three pebbles, hence now has four. For $t=0$, four pebbles and the melted $1/9$ make it clean. For $t>0$, let $T$ be the combined value after decreasing $b$ and before payment. The old merge value $v_i(M)>1$ gives $T>19/9-b_0$. Its three old pebbles and $a_0\ge b_0$ also give $T\ge16/9+a_0\ge16/9+b_0$. Splitting at $b_0=1/6$ yields $T>35/18$. After bringing $Z$ to $17/18$, the four-pebble $M$ therefore remains worth more than $1$.

If $M$ originally had two pebbles,~\eqref{eq:critical-pair-two-pebble-value} shows that exchanging $a$ for $z$ and losing less than $1/9$ in $b$ leaves its new three-pebble value above $17/18$. If $t>0$, the combined value exceeds $19/18+1-1/9=35/18$, so $M$ finishes above $1$.

Inequality~\eqref{eq:critical-value-bound} for $M$ follows from the original two-pebble holder $Y$ of $q$. Write $r,c$ for that holder's other pebble and ice item. Then
\[
    v_i(r)+v_i(c)+F(Y)=v_i(Y)-v_i(q)>11/18.
\]
The pebble $r$ stays in $M$, whose final water is its original water plus $1/9-t$. Its largest pebble plus water therefore exceeds $13/18-v_i(c)-t$. For $t=0$ this is greater than $1/2$; for $t>0$, $v_i(c)\le a_0$ and~\eqref{eq:critical-pair-payment} give the stronger bound $19/36>4/9$.

All other survivors are unchanged. Both repaired bundles have at least three pebbles, no bundle has two solid ice items, and every bundle with at least four pebbles is clean. Thus \eqref{eq:critical-bounds}--\eqref{eq:critical-value-bound} hold. The repair can create a positive three-pebble deficit, but never one exceeding $1/18$.

Validity follows from dominance deletion, whole-item exchanges, and allowed value decreases. Every ice decrease used here gives value $1/9$. If the last solid ice disappears, all non-pebble items are below $4/27$ and critical bundle values are at least $17/18$, giving \textnormal{\textsc{WLC}}.
\end{proof}

\subsubsection{Standard $\reduction{1}$ and $\reduction{2}$ updates}
\label{apdx:stage2-reductions}
\label{apdx:stage2-r2-complete}
Since $P(\bW)>2k$, the first $rk+1$ items are pebbles. The Pigeonhole Principle supplies $C$. Write its first $r+1$ pebble ranks as $j_1<\cdots<j_{r+1}$. Then
\[
    j_s\le rk-r+s=r(k-1)+s,
\]
so $T$ dominates the real $\reduction{r}$ bundle. Deleting $C$, including its ice and water, preserves validity by dominance and the allowed value decreases.

Every survivor is unchanged or gains whole pebbles. A gain of $2/9$ clears any old deficit. In the critical case, normalizing a recipient's existing ice loses less than $1/9$, leaving a net gain greater than $1/9>1/18$. Thus every recipient reaching four pebbles is nondeficient. A still-deficient bundle is unchanged, retaining its water and largest-pebble inequalities. Clean bundles remain nondeficient, and $\Delta$ cannot increase. No source ice or water needs a recipient.

Writing $k'=k-1$, the counts satisfy
\[
    P'=P(\bW)-r-1\ge2k',
    \qquad P'\ge3k'\quad\text{on a critical input}.
\]
Equality in the first inequality is possible only for $r=2$ and $P(\bW)=2k+1$. A no-ice input then has exactly one three-pebble bundle, namely $C$; every survivor is unchanged with two pebbles. Lemma~\ref{lem:stage2-equality} supplies the continuation. A clean equality output first uses that lemma's preparation; it need not already have two pebbles in every bundle.

For a continuing high-anchor input, $h$ cannot increase. If $r=1$, the deletion removes an old top-$k$ item, preserving~\eqref{eq:noice-indices}. For $r=2$, this also holds when $j_1\le k$. If $j_1>k$, only old $w_k$ newly lies after position $k-1$; direct deletion is therefore valid unless $\sigma<7/27$ and $\sigma<v_i(w_k)\le5/18$. That exception is handled in Appendix~\ref{apdx:noice-high-r2}.

\subsubsection{Special $\reduction{2}$ without ice}
\label{apdx:noice-high-r2}
We use a special update only in the no-ice high-anchor case. Its purpose is to preserve the frozen index condition when the standard deletion would leave old $w_k$ after the new first $k-1$ positions.
All ranks are fixed before the update.

\begin{algorithm}[H]
\caption{Special high-anchor $\reduction{2}$ maintenance}
\label{alg:noice-high-r2-full}
\begin{algorithmic}[1]
\Require No ice, the high-anchor inequalities, and $P(\bW)\ge2k+2$.
\State Choose $C$ containing at least three of $w_1,\ldots,w_{2k+1}$.
\State Let its first three be $c_s=w_{j_s}$, with $j_1<j_2<j_3$.
\If{$\sigma<7/27$, $\sigma<v_i(w_k)\le5/18$, and $j_1>k$}
    \State Let $Z$ contain $w_k$.
    \State Exchange $w_k$ for $C\setminus\{c_2,c_3\}$.
    \State Delete $C=\{w_k,c_2,c_3\}$.
\Else
    \State Distribute the other pebbles of $C$ whole.
    \State Delete $C$, including its remaining water.
\EndIf
\State Re-index.
\end{algorithmic}
\end{algorithm}

\begin{proof}
Since $P(\bW)>2k$, the first $2k+1$ items are pebbles.
The Pigeonhole Principle supplies $C$, with $j_s\le2k-2+s$.
Thus its selected triple dominates the real reduction positions $2k-1,2k,2k+1$.
The special triple also dominates, since $k\le2k-1$, $j_2\le2k$, and $j_3\le2k+1$.

In the internal-deletion branch, all survivors receive only gains.
Their old water is unchanged, and a received pebble clears any old deficit because $2/9>2/27$.
In particular every recipient whose count reaches at least four is nondeficient.

In the special branch, $C$ contains no old top-$k$ item, so $C\ne Z$.
Sortedness and~\eqref{eq:noice-indices} give $v_i(c_1)\le\sigma$.
If $C$ is nondeficient, its retained remainder has value at least
\[
 1-v_i(c_2+c_3)\ge1-2(5/18)=4/9.
\]
If $C$ is deficient, it has exactly three pebbles and its water inequality gives $F(C)\ge2/9$.
Its retained remainder is then $c_1$ and this water, again worth at least $4/9$.
Hence $Z$ gains at least
\[
 4/9-v_i(w_k)\ge1/6>2/27
\]
and becomes nondeficient.
It has $P(Z)+P(C)-3\ge2$ pebbles; every other survivor is unchanged.
All count-dependent deficit and water bounds follow.

The special branch deletes old $w_k$, while an internal deletion with $j_1\le k$ deletes another old top-$k$ item.
In either event all surviving old top-$k$ items lie within the new first $k-1$ positions.
If an internal deletion has $j_1>k$, the only such item becoming late is $w_k$; the branch condition ensures its value is at most $\sigma$ or greater than $5/18$, whenever the index condition is required.
Thus~\eqref{eq:noice-indices} is preserved.
No item value increases, so $v_i(w'_1)+\sigma<7/9$ persists, and dominance preserves validity.
Finally
\[
 k'=k-1,\qquad P'=P(\bW)-3,\qquad P'-2k'=P(\bW)-2k-1.
\]
Thus $P'>2k'$, and the high-anchor inequalities continue.
Every surviving bundle remains worth at least $25/27>7/9+\Delta'$, so \textnormal{\textsc{WLC}} also holds.
\end{proof}

\paragraph{The equality output.}
If initially $P(\bW)=2k+1$, the common update deletes the unique three-pebble bundle, including its water. Every survivor retains two pebbles and deficit at most $2/27$. Lemma~\ref{lem:stage2-equality} supplies the continuation to the canonical case; the frozen index condition is no longer needed.

\subsubsection{Continuation to the canonical case}
\label{apdx:stage2-equality}
\begin{proof}[Proof of Lemma~\ref{lem:stage2-equality}]
The no-ice bounds give two pebbles per bundle and value at least $4/9$ after removing its larger pebble. For a clean output, Algorithm~\ref{alg:clean-equality-preparation} gives the same conditions and \textnormal{\textsc{WLC}}.

For an ordinary real pair $\{u_1,u_\ell\}$, put
\[
    q=w_{\min\{\ell,2k\}},
\]
and let $X,Y$ contain $w_1,q$. The item $q$ is a whole pebble with index between $k+1$ and $\ell$, so $\{w_1,q\}$ dominates the real pair and $v_i(q)<7/18$. If $X=Y$, delete $X$ with its remaining water. Otherwise exchange $q$ for $X\setminus\{w_1\}$ and remove $X$. The surviving merge contains exactly two pebbles and gains more than $4/9-7/18=1/18$ in value. Each of its two one-pebble remainders was worth at least $4/9$; combining their water therefore preserves the same bound after removing the larger surviving pebble. Thus exactly two pebbles and one bundle disappear, while no positive deficit is created or increased. An internal deletion leaves all other bundles unchanged.
For $\reduction{1}$, delete a bundle containing two of $w_1,\ldots,w_{k+1}$. For $\reduction{2}$, choose a bundle containing two of $w_k,\ldots,w_{2k}$. Their sum is below $7/9$ because $\reduction{1}$ has failed; \textnormal{\textsc{WLC}} therefore supplies water of total value greater than $\Delta$ to collect $w_{2k+1}$ at equal value. Delete that dominance bundle. Both reductions leave all survivors unchanged.

Thus ordinary maintenance preserves validity, two pebbles per bundle, and \textnormal{\textsc{WLC}}, without creating or increasing a positive deficit. Once the pair test fails, $v_i(C_1)<7/9+\Delta$ gives canonical entry. Each other round reduces $k$, until agent $i$ is satisfied.
\end{proof}

\subsubsection{Clean and no-ice $\reduction{3}$ rounds}
\label{apdx:standard-r3}
At this bag-filling branch the real triple test has failed. In a clean state every bundle is worth at least $1$ and $\Delta<2/9$, so \textnormal{\textsc{WLC}} holds. In the no-ice maintenance, the value bound $25/27$ and $\Delta<4/27$ give the same conclusion. If $P(\bW)\le3k$, validity yields $v_i(C_1)<7/9$ and Lemma~\ref{lem:canonical-entry} applies before the pending reduction.

If $P(\bW)>3k$, all four standard target items are pebbles. Standard exchanges preserve the pebble count and water in each outside holder and increase its pebble values. Remainder gains preserve the clean bounds or clear the allowed no-ice deficits when a pebble is received. The failed triple test also gives $v_i(w_1)<1/3$, so only the low-anchor water inequality is needed; it is retained because the largest pebble and water of each outside holder do not decrease. This proves the standard continuation. Exactly four pebbles are deleted, leaving $P'\ge3(k-1)>2(k-1)$.

\subsubsection{Critical maintenance for $\reduction{3}$}
\label{apdx:critical-r3}

Assume~\eqref{eq:critical-bounds}--\eqref{eq:critical-value-bound} and a real $\reduction{3}$ allocation to another agent. Thus $k\ge2$ and $P(\bW)\ge3k$. The special choice below prevents an ice-for-pebble exchange from producing a deficient four-pebble survivor.

\begin{algorithm}[H]
\caption{Critical witness maintenance for $\reduction{3}$}
\label{alg:critical-r3-complete}
\begin{algorithmic}[1]
\Require A real $\reduction{3}$ round and the critical invariant and deficit bounds.
\State Decrease ice in every bundle with at least four pebbles to $1/9$, melt it, and re-index.
\State Set $T=\{w_{3k-2},w_{3k-1},w_{3k},w_{3k+1}\}$.
\If{$P(\bW)=3k$}
    \State Let $q=w_{3k+1}$.
    \If{$q$ is water}
        \State Reorganize canonically and \Return.
    \EndIf
    \State Let $B$ contain $q$.
    \If{$P(B)=3$}
        \State Set $X=B$.
    \Else
        \State Choose $X$ with at least four pebbles.
    \EndIf
\Else
    \State Choose $X$ containing four of the first $3k+1$ items.
\EndIf
\State Collect $T$ in $X$ by the standard exchanges.
\State Distribute remaining pebbles whole, then water; delete $T$ and remove $X$.
\State Normalize ice in surviving bundles with at least four pebbles and re-index.
\end{algorithmic}
\end{algorithm}

The temporary target container may contain $q$ and at least four pebbles. Do not melt this target ice: it is deleted with $T$. Normalization applies to the surviving bundles.

\begin{proof}
The preliminary ice decreases are safe by the four-pebble argument used for the standard reductions. The failed real triple test and validity give
\[
    v_i(w_1)+v_i(w_{k+1})+v_i(w_{2k+1})<7/9,
\]
so $v_i(w_1)<1/3$.

If $P(\bW)>3k$, all four targets are pebbles. Pigeonhole supplies the stated $X$, which has at least four pebbles and is ice-free. All outside exchanges are pebble-for-pebble gains: counts, ice, and water are unchanged, and largest-pebble values cannot decrease. They preserve \eqref{eq:critical-bounds}--\eqref{eq:critical-value-bound}.

Suppose $P(\bW)=3k$. The first three targets are pebbles, and $q=w_{3k+1}$ is a non-pebble. If it is water, no solid ice remains. The ice decreases give value $1/9$, so $\Delta<4/27$; the lower bundle value $17/18$ supplies \textnormal{\textsc{WLC}}. The failed triple test gives $v_i(C_1)<7/9$, and Lemma~\ref{lem:canonical-entry} applies before the reduction.

Otherwise $q$ is ice and its holder $B$ has two or three pebbles. For $P(B)=3$, choosing $X=B$ includes $q$ already, so every outside exchange is again between pebbles. For $P(B)=2$, the equality $P(\bW)=3k$ supplies an ice-free bundle $X$ with at least four pebbles. Collecting $q$ there gives $B$ an earlier pebble in exchange. The old two-pebble $B$ was clean, so its new ice-free three-pebble bundle remains clean. No deficient four-pebble survivor is created; every other outside exchange is between pebbles.

In each branch $X$ has four items among the first $3k+1$. Every missing target can therefore be exchanged for a distinct item of $X$ whose index is less than $3k-2$, exactly as in the standard reduction. The only ice that ever enters $X$ is $q\in T$, so its remaining material is pebbles and water. Its redistribution is safe by the critical $\reduction{2}$ gain argument: a pebble clears any old $1/18$ deficit, subsequent four-pebble ice normalization stays clean, and water gains preserve every still-deficient holder's inequality~\eqref{eq:critical-value-bound}. No survivor acquires a second solid ice item.

The surviving counts are at least two per bundle. Exactly four pebbles are deleted when $P(\bW)>3k$, giving $P'\ge3(k-1)$; exactly three are deleted when $P(\bW)=3k$, giving $P'=3(k-1)$. Every still-deficient ice-free three-pebble holder retained its count, water, and a nondecreasing largest pebble. The critical invariant, the deficit bounds, and \eqref{eq:critical-value-bound} therefore persist.

Finally $T$ matches the real reduction ranks and is a dominance bundle. Standard exchanges, gains, and allowed ice decreases preserve validity. If the last solid ice disappears, all remaining non-pebble items are below $4/27$, while each bundle is worth at least $17/18$, so \textnormal{\textsc{WLC}} holds. This update needs no additional canonical claim unless its hypotheses also hold.
\end{proof}

\section{Common Witness Proofs}
\label{apdx:common-witness}

\begin{proof}[Proof of Proposition~\ref{prop:remove-dominance}]
Decreasing a witness-item value cannot increase the value at any rank after sorting, so validity is preserved.
For deletion, let $u_p$ be the real item at remaining rank $j$.
Exactly $p-j$ real items with old index less than $p$ were deleted.
Their distinct matched witness items also had old indices less than $p$ and were deleted.
Thus at most $j-1$ of the first $p-1$ witness items remain.
The witness item at remaining rank $j$ therefore had an old index $q\ge p$, and
\[
 v_i(w_q)\le v_i(u_q)\le v_i(u_p).
\]
This is validity after re-indexing. Extra witness items in the deleted bundle can only strengthen the conclusion; zero-valued dummies cover missing ranks.
\end{proof}

\begin{proof}[Proof of Lemma~\ref{lem:canonical-entry}]
Let $V$ be the total witness value. By \textnormal{\textsc{WLC}}, $V>kL$.
For every $r\in[k]$, sortedness gives
\[
 v_i(C_r)\le v_i(C_1)\le L<V/k.
\]
Place the items of each $C_r$ whole in $W_r$.
Every other item has index greater than $3k$ and is a non-pebble, since $P(\bW)\le3k$.
Melt any ice among these other items without changing its value, and distribute the resulting water so that every bundle has value $V/k$.
This is possible because the required added values are positive and sum to the value of exactly these other items.
No ordered item value changes, so validity is preserved.
Every bundle contains its triple $C_r$ and has value $V/k>L$, proving canonicality.
\end{proof}

\begin{proof}[Proof of Lemma~\ref{lem:canonical-packet}]
Every item outside the triples $C_r$ has index greater than $3k$ and is a non-pebble.
Melt any ice among these items without changing values.
In particular, all of $W_1\setminus C_1$ is water.

Only the ordinary bag-filling branch can return failure.
The initial real triple together with all real items of index greater than $3k$ has value at least
\[
 v_i(C_1)+\sum_{j>3k}v_i(w_j)
 \ge v_i(W_1)>7/9.
\]
Hence this branch cannot exhaust the available items without finding a satisfactory bundle for agent $i$.

Now suppose that another agent is served, so $k\ge2$.
For an ordinary pair, $\ell\ge k+1$, and for an ordinary triple, $\ell\ge k+1$ and $t\ge2k+1$.
Thus $C_1$ dominates either real bundle.
It also dominates $\reduction{0}$, $\reduction{1}$, and $\reduction{2}$, using its first one, two, and three items, respectively: the ranks $1,k+1,2k+1$ are no later than the corresponding real ranks.
In these cases, distribute $W_1\setminus C_1$ among the other bundles, then delete $C_1$ and remove $W_1$.

Two cases require collecting further items in $W_1$.
For $\reduction{3}$, collect $w_{3k+1}$.
The failed real triple test and validity give
\[
 v_i(C_1+w_{3k+1})<7/9+\Delta=L.
\]
The four ranks $1,k+1,2k+1,3k+1$ dominate the real reduction ranks.
For an ordinary bag-filling allocation whose last added item is $u_s$, collect all of $w_{3k+1},\ldots,w_s$.
Immediately before $u_s$ was added, the real bundle was unsatisfactory for every active agent, including $i$.
Consequently validity gives
\[
 v_i(C_1)+\sum_{j=3k+1}^{s}v_i(w_j)
 <7/9+v_i(w_s)\le7/9+\Delta=L.
\]
These witness items have exactly the ranks of the allocated real items.

Both collections are feasible by equal-value water exchanges.
The water in $W_1\setminus C_1$ is worth more than $L-v_i(C_1)$, whereas all required additional items together are worth at most $L-v_i(C_1)$.
Subtracting the value already assigned to $W_1$ from both bounds leaves enough other water in $W_1$ to pay for the missing parts.
Collect each required item whole, paying every other bundle the same water value that it gives up.
Keep $C_1$ and the collected items whole in $W_1$, distribute only its other water among the surviving bundles, and delete $W_1$.
The other bundles retain their values or gain water.

In every case, the deleted bundle dominates the real allocation, so Proposition~\ref{prop:remove-dominance} preserves validity.
Only $C_1$ is deleted from the first $3k$ witness items; all additional deleted items have later indices.
After re-indexing, the old $C_2,\ldots,C_k$ are therefore exactly the new triples for $k-1$ bundles, and each remains whole in its bundle.
All pebbles were among the old first $3k$ items, so the remaining pebbles lie among the other $3k-3$ items and
\[
 P(\bW['])\le3(k-1).
\]
Finally $\Delta'\le\Delta$, while every surviving bundle still has value greater than $L=7/9+\Delta$.
Thus the remaining witness allocation is valid and canonical, as required.
\end{proof}

\subsection{Entry to Stage~3}
\label{apdx:stage3-entry}

\begin{proof}[Proof of Lemma~\ref{lem:stage3-entry}]

Consider the first ordinary bag-filling round, with $k$ active agents. Both the triple test and $\reduction{3}$ have failed. Validity therefore gives
\[
    v_i(C_1)<7/9,
    \qquad
    v_i(w_{3k-2}+w_{3k-1}+w_{3k}+w_{3k+1})<7/9.
\]
The second inequality implies $v_i(w_{3k+1})<7/36<2/9$, so $P(\bW)\le3k$. The trivial dense case cannot apply: its count bounds would make this reduction target worth at least $22/27$, a contradiction. If the witness is already canonical, there is nothing to prove.

We verify \textnormal{\textsc{WLC}} for every remaining maintenance case. In the clean maintenance, every bundle has value at least $1$ and $\Delta<2/9$, hence $v_i(W)>7/9+\Delta$. In the no-ice maintenance, every bundle has value at least $25/27$ and $\Delta<4/27$, giving the same strict inequality. Lemma~\ref{lem:stage2-equality} also retains \textnormal{\textsc{WLC}} during the equality continuation and permits canonical entry at this non-pair round.

In the critical maintenance, $P(\bW)\ge3k$, so the established upper bound gives $P(\bW)=3k$. Consequently the first three items in the reduction target are pebbles, while $w_{3k+1}$ has value $\Delta$. The failed reduction therefore gives
\[
    2/3+\Delta
       \le v_i(w_{3k-2}+w_{3k-1}+w_{3k}+w_{3k+1})
       <7/9,
\]
and thus $\Delta<1/9$. Every critical bundle has value at least $17/18>8/9>7/9+\Delta$, so \textnormal{\textsc{WLC}} holds here as well.

The witness is valid in all cases. We have verified \textnormal{\textsc{WLC}}, $v_i(C_1)<7/9\le7/9+\Delta$, and $P(\bW)\le3k$. Lemma~\ref{lem:canonical-entry} now gives a valid canonical reorganization.
\end{proof}

\section{Proofs for the Strongly Polynomial-Time Algorithm}
\label{apdx:exact-search}

\subsection{Safe threshold updates}
\label{apdx:exact-search-safe-update}
\begin{proof}[Proof of Lemma~\ref{lem:exact-search-safe-update}]
Fix a call that returns $(\textsc{Failure},F)$ and an agent $i\in F$.
We have $b_i<\alpha_i$ by the definition of $b_i$, and $\MMS_i<\alpha_i$ by Theorem~\ref{thm:7/9}.

Suppose that $b_i<\MMS_i$.
We consider another run of Algorithm~\ref{alg:pebble-and-water} on the original instance, using threshold $\MMS_i$ for agent $i$ and threshold $\alpha_j$ for every agent $j\ne i$.
If a bundle $T$ tested for $i$ in the first call was unsatisfactory, then
\[
v_i(T)\le7b_i/9<7\MMS_i/9,
\]
so it remains unsatisfactory under the new threshold.
If it were satisfactory, then
\[
v_i(T)\ge7\alpha_i/9>7\MMS_i/9,
\]
so it remains satisfactory.
For every other agent, we preserve all comparison outcomes because we use the same threshold as in the first call.

By induction over the steps of the algorithm, using the prescribed scanning and recipient-selection rules, we obtain the same tested bundles, allocations, and failed set.
Thus agent $i$ still fails, contradicting Theorem~\ref{thm:7/9}, since her new threshold equals her MMS.
We conclude that $\MMS_i\le b_i$.
\end{proof}

We apply the lemma to each failed agent using the thresholds from the completed call, so we can update all failed agents' thresholds simultaneously.
Starting from the proportional shares, we therefore maintain $\alpha_i\ge\MMS_i$ throughout the search.
On a successful return, we give every agent value at least $7\alpha_i/9\ge7\MMS_i/9$.
By nonnegativity, we preserve this guarantee when assigning the remaining goods.

\subsection{Changes before tail filling}
\label{apdx:exact-search-call-groups}
\begin{proof}[Proof of Lemma~\ref{lem:exact-search-call-groups}]
For each agent $i$, we consider the values $9v_i(S)/7$ for subsets $S\subseteq M$ of either of the following forms:
\begin{enumerate}
    \item $S$ contains at most four goods, all among the first $\min\{4n,m\}$ original goods;
    \item $S$ contains at most three goods, including at least one of the first $\min\{4n,m\}$ original goods.
\end{enumerate}
For each agent, we count the different possible values of $9v_i(S)/7$.
We add these counts over all agents and denote the total by $Q$.
For the first form, we choose at most four goods from $O(n)$ positions.
For the second form, we choose one good from these positions and at most two other goods.
Multiplying by the number of agents, we obtain
\[
Q=O(n^5+n^2m^2)=O(n^3m^2),
\]
where we use $n\le m$ for the last bound.

We next show that these forms cover every bundle tested before a call first reaches tail filling.
Consider such a test in a round with $k$ active agents.
At this point, $n-k$ agents have received bundles, each containing at most four real goods.
Thus a remaining real good at current rank $j$ has original rank at most $j+4(n-k)$.
In a reduction test, we use at most four goods at current ranks at most $3k+1$, whose original ranks are therefore at most
\[
3k+1+4(n-k)\le4n.
\]
In a pair or triple test, we use the current first good and at most two other goods.
The first good has original rank at most $1+4(n-k)\le4n$.
We omit zero-valued dummy goods when identifying a tested bundle with a subset of $M$.
If the first indexed good is a dummy, the corresponding real-good subset is empty and is covered by the first form.
We have therefore counted every possible comparison value before a call first reaches tail filling.

Note that Algorithm~\ref{alg:pebble-and-water} returns failure only during tail filling.
Consider failed calls in which some updated value $b_i$ is attained by a comparison made before the first tail-filling step.
In each such call, we update at least one agent's threshold to one of the values counted above.
By Lemma~\ref{lem:exact-search-safe-update}, we strictly decrease every updated threshold, so each agent can reach each counted value at most once.
We therefore have at most $Q$ failed calls of this kind over the entire search.

Now consider any other failed call.
For each updated agent $i$, every unsatisfactory bundle $T$ tested before the first tail-filling step satisfies $9v_i(T)/7<b_i$, since no comparison before that step attains $b_i$.
After setting $\alpha_i=b_i$, we therefore still find each such bundle unsatisfactory.
We also preserve every satisfactory comparison because we only lower thresholds.
For agents whose thresholds are not updated, we preserve all comparison outcomes.
Since the comparison outcomes are unchanged, we assign the same bundles to the same agents before the first tail-filling step, following the fixed scanning and recipient-selection rules.
We therefore reach that step with the same active agents and remaining goods as in the previous call.

Consequently, whenever two consecutive failed calls reach their first tail-filling steps with different active agents or remaining goods, the first of these calls must be one of the at most $Q$ calls counted above.
Over the entire search, we therefore have at most $Q=O(n^3m^2)$ such changes.
\end{proof}

\subsection{Candidate values during tail filling}
\label{apdx:exact-search-bag-candidates}
\begin{proof}[Proof of Lemma~\ref{lem:exact-search-bag-candidates}]
Recall that all calls in this group have the same active agents and remaining goods when they first reach tail filling.
Fix an agent $i$.
In each call with $i\in F$, we test the final bundle for $i$ during tail filling and find it unsatisfactory.

We first bound the number of possible tested or selected bundles in an arbitrary round.
Once the remaining goods and the number $k$ of active agents are specified, we have at most four reduction bundles, $O(m)$ candidate pairs and triples, and $O(m)$ partial bundles obtained by appending tail goods in order.
We therefore have $O(m)$ possible tested or selected bundles in that round, even if the active agents or their thresholds differ.
When counting these bundles, we ignore zero-valued dummy goods.

We first consider rounds with at least three active agents.
Throughout this argument, we refer to positions in the common ordered goods list at the first tail-filling step, using zero-valued dummies for missing positions.
We prove by induction that every round from that step onward with $k\ge3$ active agents reaches tail filling and has the following structure.
Its first $3k$ goods occupy positions
\[
K-k+1,\ldots,K;\qquad
2K-k+1,\ldots,2K;\qquad
3K-k+1,\ldots,3K
\]
in that initial list.
All other remaining goods form an unconsumed suffix of the tail after position $3K$.

When $K\ge3$, we obtain the base case $k=K$ from the first tail-filling step.
In a successful tail-filling round with $k\ge3$, we delete the goods at fixed positions
\[
K-k+1,\qquad 2K-k+1,\qquad 3K-k+1,
\]
together with a nonempty prefix of the remaining tail.
We must append at least one real tail good because the initial triple is unsatisfactory to every active agent.
Thus all of the first $3k$ positions are real goods whenever we perform this deletion.
If no real tail remains, the call fails without deleting any goods.

Suppose that a successful tail-filling round has $k\ge4$ active agents, so the next round still has at least three active agents.
In the next round, the initial triple uses fixed positions
\[
K-k+2,\qquad 2K-k+2,\qquad 3K-k+2.
\]
By orderedness, every remaining agent values each of these goods no more than the corresponding good in the preceding initial triple.
Thus every remaining agent still finds the initial triple unsatisfactory.
By the implications established after Algorithm~\ref{alg:pebble-and-water}, the algorithm skips $\reduction{0},\reduction{1},\reduction{2}$ and the pair and triple branches.

We also reject $\reduction{3}$.
In both rounds, its first three goods occupy fixed positions $3K-2,3K-1,3K$.
Since $k\ge4$, the largest fixed position in the deleted triple is $3K-k+1\le3K-3$, while all appended tail goods lie after position $3K$.
Thus none of these three goods is deleted.
The fourth good of the new $\reduction{3}$ bundle is the first remaining tail good, or a zero dummy if the tail is empty.
By orderedness, every remaining agent values this good no more than the fourth good of the preceding $\reduction{3}$ bundle.
Since that bundle was unsatisfactory to every remaining agent, the new one is also unsatisfactory.
We therefore reach tail filling again.
By induction, we obtain the displayed positions and reach tail filling in every subsequent round with at least three active agents.

We can now determine each remaining goods set with $k\ge3$ from $k$ and the position of the first unconsumed tail good in the initial list.
We have at most $K$ choices for $k$ and at most $m+1$ possibilities for the remaining tail, including the empty tail.
For each choice of $k$ and the remaining tail, we have $O(m)$ possible tested bundles.
We therefore obtain $O(Km^2)$ possible comparison values for agent $i$ from these rounds.

We next consider rounds with at most two active agents.
If $K\le2$, we start from the common remaining goods set at the first tail-filling step.
If $K\ge3$, a call can reach two active agents only after $K-2$ successful tail-filling rounds.
At that point, the remaining goods from the first $3K$ positions of the initial list occupy positions
\[
K-1,K;\qquad 2K-1,2K;\qquad 3K-1,3K.
\]
Different calls may have consumed different numbers of tail goods before reaching two active agents.
Since tail goods are consumed in order, the remaining tail is always a suffix of the initial tail.
We have at most $m+1$ possible suffixes, including the empty tail.
We therefore have $O(m)$ possible goods sets at this point.

For each such goods set with two active agents, we have $O(m)$ possible bundles that can be tested or assigned.
Once we specify the assigned bundle, we know exactly which goods remain for the one-agent round.
Thus, each two-agent goods set leads to at most $O(m)$ possible goods sets for the last agent.
We next bound the possible maximum values from the one-agent rounds.
If a call with $i\in F$ reaches such a round, its only remaining agent is $i$.
The initial triple then uses the first three positions in the current goods list, and the tail contains all later goods.
Since the call fails, we append the entire tail and test the full remaining goods set $U$, which is unsatisfactory to $i$.
Every bundle tested earlier in this round is a subset of $U$.
By nonnegativity, we therefore need only count $9v_i(U)/7$ as a possible value of the maximum from this round.

Hence, for each of the $O(m)$ goods sets counted above, we have $O(m)$ possible tested bundles in a two-agent round, if such a round occurs.
From each starting goods set, we also obtain at most $O(m)$ possible goods sets $U$ for a one-agent round, each contributing only $9v_i(U)/7$.
We therefore obtain $O(m^2)$ possible values that can attain the maximum from rounds with at most two active agents.

In every failed call with $i\in F$, the maximum from the first tail-filling step onward is attained either in a round with at least three active agents or in a round with at most two.
Combining the two counts and using $K\ge1$, we obtain
\[
O(Km^2)+O(m^2)=O(Km^2)
\]
possible values of this maximum for agent $i$.
\end{proof}

\clearpage
\section{Running Example}
\label{apdx:running-example}

\begin{example}
\label{ex:running-witness}
Consider an ordered instance with three agents and 17 goods $g_1,\ldots,g_{17}$.
Let $\alpha_1=\alpha_2=\alpha_3=1$.
The item values, in units of $1/54$, are
\begin{equation*}
\begin{array}{c|rrrrrrrrrrrrrrrrr}
 &g_1&g_2&g_3&g_4&g_5&g_6&g_7&g_8&g_9&g_{10}&g_{11}&g_{12}&g_{13}&g_{14}&g_{15}&g_{16}&g_{17}\\ \hline
54v_1=54v_2&35&21&20&20&6&6&6&6&6&6&6&6&6&5&5&1&1\\
54v_3       &31&21&20&20&10&8&8&8&8&6&5&5&4&3&2&2&1
\end{array}
\end{equation*}

\paragraph{Initial state.}
We write $W_r^a$ for the $r$th witness bundle private to agent $a$ and $\bW[a]$ for her entire witness allocation.
In each private witness allocation, $g_r$ denotes the witness item corresponding to real good $g_r$.
For $a\in\{1,2\}$, the initial witness bundles are
\begin{align*}
W^a_1&=\{g_1,g_5,g_6,g_7,g_{16}\}, \qquad W^a_2 =\{g_4,g_8,g_9,g_{10},g_{11},g_{14},g_{15}\},\\
W^a_3&=\{g_2,g_3,g_{12},g_{13},g_{17}\}.
\end{align*}
For agent~3, we use
\begin{align*}
W^3_1&=\{g_1,g_5,g_6,g_{11}\}, \qquad W^3_2=\{g_4,g_7,g_8,g_9,g_{10},g_{13}\},\\
W^3_3&=\{g_2,g_3,g_{12},g_{14},g_{15},g_{16},g_{17}\}.
\end{align*}
Each agent $a$ values every bundle in her partition at one, so from $v_a(M)=3$ we conclude that this is an MMS partition with $\MMS_a=1$.

\paragraph{Round 1: the real allocation.}
We find none of $\reduction{0}$, $\reduction{1}$, and $\reduction{2}$ applicable and obtain the pair condition.
The pair $\{g_1,g_4\}$ is the satisfactory candidate with the largest partner index.
Agents~1 and~2 value this pair at $55/54$, and agent~3 values it at $51/54$.
Suppose that $S_1=\{g_1,g_4\}$ is allocated to agent~1.
Figure~\ref{fig:running-witness} shows how we regroup the witness items for agents~2 and~3 after this allocation.

\begin{figure}[t]
  \centering
  \begingroup
  \begin{tikzpicture}[
      x=1cm,
      y=1cm,
      font=\normalfont\normalsize,
      line cap=round,
      line join=round,
      rw/frame/.style={
          draw=black!75,
          line width=.55pt
      },
      rw/solid/.style={
          draw=black!65,
          fill=black!10,
          line width=.4pt
      },
      rw/fluid/.style={
          pattern=north east lines,
          pattern color=black!25
      },
      rw/arrow/.style={
          -{Stealth[length=2mm,width=1.4mm]},
          line width=.65pt
      },
      rw/unit/.style={
          densely dashed,
          draw=black!65,
          line width=.45pt
      },
      rw/header/.style={
          font=\normalsize\bfseries
      },
      rw/item/.style={
          font=\normalsize,
          inner sep=0pt
      },
      rw/amount/.style={
          fill=white,
          inner sep=1.5pt,
          font=\normalsize
      }
  ]

  \def\rwH{2.25}
  \def\rwB{4.10}

  \newcommand{\rwInitial}[2]{%
    \begin{scope}[yshift=#2cm]
      \node[rw/header,anchor=west]
        at (.40,2.63) {Agent $#1$};

      \foreach \xx/\rr in {.45/1,1.35/2,2.25/3}{%
        \draw[rw/frame]
          (\xx,0) rectangle ++(.65,\rwH);
        \node at ({\xx+.325},1.125) {$1$};
        \node at ({\xx+.325},-.28) {$W^{#1}_{\rr}$};
      }
    \end{scope}%
  }

  \newcommand{\rwWitness}[7]{%
    \begin{scope}[shift={(#1,#2)}]
      \pgfmathsetmacro{\rwF}{\rwH*#3/54}%
      \pgfmathsetmacro{\rwT}{\rwH*#4/54}%
      \pgfmathsetmacro{\rwS}{#4-#3}%

      \fill[white]
        (0,0) rectangle (\rwB,\rwF);
      \path[rw/fluid]
        (0,0) rectangle (\rwB,\rwF);

      \foreach \ii/\vv
        [remember=\rwNext as \rwPrev (initially 0)]
        in {#5}{%
          \pgfmathsetmacro{\rwNext}{
              \rwPrev+\rwB*\vv/\rwS
          }%
          \path[rw/solid]
            (\rwPrev,\rwF) rectangle (\rwNext,\rwT);
          \node[rw/item]
            at ({(\rwPrev+\rwNext)/2},{(\rwF+\rwT)/2})
            {$g_{\ii}$};
      }

      \draw[rw/frame]
        (0,0) rectangle (\rwB,\rwT);

      \node[rw/amount]
        at ({\rwB/2},{\rwF/2}) {fluid $#3/54$};
      \node at ({\rwB/2},-.28) {$#6$};
      \node at ({\rwB/2},2.63) {$#7$};
    \end{scope}%
  }

  \node[
      draw=black!35,
      fill=black!3,
      rounded corners=1.5pt,
      minimum width=15.90cm,
      minimum height=.62cm,
      inner sep=5pt
  ] at (7.95,8.60) {
      \textbf{Real allocation:}\quad
      $S_1=\{g_1,g_4\}$ is allocated to agent $1$.
  };

  \node[rw/header,text width=3.3cm,align=center]
    at (1.68,7.70) {Initial MMS partition};

  \node[rw/header,text width=2.8cm,align=center]
    at (5.30,7.70) {Remove $g_1,g_4$ and repack};

  \node[rw/header,text width=8.6cm,align=center]
    at (11.10,7.70) {Illustrative private witnesses};

  \rwInitial{2}{4.40}

  \rwWitness
    {6.75}{4.40}{13}{54}
    {2/21,3/20}
    {W^2_1}
    {54/54=1}

  \rwWitness
    {11.20}{4.40}{11}{53}
    {5/6,6/6,7/6,8/6,9/6,10/6,11/6}
    {W^2_2}
    {53/54\;\text{(deficit }1/54\text{)}}

  \draw[rw/arrow]
    (3.97,5.55) -- (6.30,5.55);

  \node at (5.135,5.97) {$v_2(S_1)=55/54$};

  \node[align=center] at (5.135,4.93) {
      Remaining total:\\[2pt]
      $107/54<2$
  };

  \draw[rw/unit]
    (6.40,6.65) -- (15.50,6.65);

  \node[anchor=east,inner sep=2pt]
    at (6.41,6.65) {$1$};

  \draw[black!18,line width=.4pt]
    (.15,3.68) -- (15.50,3.68);

  \rwInitial{3}{.72}

  \rwWitness
    {6.75}{.72}{13}{54}
    {2/21,3/20}
    {W^3_1}
    {54/54=1}

  \rwWitness
    {11.20}{.72}{15}{57}
    {5/10,6/8,7/8,8/8,9/8}
    {W^3_2}
    {57/54}

  \draw[rw/arrow]
    (3.97,1.87) -- (6.30,1.87);

  \node at (5.135,2.29) {$v_3(S_1)=51/54$};

  \node[align=center] at (5.135,1.25) {
      Remaining total:\\[2pt]
      $111/54>2$
  };

  \draw[rw/unit]
    (6.40,2.97) -- (15.50,2.97);

  \node[anchor=east,inner sep=2pt]
    at (6.41,2.97) {$1$};

  \begin{scope}[xshift=-.20cm]
    \path[rw/solid]
      (2.00,-.18) rectangle ++(.40,.22);
    \node[anchor=west,inner sep=3pt]
      at (2.44,-.07) {whole solid item};

    \path[rw/fluid]
      (6.33,-.18) rectangle ++(.40,.22);
    \draw[rw/frame]
      (6.33,-.18) rectangle ++(.40,.22);
    \node[anchor=west,inner sep=3pt]
      at (6.77,-.07) {recorded fluid amount};

    \draw[rw/unit]
      (10.85,-.07) -- (11.39,-.07);
    \node[anchor=west,inner sep=3pt]
      at (11.44,-.07) {value one};
  \end{scope}

  \end{tikzpicture}%
  \endgroup

  \caption{
  After $S_1=\{g_1,g_4\}$ is allocated to agent~1, we remove the corresponding witness items of agents~2 and~3 and redistribute the remaining ones.
  All bundle heights use the same value scale, and the dashed line marks value one.
  We record a deficit of exactly $1/54$ for $W^2_2$.
  For illustration, some small witness items are shown as fluid. Solid and fluid describe an item's state, independently of the value classes in Definition~\ref{def:witness-value-classes}.}
  \label{fig:running-witness}
\end{figure}

\paragraph{Round 1: the remaining agents' private witnesses.}
For each $a\in\{2,3\}$, keep the initial $W^a_3$ unchanged and relabel it as the new $W^a_1$.
Remove $g_1$ and $g_4$ from the initial $W^a_1$ and $W^a_2$, respectively, and merge their remainders into the new $W^a_2$.
The resulting bundle values are $1$ and $53/54$ for agent~2, and $1$ and $57/54$ for agent~3.
In Figure~\ref{fig:running-witness}, the fluid items assigned to agent~2's new $W^2_1,W^2_2$ are $\{g_{12},g_{13},g_{17}\}$ and $\{g_{14},g_{15},g_{16}\}$, with total values $13/54$ and $11/54$.
For agent~3, the corresponding fluid items are $\{g_{12},g_{14},g_{15},g_{16},g_{17}\}$ and $\{g_{10},g_{11},g_{13}\}$, with total values $13/54$ and $15/54$.

\paragraph{Round 2.}
The residual instance has $A=\{2,3\}$, $U=\{g_2,g_3,g_5,\ldots,g_{17}\}$, and $k=2$.
Again, we find none of $\reduction{0}$, $\reduction{1}$, and $\reduction{2}$ applicable.
We reject both candidates in their respective tests: agents~2 and~3 value the candidate pair at $27/54$ and $31/54$, respectively, and the candidate triple at $33/54$ and $39/54$, respectively.
We also find rule $\reduction{3}$ inapplicable: its four real goods are worth $24/54$ to agent~2 and $32/54$ to agent~3.
Hence the algorithm starts the bag-filling branch with $\{g_2,g_5,g_7\}$.
After $g_9$ is added, the resulting bundle has value $39/54$ to agent~2 and $47/54$ to agent~3.
The algorithm therefore allocates $S_2=\{g_2,g_5,g_7,g_9\}$ to agent~3.
In agent~2's witness allocation, we remove the items corresponding to the four goods in $S_2$ and combine the remaining items into one witness bundle of value $68/54$.

\paragraph{Round 3.}
Only agent~2 remains; we find every reduction rule inapplicable and reject both the pair and triple candidates.
The algorithm again starts the bag-filling branch with $\{g_3,g_6,g_8\}$ and successively adds $g_{10}$ and $g_{11}$.
It produces
\begin{equation*}
    S_3=\{g_3,g_6,g_8,g_{10},g_{11}\},
    \qquad v_2(S_3)=44/54.
\end{equation*}
The algorithm allocates $S_3$ to agent~2 and terminates.

The three real bundles received by agents~1, 3, and~2 have respective values $55/54$, $47/54$, and $44/54$ to their recipients, so every agent receives at least the threshold $42/54=7/9$.
The unallocated goods $g_{12},\ldots,g_{17}$ may be added arbitrarily to these bundles.
\end{example}

\clearpage
\section*{AI Methodology Statement}

An initial version of this work was posted on arXiv in late 2025.
It established the existence of $7/9$-MMS allocations and provided an FPTAS for computing $(7/9-\varepsilon)$-MMS allocations. 
The algorithms and proofs in that version were developed entirely by the authors without AI assistance.

In subsequent revisions, the authors interacted with models including GPT-6 Astra and GPT-6 Pro to identify potential logical issues and reorganize the proofs. 
These interactions helped identify bugs in the initial version, which the authors subsequently corrected.
The authors later explored, through interactions with GPT-6.1 Pro, the possibility of replacing the FPTAS with a strongly polynomial-time algorithm. GPT-6.1 Pro proposed the algorithmic construction used in the current version. 
The authors verified the algorithm and proofs and assume full responsibility for their content.
The authors also interacted with GPT-5.6 Pro, GPT-6 Pro, and GPT-6.1 Sol to develop the accompanying Lean formalization.

\clearpage
\phantomsection

\end{document}